# Spitzer Mid-Infrared Observations of Seven Bipolar Planetary Nebulae

J.P. Phillips, G. Ramos-Larios,

Instituto de Astronomía y Meteorología, Av. Vallarta No. 2602, Col. Arcos Vallarta, C.P. 44130 Guadalajara, Jalisco, México   e-mail : jpp@astro.iam.udg.mx

**Abstract**

We have investigated the MIR and visual structures of seven bipolar planetary nebulae (BPNe), using imaging and spectroscopy acquired using the *Spitzer Space Telescope* (*SST*), and the Observatorio Astronomico Nacional in Mexico. The results show that the sources are more extended towards longer MIR wavelengths, as well as having higher levels of surface brightness in the 5.8 and 8.0 $\mu$m bands. It is also noted that the 5.8$\mu$m/4.5$\mu$m and 8.0$\mu$m/4.5$\mu$m flux ratios increase with increasing distance from the nuclei of the sources. All of these latter trends may be attributable to emission by polycyclic aromatic hydrocarbons (PAHs) and/or warm dust continua within circum-nebular photo-dissociation regions (PDRs). A corresponding decrease in the flux ratios 8.0$\mu$m/5.8$\mu$m may, by contrast, arise due to changes in the properties of the PAH emitting grains. We note evidence for 8.0 $\mu$m ring-like structures in the envelope of NGC 2346, located in a region beyond the minor axis limits of the ionized envelope. An analysis of the inner two rings shows that whilst they have higher surface brightnesses at longer MIR wavelengths, they are relatively stronger (compared to underlying emission) at 3.6 and 4.5 $\mu$m. There is also evidence for point reflection symmetry along the major axis of the outflow. Finally, the fall-off in surface brightness along the nebular minor axis suggests that progenitor mass-loss rates were more-or-less constant. NGC 6905 shows evidence for a centrally located and unresolved MIR emission excess. We show that whilst flux ratios are inconsistent with stellar and ionized gas components of emission, the emission can be explained in terms of shock excited $H_2$, or grain continuum and PAH emission bands. We provide the deepest image so far published of the outer visual structure of NGC 6445, and compare these results with Spitzer imaging at 8.0 $\mu$m. The distributions are quite distinct, and may indicate collimation of the outer ionized emission by a cylindrical PDR. We finally note that the MIR colours of the lobes and nuclei of the sources are radically different. It is also shown that 3.6 $\mu$m emission in BPNe differs from that for the generality of planetary nebulae. We discuss various mechanisms which may contribute to these differences of colour.

# 1. Introduction

Planetary nebulae (PNe) are characterised by a broad variety of morphologies, ranging from simple classical circular shells (~ 23 % of all PNe; Phillips 2001a), through to elliptical, irregular and bipolar planetary nebulae (BPNe). It has recently become apparent that the latter category of source represents a reasonably large fraction of PNe (of order ~13% (Phillips 2001a), and perhaps even larger (Phillips 2001b)), and that they are frequently associated with high N/O and He/H abundances (Peimbert & Torres-Peimbert 1983; Phillips 2005), high central star temperatures (Corradi & Schwarz 1995; Pottasch et al. 1996; Phillips 2003a), and low mean galactic latitudes (e.g. Corradi & Schwarz 1995; Phillips 2001c); all of which characteristics are consistent with high progenitor masses.

The mechanisms for the formation of such structures are a matter of continuing speculation and uncertainty, however. One of the more popular models for the formation of BPNe involves the development of common-envelope (CE) structures in close binary systems, a phase which typically occurs where the radius $R_P$ of the primary star is > 0.5a (where a is the orbital separation); the mass $M_S$ of the secondary is < 1.7$M_P$; and where 1.2 $\leq$ $M_P$/$M_S$ $\leq$ 1.8 (Pastetter & Ritter 1989). For the progenitors of PNe this then implies that $M_P \geq 1.4$ $M_\odot$ (Soker & Livio 1994). The mass-loss, in such circumstances, tends to be biased towards the orbital plane of the binary, whilst subsequent interaction with central star winds results in a collimated bubble of shocked, higher temperature gas ($T_e \approx 10^6$K), and the development of bipolar lobes.

The success of this and related models is attested by the simulations of Icke, Balick & Frank (1992), Mellema & Frank (1995), Dwarkadas & Balick (1998), García-Segura et al. (1999), Mellema (1997) and others, and has led to a variety of searches for binaries in BPNe (Schaub & Hillwig 2009; Miszalski et al. 2009; Bond et al. 2000 and references therein; see also the summary of known close binary systems provided by De Marco, Hillwig & Smith 2008). This is not, however, the only possible explanation for such sources.

It has been suggested for instance that non-CE (detached) binary systems may also lead to bipolarity (Morris 1981, 1987; Soker & Livio 1994; Mastrodemos & Morris 1998), whilst aspherical winds may develop from single AGB stars (Soker & Harpaz 1992; Frank 1995; Dwakadas & Owocki 2003). It is relevant, in this context, to note the evidence relating to V Hydrae, where the rotational velocity appears close to the break-up speed of the star (e.g. Barnbaum et al. 1995). A comparable situation may occur in many AGB stars as well, just prior to the formation of proto-planetary nebulae (PPNe) (Garcia-Segura et al. 1999). Finally, it has been noted that there is evidence for rotation in the recently ejected envelopes of CRL 2688 (Bieging & Nguyen-Q-Rieu 1988, 1996) and the red-rectangle (Jura et al. 1996). It is therefore conceivable that rotation of the progenitors leads to non-spherically symmetric mass-loss, and results in structures similar to those predicted for the CE models described above.

Although such non-spherical distributions may, yet again, interact with post-AGB winds, and lead to the bilobal structures seen in many BPNe, it is also conceivable that material is transferred from these disks to the poles of the stars, in a manner similar to that occurring in young stellar objects (YSOs; see e.g. Königl 1991). This may, in turn, lead to the development of so-called "X-type" outflows, in which a high density wind is confined to the axis of the outflow, and a lower density (and less collimated) wind arises outside of this regime (Shang et al. 1998; Arce & Sargent 2006; Shu et al. 1988, 1994, 2000; Ostriker & Shu 1995). Comparable trends may also arise as a result of the disk (D) wind models described by Matt et al. (2003), and in the review of Pudritz et al. (2007).

Finally, it is worth noting that the formation of BPNe may be influenced by magnetic fields, and there is indeed evidence for such fields in a variety of PPNe (Zijlstra et al. 1989; Kemball & Diamond 1997; Vlemmings et al. 2005, 2006; Bains et al. 2003; Etoka & Diamond 2004; Szymczak & Gérard 2005; Herpin et al. 2006; Vlemmings & van Langevelde 2008) and PNe (Miranda et al. 2001; Greaves 2002; Jordan et al. 2005; Gomez et al. 2009). It has been argued that toroidal fields may lead to self-confinement of the flows, and the development of structures similar to those we will be describing below (see e.g. Chevalier & Luo 1994; Washimi et al. 2006), whilst magneto-centrifugal models (such as that of Blackman et al. 2001) involve both accretion disks and stellar magnetic fields. Such fields may, in addition, explain a range of other phenomena in these sources, including the formation of jets, ansae, and point-symmetric structures (Garcia-Segura 1997; García-Segura & López 2000).

Certain BPNe also possess very narrow waists; a situation which is difficult to explain in terms of CE modelling, and appears to be particularly prevalent in symbiotic nebulae (see e.g. Corradi & Schwarz 1995). It is possible, for these cases, that mass is being transferred from the AGB star to a main-sequence or white-dwarf companion, and that this leads to the formation of compact accretion disks, and higher levels of outflow collimation (Soker & Rappaport 2000).

It is therefore apparent that a variety of processes may be responsible for creating BPNe (see e.g. Balick & Frank (2002) for a summary of several of these models) – and indeed, the large range of properties of these sources makes it likely that more than one mechanism is at work. Several characteristics, however, are observed to be common to many of the outflows.

In addition to the high N and He abundances, and the large central star temperatures cited above, many BPNe possess higher levels of CO and $H_2$ emission (Huggins et al. 1996; Kastner et al. 1996), nuclear regions with large emission measures, and enhanced dust emission/extinction (e.g.; Phillips & Mampaso 1988; Phillips & Ramos-Larios 2008a). Similarly, observed gas outflow velocities are of order $18.2 \pm 1.1$ km s$^{-1}$ - a value at the lower end of the planetary nebula range; spherical nebulae, for instance, have values $V_{EXP} \cong 24.8 \pm 1$ km s$^{-1}$ (Phillips 2002). The wings of the lines often show velocities extending up to ~ 370 km s$^{-1}$, however (Corradi & Schwarz 1995). Finally, although the sizes of the central condensations are comparable to those of other categories of PNe, the lobal structures can be extraordinarily large (Corradi & Schwarz 1995; Phillips 2003b). Both the kinematic properties and sizes of the sources appear to be highly unusual, therefore, and distinguish these outflows from other categories of PNe.

We report the mapping of eight of these outflows in the mid-infrared (MIR), making use of observations taken with the Spitzer Space Telescope, and the Observatorio Astronomico Nacional (OAN) in Mexico. We shall, as a result, note the presence of probable photo-dissociation regions (PDRs) about many of the sources – one of which may be responsible for collimating the large scale structure in NGC 6445; find strong (and unresolved) longer wave emission in the core NGC 6905, suggesting the presence of a dust envelope very close to the central star; discern point-reflection symmetry, and possible ring structures in the PDR of NGC 2346.

We shall also note that the colour indices of these sources are quite different from those for the generality of PNe, and that the indices of the lobes and nuclei appear at variance with one another.

**2. Observations**

We shall, in the following, be analysing observations of seven bipolar planetary nebulae based on narrow and broad band visual imaging (NGC 6440); and imaging with the Infrared Array Camera (IRAC) of the *SST* (JnEr 1, Mz 1, NGC 2346, NGC 2818, NGC 3699, NGC 6445, and NGC 6905).

Visual observations of NGC 6445 were obtained at the Observatorio Astronomico Nacional (OAN-SPM), based in the Sierra de San Pedro Martir, Baja California, Mexico. Narrow-band H$\alpha$ $\lambda$6563Å and [NII] $\lambda$6583Å images, as well as direct imaging without a filter, were taken on August 20 2009, using a Photometrics SITe1 CCD camera with 1024 × 1024 array detector mounted at the f/13.5 focus of the 1.5-m telescope. A total of three exposures were undertaken in each of the narrow band filters, both of which were mounted in the RUCA filter wheel (Zazueta et al. 2000), yielding exposure periods of 1200 s for H$\alpha$ and [N II]. A total of four images were also obtained without use of a filter, each with an exposure period of 10 s. The data were then bias-subtracted and flat-fielded using twilight flats, all of which operations were performed using standard IRAF software. The plate scale was 0.5 arcsec per pixel after employing 2×2 binning, whilst the field size was 4.2 x 4.2 arcmin$^2$. Finally, the full width at half-maximum of nearby field stars indicated a seeing of 1.6 arcsec.

The MIR imaging products for NGC 3699 and JnEr 1 derive from *SST* program 30285 ("Spitzer observations of planetary nebulae 2"), whilst imaging for the remaining sources derives from program 68 (Studying Stellar Ejecta on the Large Scale using SIRTF-IRAC). The Principal Investigator for both programs was Giovanni Fazio. Observations were undertaken using the Infrared Array Camera (IRAC; Fazio et al. 2004), and employed filters having isophotal wavelengths (and bandwidths $\Delta\lambda$) of 3.550 $\mu$m ($\Delta\lambda$ = 0.75 $\mu$m), 4.493 $\mu$m ($\Delta\lambda$ = 1.9015 $\mu$m), 5.731 $\mu$m ($\Delta\lambda$ = 1.425 $\mu$m) and 7.872 $\mu$m ($\Delta\lambda$ = 2.905 $\mu$m). The normal spatial resolution for this instrument varies between ~1.7 and ~2 arcsec (Fazio et al. 2004), and is reasonably similar in all of the bands, although there is a stronger diffraction halo at 8 $\mu$m than in the other IRAC bands. This leads to differences between the point source functions (PSFs) at ~0.1 peak flux. The observations have a spatial resolution of 1.2 arcsec/pixel.

Details of the processing and reduction of these data have been described in the IRAC data handbook (available at http://ssc.spitzer.caltech.edu/irac/dh/iracdatahandbook_3.0.pdf), and in previous papers by ourselves (e.g. Phillips et al. 2009; Ramos-Larios & Phillips 2009) and others, and involves the creation of Basic Calibrated Data (BCD) files, followed by post-BCD processing; the latter permitting removal of defects; refinement of the pointing, and the production of mosaics based on multiple AORs (Astronomical Observation Requests).

We have used these data to produce colour coded combined images in three of the four IRAC bands, where 3.6 $\mu$m results are represented as blue, 4.5 $\mu$m as green, and 8.0 $\mu$m fluxes are indicated by red (see Fig. 1). The 5.8 $\mu$m results were not used for this purpose given that they appear to be significantly more noisy. Several of these results have also been processed using unsharp masking techniques, whereby a blurred or "unsharp" positive of the original image is combined with the negative. This leads to a significant enhancement of higher spatial frequency components, and an apparent "sharpening" of the image (see e.g. Levi 1974).

The background emission appears be reasonably uniform, and consists of both instrumental noise, and large scale emission from diffuse interstellar gas and HII regions. In certain cases, particularly towards shorter wavelengths, we also note that the stellar background can become significant. This leads to a fluctuating veil of weak point source functions which may, in certain cases, be almost spatially

continuous – the individual PSFs overlap within the relevant image planes.

Contour mapping is provided for all of the sources, for which we have set the lowest mapping contours at $3\sigma$ background noise levels or greater. The intrinsic flux $E_n$ corresponding to contour $n$ takes a value

$$E_n = A\,10^{(n-1)\Delta C} - B \quad \text{MJy sr}^{-1} \quad \ldots\ldots(1)$$

where A is a constant, $n = 1$ corresponds to the lowest (i.e. the outermost) contour level, B is the background, and $\Delta C$ is the logarithmic separation between contours. The contours are therefore defined in terms of A, $\Delta C$, and B, and these values are specified in the captions to the figures.

Profiles through the sources have been produced in all of the IRAC bands; a process which required the elimination of the background described above, and any associated spatial gradients (particularly important in the 5.8 and 8.0 µm bands). These results were then processed so as to indicate the variation of 3.6µm/4.5µm, 5.8µm/4.5µm and 8.0µm/4.5µm ratios with distance from the nebular nuclei. The rationale behind this is based on the fact that many PNe possess strong polycyclic aromatic hydrocarbon (PAH) emission bands at 3.3, 6.2, 7.7 and 8.6 µm, located in the 3.6, 5.8 and 8.0 µm IRAC bands. Given that the 4.5 µm band is usually dominated by the bremsstrahlung continuum, as well as by a variety of molecular and ionic transitions, such ratios enable us to gain an insight into the importance of the PAH carrying grains and their distributions within the shells (see e.g. Phillips & Ramos-Larios 2008a, 2008b).

We have also obtained contour mapping of the 8.0µm/4.5µm and 8.0µm/5.8µm flux ratios for the cases of NGC 2818 and NGC 6905. This was undertaken by estimating the levels of background emission, removing these from the 4.5, 5.8, and 8.0 µm images, and subsequently setting values at $< 3\sigma_{rms}$ noise levels to zero. The maps were then ratioed on a pixel-by-pixel basis, and the results contoured using standard IRAF programs. Contour levels are given through $R_n = A\,10^{(n-1)B}$, where the parameters (A, B) are, yet again, provided in the captions to the figures.

These flux ratios are open to uncertainties associated with scattering between the detector and multiplexer (Cohen et al. 2007); a problem which requires one to apply the flux corrections described in Table 5.7 of the handbook. These appear to be of maximum order 0.944 at 3.6 μm, 0.937 at 4.5 μm, 0.772 at 5.8 μm and 0.737 at 8.0 μm, although the correction also depends upon the surface brightness distribution of the source under consideration. The handbook concludes that "this remains one of the largest outstanding calibration problems of IRAC".

We have, in the face of these uncertainties, chosen to leave the flux ratio mapping and profiles unchanged. The maximum correction factors for the ratios are likely to be > 0.8, but less than unity, and ignoring this correction has little effect upon our interpretation of the results.

We have, finally, provided photometry for all of these sources, as well as estimates of flux for the centres and lobes of NGC 2818, NGC 6072, and NGC 6905. These values were estimated by integrating within circular or polygonal apertures centred upon the sources; estimating the levels of background at various regions about the nebulae; and applying the instrumental scattering corrections described above. We have been careful to remove the contribution of central and field stars (particularly problematic in the cases of NGC 2346 and Mz 1, although also contributing to fluxes in NGC 6445 and NGC 3699), and determined magnitudes using the α Lyrae calibration of Reach et al. (2005). The results are presented in Table 1.

## 3. The Structures of Bipolar Nebulae in the Mid-Infrared

### 3.1 NGC 2346

#### 3.1.1 The Structure of the Source in the MIR

The planetary nebula NGC 2346 possesses what is in many ways a classic bipolar morphology (see e.g. Balick 1987). The wings or lobes of the structure, for instance, have an ellipsoidal configuration; a structure which is also noted in the position-velocity (PV) diagrams of Walsh, Meaburn & Whitehead (1991). Such diagrams suggest that the lobes are closed at the radial extremities of the optical emission, located at a distance of ~1 arcmin from the central star. The MIR lobes will be shown to extend to even greater distances, however, whilst deep optical imaging by Corradi & Schwarz (1994) shows what appear to be diffuse extensions to the structures, extending to r >1.8 arcmin from the central star. These latter observations may correspond to scattering within pre-existing neutral structures, and they are certainly comparable, in size and orientation, to certain of the MIR features to be described below. It is therefore possible that while a bubble of hot, shocked stellar wind has progressed part of the way into the lobes, the lobes may represent pre-formed structures created during an earlier mass-loss phase.

It has been suggested that many bipolar structures arise as a result of CE mass-loss in binary systems (see Sect. 1), and this source offers further evidence in favour of this hypothesis. The central star has a spectral type A which is too cool to ionise the high excitation envelope, and also shows evidence for being one component of a single-line spectroscopic binary (Mendez 1978; Mendez & Niemela 1981). It is therefore clear that the companion of the A-type star is likely to represent the central ionising star of the nebula. There is, apart from this, also evidence that the nebula is very dusty indeed, and that it contains a high mass fraction of molecular material. Thus for instance, the source represents the only PN in which the ejection and dispersion of a dust cloud has been observed over secularly short periods of time (~ 4 yrs), leading to periodic eclipses with a depth of $\Delta V \sim 2$ mag (see e.g. the visual observations of Kohoutek 1982, 1983; Mendez et al. 1982, 1984; Acker & Janiewicz 1985, Schaeffer 1985; Costero et al. 1986; Jasniewicz & Acker 1986; the UV observations of Feibelman & Aller 1983; and the NIR observations of Roth et al. 1984). The presence of dust is also testified through the IRAS observations of Pottasch et al. (1984), and more recently, through the 24 $\mu$m and 70 $\mu$m MIPS imaging of Su et al. (2004), which shows enhanced 24 $\mu$m emission from the unresolved central core, and strong 70 $\mu$m emission from a circumstellar torus

Similarly strong $H_2$ S(1) and CO emission has been noted from the inner toroidal regime (i.e. the minor axis regions within the optical limits of the outflow), as well as along outer portions of the nebular lobes (see e.g. Zuckerman & Gatley 1988; Kastner et al. 1996; Vicini et al. 1999; and Arias et al. 2001 for spectroscopy and mapping of the $H_2$, and Huggins & Healy 1986; Healy & Huggins 1988, and Bachiller 1989 for observations of the J=1-0 and J=2-1 transitions of CO). Analysis of these results suggests shocked $H_2$ masses ~1.3 $10^{-4}$ $M_\odot$ (Arias et al. 2001), total masses of molecular gas ~0.34→1.85 $M_\odot$, and corresponding molecular-to-ionised mass ratios ranging from ~0.4 to ~80 (Arias et al. 2001; Healy & Huggins 1988; Huggins et al. 1996). The CO and $H_2$ densities appear to be in the region of < $10^4$ $cm^{-3}$ (Bachiller et al. 1989; Vicini et al. 1999; Arias 2001). Finally, although the CO and $H_2$ emission regimes are somewhat different, and the transitions of these molecules arise due to differing excitation mechanisms, it is clear that the two regimes have closely similar kinematics (Bachiller et al. 1989; Arias et al. 2001).

The present Spitzer results are illustrated in Fig. 2 and reveal the presence of comparable structures in all four of the IRAC bands. There is evidence however for an evolution in source dimensions, with the size at 8.0 $\mu$m being significantly greater than at shorter IRAC wavelengths. This apparent increase in size may, in part, be attributable to the limiting sensitivity of the Spitzer imaging, and the greater surface brightnesses prevailing at longer IRAC wavelengths. Put at its simplest, where the outer parts of the envelope are too faint to detect at 3.6 and 4.5 $\mu$m, but are seen at longer wavelengths as a result of increased levels of surface brightness, then this of itself may result in apparent increases in size. Such an explanation is likely to represent only one of the reasons for the variations observed here, however.

Thus, optical images of the source show that the minor axis limits are very sharply defined indeed; a feature which suggests the presence of unresolved ionisation fronts, and an enveloping neutral regime. Similar minor axis limits, with a V-shaped morphology, are to be noted in our 3.6 and 4.5 $\mu$m imaging, wherein it is apparent that very little emission is present at distances > 24 arcsec from the nucleus. This conclusion is further reinforced by profiles through the nucleus of the source (Figs. 3 & 4), whence it is apparent that both the 3.6 and 4.5 $\mu$m trends are very similar, and fall-off steeply beyond the optical minor-axis limits of the source.

Such behaviour at shorter wavelengths is broadly as would be expected for this source, given that the bands are frequently dominated by Bremsstrahlung emission, and various ionic transitions – lines such as Brα λ4.052 μm, [Mg IV] λ4.49 μm, and [ArVI] λ4.53 μm may be important in higher excitation PNe, for instance. It is also possible that fluxes are affected by the v = 1-0 O(5), and v = 0-0 S(6)→S(13) transitions of $H_2$, as well as a 3.3 μm band deriving from PAHs. The latter component is normally quite weak in Galactic PNe, however.

As one progresses to larger IRAC wavelengths, then it is clear that surface brightnesses increase (Figs. 2, 3, 4), envelope dimensions appear larger, and emission occurs beyond the optically defined limits of the source. This is apparent for instance in the 5.8 and 8.0 μm contour maps, where emission is seen to greater distances along the major and minor axes of the source, and in the profiles in Fig. 3, where it is apparent that 5.8 and 8.0 μm trends are similar, and extend over at least ±100 arcsec.

Finally, the extension of MIR emission outside of the visual ionised structure is evident in Fig. 5, where we have superimposed an image from the Hubble Space Telescope (HST) over contouring of the present 8.0 μm results. The contour parameters are defined so as to emphasise weaker components of emission, whilst the HST image is taken from http://hubblesite.org/gallery/album/entire/pr1999035d/. In this latter case, blue corresponds to filter F502N ([OIII]), green results are for filter F656N (Hα), and red emission is for filter F658N ([NII]).

The presence of enhanced longer wave emission, and MIR components of flux outside of the ionised regime, is reminiscent of the trends observed in other PNe and HII regions (see e.g. Phillips & Ramos-Larios 2008a,b,c; Phillips & Perez-Grana 2009), where they are normally considered as arising in photo-dissociative regimes (PDRs), and to signal the presence of the 6.2, 7.7 and 8.6 PAH band features. Such a circumstance will only arise where C/O > 1, however; a situation which is likely to occur where progenitor masses are large and of order ≈2-4 $M_\odot$. Under these circumstances, the third dredge-up of materials from the cores of AGB stars tends to increase the surface abundances of He, C and s-process elements, and lead to values of C/O which are greater than unity (see e.g. Renzini & Voli 1981; Marigo et al. 1998; van den Hoek & Groenewegen 1997). Given that most BPNe are likely to arise from precisely this category of star (see e.g.

Phillips 2005, and Sect. 1), then it seems possible that C/O ratios are greater than unity in a good fraction of the present sources. More concrete evidence for such abundances appears to be a little hard to come by, although we note that C/O ratios are ~1.6 in NGC 2818 (Phillips 2003c, & references therein), and ~1 in NGC 6440 (van Hoof 2000). There appears to have been no prior detection of IR band features in any of these sources (Casassus et al. 2001).

As stellar masses exceed ~4 $M_\odot$, on the other hand, then O-rich stars may dominate at higher metallicities ([Fe/H] > 0.01) (see e.g. Marigo 2008; Herwig 2005). For a normal stellar mass function $\xi(M) \propto M^{-2.7}$ (Kroupa et al. 1993; appropriate for 1 < $M/M_\odot$ < 100), and allowing for an upper limit progenitor mass of ~8 $M_\odot$ (higher mass stars are believed to form supernovae), then this may imply that ~30 % of stars having M > 2 $M_\odot$ have a C/O ratio which is less than unity – a proportion which is likely be typical, as well, for the progenitors of the BPNe.

Dust continua with temperatures 50-150 K appear to dominate the FIR continua of the majority of PNe (e.g. Stasinska & Szczerba 1999), and may lead to increasing fluxes to larger MIR wavelengths; although the thermal balance of normal sized grains (radii ≈ 0.1 $\mu$m) is unlikely to result in comparably extended emission – particularly where they are heated by the central star radiation field (see e.g. Phillips & Ramos-Larios 2008a). It is conceivable however that very small grains with low thermal capacity, in which the absorption of individual photons causes large excursions in temperature (Draine 2003), could lead to extended continua similar to those observed here (see e.g. Phillips & Ramos-Larios 2008a). Such grains may also be responsible for the extended NIR halos around several PNe (see e.g. Phillips & Ramos-Larios 2005, 2006, and references therein). Note that higher temperature grain continua, where they exist, are likely to be very weak at MIR wavelengths, or concentrated close to the central star. They may lead to stronger shorter wave emission in a very few PNe.

It has been noted that the 8.0$\mu$m/4.5$\mu$m and 5.8$\mu$m/4.5$\mu$m flux ratios of many PNe increase with increasing radial distance from the nuclei; a situation which is confirmed along the minor axis of the present source (Fig. 3), but is a little less obvious along orthogonal directions (Fig. 4). There may, in addition, be evidence for particularly strong 8.0μm/4.5μm ratios close to the nucleus (at relative position RP ~ 7.4 arcsec) (see Fig. 3), although some care must be taken in interpreting these results. The nucleus of the source is saturated, and this, together with the slightly differing PSFs of the IRAC bands; possible errors in the registration of the images; non-symmetric diffraction patterns about the central A-type star; and the steepness in the fall-off in surface-brightness, may lead to precisely the effects which are noted in Fig. 3.

These large scale variations in 8.0μm/4.5μm and 5.8μm/4.5μm ratios can be explained in terms of increasing fluxes due to the PAH emission features. Thus, whilst the 5.8 μm band contains the 6.2 μm PAH feature, and fluxes at 8.0 μm are affected by the 7.7 and 8.6 μm bands, the 4.5 μm band is devoid of any such emission. The 8.0μm/4.5μm and 5.8μm/4.5μm ratios therefore give a good indication of the importance of the PAH emission bands, particularly at larger distance from the nuclei where grain continuum components may be reduced. Similarly, whilst the central portions of the source have the highest levels of line and bremsstrahlung emission, which likely dominate the fluxes at 4.5 μm, the outer portions will be more affected by the cooler, more neutral regions of emission – including the PDRs enveloping the ionised regimes. One therefore anticipates that both of these ratios should increase with increasing radial distance as a result of decreases in the 4.5 μm flux, and increasing proportions of PAH band emission.

An alternative explanation for such trends may be that we are observing variations in grain temperatures – that grains have a lower temperature $T_{GR}$ the further away they are from the nucleus of the source. Simple irradiation by the central star would be expected to lead to a variation $T_{GR} \propto R^{-0.3}$ where the grains are composed of carbon, and have dimensions $a$ ~ 0.1 μm; although similarly shallow gradients are to be expected for other sizes and compositions as well. This, combined with the possibility that we are dealing with the Wien portion of the emission spectrum, may be sufficient to lead to marked variations in MIR emission ratios.

Finally, it is worth noting that the fact that such increases are less strong along the major axis of the source (i.e. along the lobes) is presumably indicative of increased levels of ionised emission, and ionisation and compression of the AGB mass-loss envelope.

The mapping at MIR wavelengths is therefore consistent with trends at longer (24 and 70 μm) MIPS wavelengths (Su et al. 2004), where it has been noted that emission extends well beyond the ionised boundary, and has a somewhat boxy overall appearance.

**3.1.2 The Radial fall-off of Emission in the Envelope**

The longer wave MIPS results of Su et al. (2004) reveal that surface brightnesses decline at differing rates with increasing distances from the nucleus. Thus, for a radial distance of between 20 and 36 arcsec they find that surface brightnesses decline as $\sim r^{-1}$; between 36 and 80 arcsec it is more like $\sim r^{-2}$; and for larger radial distances it is even steeper. Such results are similar to the 8.0 and 5.8 μm fall-offs observed here, although it is of interest to note that radial variations in surface brightness depend upon the regions of source which are considered. Thus for instance, whilst the 8.0 μm emission within the outer parts of the lobes (r > 14 arcsec) varies as approximately $\sim r^{-1.5}$, the fall-off for distances > 25 arcsec along the minor axis varies as $\sim r^{-2.9}$. If the emission is optically thin, and the minor axis contribution is assumed to arise from a spherically symmetric cloud (or more correctly, from a cloud whose transverse line-of-sight structure is circularly symmetric along the minor axis of the source), then this would imply a fall-off in volume emissivity of close to $\sim r^{-1.9}$. Pursuing this argument still further, one can say that where the emission derives from grains; where excitation mechanisms and efficiencies are invariant with radius; and where grain properties and sizes are similar, and dust/gas mass ratios are invariant, then the variation in volume emissivity would suggest a comparable fall-off in gas densities. It is important to list these assumptions explicitly, since previous analyses have often failed to do so (e.g. Su et al. (2004) and Ueta (2006)), and it is entirely possible that grain properties evolve with both time and position, depending upon the characteristics of the progenitor grain-formation regime, and their environment within the AGB mass-loss envelopes. Where this occurs, then such simple analyses of dM/dt would fail.

Where these conditions *are* satisfied, however, then it is apparent that progenitor mass-loss rates may have been reasonably constant: a result which differs from the 160 μm analysis of Su et al. (who suggest an $\sim r^{-3}$ density fall-off), and the similar analyses for NGC 650 (Ueta 2006; Ramos-Larios, Phillips & Cuesta 2008), where it is argued that the density fall-off is likely to be even steeper. This difference may arise for two primary reasons. In the first place, the 160 $\mu$m fluxes of Su et al. are likely to derive from a single population of grains having temperatures ~25 K. This may permit a more reliable analysis than is possible in the MIR, where multiple contributions (grain continua/PAH bands/ionic and molecular transitions) occur. On the other hand, the Su et al. trends extend out to 100 arcsec from the centre, and have a spatial resolution of 40 arcsec. It is therefore possible that they are affected by the ionised lobes (which extend out to ±60 arcsec from the nucleus), and that their gradients in surface brightness differ from those of the undisturbed superwind material.

### 3.1.3 The Detection of Possible Faint Structural Characteristics

It is finally worth noting that although Walsh, Meaburn & Whitehead (1991) see closure of the lobes in their various spatio-kinematic diagrams, there is no evidence that the walls of these structures converge at large radial distances. Rather, the present 8.0 $\mu$m results, and particularly those presented in Fig. 6, show that both of these lobes end with rather ragged terminations, extending ~ 0.5 arcmin beyond the kinematic closure radius of Walsh, Meaburn & Whitehead (1991).

There is also evidence for emission at even greater distances along the major axis, at separations from the nucleus of order r ~2.5 arcmin (see e.g. Fig. 5). These latter features, should they be confirmed, may be similar to those observed in M 2-9 ((Schwarz et al. 1997) and other BPNe. However, the structures are only partially detected at the limits of the imaging planes, and have a low S/N. Further deeper [NII] $\lambda$6584 Å observations may be useful in confirming these features, and defining their characteristics

A further weak emission characteristic of interest is also noted in Fig. 5, where it will be noted that there are elongations of the 8.0 $\mu$m contours to the upper left-hand, and lower right-hand sides of the shell – "tentacles" which extend out to ~2.4 arcmin from the nucleus. Although these features are, yet again, at the limits of detection, they appear to be real, and offer evidence for point-reflection symmetry in the (possibly neutral) outer envelope. The contour associated with this feature (1.2098 MJy sr$^{-1}$) has a level 3.1 $\sigma_{rms}$, where $\sigma_{rms}$ is the root mean square background noise.

Finally, we have detected what appear to be faint ring-like structures in the 8.0 μm envelope, evident in extended emission outside of the optical minor-axis limits of the source - and possibly, in addition, extending through the main lobal structures as well. These structures are evident in Fig. 6, where the 8.0 μm image has been processed using unsharp masking to reveal faint and finer features of the nebular envelope (see Sect. 2). The suggested ring-features are identified in the left-hand panel using superimposed circles. Note that the separation of these rings (~10 arcsec) is ~5 times greater than would be predicted for the Airy diffraction pattern of the central star.

It is difficult to obtain reliable profiles through these rings, given the problem of background stellar emission (which affects the 3.6 and 4.5 μm channels in particular), and the necessity of having rather large profile widths in order to increase the levels of S/N. Profiles for the inner two rings are nevertheless indicated in Fig. 7, where we show absolute surface brightnesses (in the upper panel), and the ratio between ring and total surface brightnesses (in the lower panel). The traverse has a width of 8.0 arcsec, crosses the nucleus, and extends along a PA of 77°. Ring surface brightnesses have been determined by fitting the underlying (and smoother) component of emission using a sixth order least-squares polynomial fit, and removing this from the total surface-brightness fall-off.

Two aspects of ring emission will be immediately apparent. The first is that the longer wavelength channels are stronger in terms of absolute surface brightnesses (upper panel), and the second is that they are relatively much weaker when compared to total emission levels (lower panel). Such trends are identical to those noted in NGC 3242 and NGC 7354 (Phillips et al. 2009), and may arise as a result of differences between the properties of grains within the rings and surrounding envelope, and/or in the relative incidence of PAH carrying particles. Radiative acceleration of the grains can also lead to spatial disconnects between the dust and gaseous structures, and differences in the positioning of these features in the various IRAC bands (see e.g. Simis, Icke & Dominik 2001; Mastrodemos & Morris 1999; and Phillips et al. 2009). Care should be taken in interpreting the shifts in Fig. 7, however, given the low S/N of the present results.

Should these results be confirmed, then this would represent the second PN in which such rings have been discovered using the *SST* (Phillips 2009). Indeed, it is possible that the *SST* is uniquely equipped to discover such rings in the extended neutral halos of evolved PNe - regions which are very difficult to observe in the visible or NIR. The origins of such rings are still unclear, although various mechanisms have been summarised by Corradi et al. (2004), Phillips et al. (2009) and others.

### 3.2 NGC 2818

NGC 2818 represents one of the few PN which is verifiably located within a Galactic cluster, and appears to represent a Type I outflow with large He & N abundances (Dufour 1984). The envelope possesses a complex filamentary structure in lower excitation ionic transitions (Phillips & Cuesta 1998), and in the S(1) v = 1-0 $\lambda 2.122$ $\mu$m transition of $H_2$ (Kastner et al. 1996; Schild 1995), as well as having sharp and correlated variations in density (Phillips & Cuesta 1998).

Estimates for the volume filling factor $\varepsilon$ have varied from 0.03 to 0.22 (Boffi & Stanghellini 1994; Kohnoutek et al. 1996; Dufour 1984); and although the latter results are uncertain, they are consistent with the fragmentary appearance of the source. Finally, it has been noted that optical line ratios, and the properties of $H_2$ emission are consistent with shock excitation of the molecular/ionic gas, and imply shock velocities of order 10-30 km s$^{-1}$ for the $H_2$ (Schild 1995) and >110 km s$^{-1}$ for the ionised gas (Phillips & Cuesta 1998).

The present MIR results are illustrated in Fig. 8, and reveal closely similar structures in all of the MIR bands – morphologies which are also comparable to those observed for the visual and $H_2$ S(1) transitions (Schild 1995; Phillips & Cuesta 1998). There is, yet again, as noted in Sect. 3.1, evidence for increasing surface brightnesses towards longer MIR wavelengths (see e.g. the profiles in Fig. 9), and corresponding increases in the dimensions of the source (Fig. 8), both of which may be indicative of increasing PAH emission, and the presence of neutral PDRs about the ionised regime. It is finally worth noting that the 8.0μm/4.5μm and 8.0μm/5.8μm band ratios show systematic trends throughout the envelope (see Fig. 10); the former ratio being weaker at the centre of the source, and increasing radially outwards in all directions, and the latter having its largest value at the centre and decreasing to larger radial distances. The former changes in 8.0μm/4.5μm ratios may, as in the case of NGC 2346, indicate the increasing importance of PDR emission towards the periphery of the envelope, and decreasing levels of ionised (bremmstrahlung and ionic) emission at 4.5 μm. It is also possible that certain of the trends are influenced by $H_2$ emission close to the HI/HII interfaces, where shocks or fluorescence lead to enhanced MIR fluxes. In the case of shock excited $H_2$ fluxes, for instance, the 8.0μm/4.5μm flux ratios are expected to be of order ~1.9 (Reach et al. 2006), although differing shock conditions may result in variations in this ratio. Increasing relative contributions by $H_2$ emission may therefore lead to increases in the 8.0μm/4.5μm ratios. This is likely to be particularly prevalent towards the edges of the source, where the HI/HII shock interface is viewed tangentially to the line of sight.

The variation in the 8.0μm/5.8μm ratio, on the other hand, is somewhat enigmatic, but may occur as a result of variations in the properties of the PAH emitting particles. Where ionisation is important, for instance, then this will tend to enhance C-C stretching vibrations, such as are principally responsible for the 6.2 μm PAH band feature (see e.g. Peeters et al. 2002, and the related laboratory work of Allamandola et al. 1999). Changes the levels of PAH ionisation may therefore create a bias towards the 6.2 μm fluxes, compared to emission associated with the 7.7 and 8.6 μm bands (where fluxes depend upon C-H bending modes as well). These, and other more-or-less complex mechanisms, may possibly explain the small (but significant) changes in 8.0μm/5.8μm ratios noted above.

### 3.3 NGC 3699

The unusual source NGC 3699 appears, up to the present, to have been the subject of very few investigations in any wavelength regime. The best optical images of the nebula appear to be those of Hua et al. (1998), and show the central regions to consist of two convex emission structures located on either side of a low emission-measure "void".

Such a structure is suggestive of a possible high extinction disk oriented almost edge-on to the line-of-sight, with the convex arcs arising from surface illumination and/or ionisation of the disk; a type of morphology which is more commonly associated with YSOs. However, were this to be the case, then one might anticipate that any bilobal emission would be oriented perpendicular to the disk, and extend in a N-S direction.

The imaging of Hua et al. (1998) is somewhat confusing in this regard, since although it shows the presence of extended emission about the central core, the outer structure is complex, diffuse and irregular. The most convincing evidence for outer bipolarity is to be found in their [NII] and [OIII] images, from which we see evidence for outer symmetric arcs along a NE-SW direction – that is, in a direction which is not orthogonal to that of the central core.

Our present MIR image is shown in Fig. 1, where the convex features described above represent the most clearly defined structures. We see evidence that the inner portions of these features (i.e. those on either side of the central "void") appear to have a yellow appearance, testifying to relatively higher levels of emission in the shorter-wave IRAC bands. The outer portions of the arcs are by contrast red in appearance, and this red (i.e. longer wavelength) emission extends somewhat away from the arcs, in a N-S direction.

A straightforward perusal of the combined IRAC colour images therefore suggests that the inner portions of the structure are dominated by ionised emission, which is particularly important in defining shorter wave MIR fluxes, whilst regions beyond the arcs may be neutral, and stronger in the (PAH dominated) 5.8 and 8.0 $\mu$m bands.

Contour maps of the structure are illustrated in Fig. 11, although it is difficult to make too much sense of the 3.6 $\mu$m results, dominated as they are by unrelated field stars. It's clear however that the structure of the source is similar in all of the bands, with little evidence for changes in morphology – although the 8.0 $\mu$m image, as for other sources in this study, shows evidence for significantly greater extension.

Profiles through the source are illustrated in Fig. 9, and these again show the influence of field stars in distorting the 3.6 and 4.5 $\mu$m trends.

We nevertheless see clear evidence for the two internal arcs; a systematic increase in surface brightness with increasing IRAC wavelength; and evidence for an extension of 8.0 μm emission out to at least ~80 arcsec from the nucleus.

### 3.4 NGC 6445

NGC 6445 is one of the best observed of the present crop of BPNe, and has He/H and N/O abundances consistent with a Type I classification (Perinotto 1991; Aller et al. 1973; Peimbert & Torres-Peimbert 1983). Zanstra temperatures $T_Z(HI)$ and $T_Z(HeII)$ are also appreciable, and very closely similar, taking the respective values 182.8 ±3.5 and 187.7 ± 5.8 kK (see Phillips (2003) and references therein); values which again suggest high central star temperatures, and a shell which is opaque to $H^0$ and $He^+$ ionising radiation.

Optical and $H_2$ emission maps show that the predominant emission is located in a rectangular ring (see e.g. Cuesta & Phillips 1999; Kastner et al. 1996), within which extinction appears to be enhanced (Cuesta & Phillips 1999), whilst CO measures suggest a large mass fraction of molecular gas (Huggins et al. 1996). Deeper images also show the presence of lobes extending to the east and west, and possessing a size of 3.1 x 1.9 arcmin$^2$ (see e.g. Schwarz et al. 1992), although the properties and structure of this region are relatively ill-defined.

The Spitzer colour imaging in Fig. 1 shows the central regions of the source, wherein the oblong ring appears to be yellow, and contains appreciable fractions of shorter and longer wave emission, whilst regions immediately outside of the ring appear red, and are dominated by 8.0 μm emission. This suggests that the regions immediately outside of the core may contain appreciable PAH emission, and associated neutral gas.

The larger-scale structure of the region is illustrated in Fig. 12, where we present unsharp masked imaging of the source taken with the 1.5 m telescope in San Pedro Martir (see Sect. 2 for details). For this case, the [NII] λ6583 Å results are indicated as red, the Hα λ6563 imaging is green, and the direct imaging (without filter) is blue. This is, we believe, the best representation of the outer structure to be published so far, and shows outer envelope emission to be dominated by [NII].

Comparative Spitzer imaging is shown in the lower panel of this figure, and is clearly much more noisy, and significantly less "deep". For this latter case, 3.6 µm emission is represented as blue, 4.5 µm emission is green, and 8.0 µm fluxes are indicated as red. Despite the inadequacies of this latter image, several differences are apparent from the results obtained in the visible. It is apparent for instance that the MIR emission (primarily that at 8.0 µm) extends further north and south (i.e. towards the top and bottom) than is the case for [NII] – and may, indeed, extend outside of the limits of the image. On the other hand, it is clear that the optical emission is more extended in an E-W (left-right) direction. It is therefore conceivable that we are seeing MIR emission from a sleeve or tube of neutral material which surrounds, and is perhaps collimating the E-W ionized lobes.

Contour maps are illustrated in Fig. 13, and show the predominance of 3.6-5.8 µm emission within the inner ring, and evidence for 8.0 µm emission outside of the ring, whilst profiles (in Fig. 14) show an interesting tendency for shorter wave emission to be concentrated in the inner portions of the ring (major axis peak-to-peak diameters are of order 30.0 arcsec at 3.6 µm), and 5.8 and 8.0 µm emission to be stronger in the outer portions (peak-to-peak diameters are ~36.5 arcsec at 8.0 µm). As for the image presented in Fig. 1, therefore, it is clear that MIR emission is highly stratified throughout this structure, with the inner parts likely to be ionised, and the outer portions being more dusty, and possibly more neutral.

### 3.5 NGC 6905

The source NGC 6905 has a bright ellipsoidal central shell surmounted by conical extensions, and associated ansae; features which appear to be enhanced in the λ6584 Å transition of [NII] (Cuesta, Phillips & Mampaso 1993). The kinematics of this source have been studied by Cuesta, Phillips & Mampaso (1993) and Sabbadin & Hamzaoglu (1982), whence it is apparent that the inner (and brighter) sectors of the envelope can be modelled in terms of ellipsoidal shell expansion, whilst the lobes may represent regions where the stellar wind is being shock refracted.

The present Spitzer results are illustrated in Fig. 1, where it is interesting to note the marked differences in colour between the nucleus and lobes. This suggests that much of the lobal emission arises at 5.8-8.0 $\mu$m, as is also evidenced from the contour maps in Fig. 15.

Profiles through the major and minor axes of the shell are provided in Fig. 16, whence it is clear that the central star appears to be detected in all of the photometric bands (see the upper panel in particular). Comparison with the stellar profile at RP ~ -39 arcsec (upper panel) implies that the central emission is barely, if at all resolved, and implies that it must be spatially extremely compact. The relative fluxes in Janskys for channels 1, 2, 3 and 4 (i.e. in the channels centred close to 3.6, 4.5, 5.8 and 8.0 $\mu$m) are found to be given by 0.45/0.43/0.45/1.0, from which it appears that fluxes are mostly invariant in the lowest three IRAC bands. It is unclear however what single mechanism could give rise to such a trend. It is apparent for instance, from the analysis of Reach et al. (2006), that such ratios would be inconsistent with PAH band and ionised emission components, whilst the hot central star continuum would lead to ratios closer to 4.9/3.1/1.9/1.0. The nearest possibility for explaining these trends in terms of a single emission mechanism appears to be through shock excited $H_2$ emission (see e.g. the diagnostic diagram of Reach et al. (2006)), although it is always possible that PAH band emission and a grain continuum peaking close to ~5 $\mu$m might, in combination, lead to similar results.

Finally, the IRAC band ratio maps are illustrated in Fig. 17, where it is apparent that there is little discernable variation in 8.0$\mu$m/5.8$\mu$m ratios. This is not however the case for the 8.0$\mu$m/4.5$\mu$m ratios, which change very strongly indeed, and are significantly weaker towards the minor axis limits. The impression obtained from the mapping is that we may be detecting a toroidal or cylindrical edge-on structure; the type of feature which may very well be implicated in the formation of the conical lobes. It is also interesting to note that similar features have been observed in bipolar structures about YSOs (Phillips & Perez-Grana 2009).

**3.6 JnEr 1**

JnEr 1 has a fascinating optical structure, which takes the form of a broad elliptical ring with two inner symmetrically located condensations (see e.g. Machado et al. 1996). The source has, as a result, been classified as having an elliptical morphology, although Bohigas (2001) argues that it may have a slightly tilted bipolar structure similar to those suggested for NGC 6720 (Bryce, Balick & Meaburn 1994) and NGC 6781 (Ramos-Larios, Phillips & Guerrero, in preparation) – sources in which the bipolar lobes are likely to be oriented close to the line-of-sight.

The chemical abundances of Kaler (1983) suggest that N/O $\cong$ 1.2, and that it is a Type I PNe. Bohigas (2001) however determines the abundances N/O $\cong$ 0.39 and He/H $\cong$ 0.165; the latter value consistent with a Type I source, the former in conflict with such a classification. Bohigas suggests, on the basis of these anomalies, that the shell may be chemically inhomogeneous.

The Spitzer imaging for this source is illustrated in Fig. 1, and shows that emission is again dominated by the two interior lobes; features which are highly fragmented, and strongest at 5.8 and 8.0 $\mu$m. The connecting ring, so clearly evident in the visible, is by contrast is very much fainter at MIR wavelengths, and barely visible in this image. Most of the fragments in the ring appear to have blue-green hue indicative of stronger 3.6-4.5 $\mu$m emission. It is interesting, in this context, to note the variation in structure evident in the contour maps in Fig. 18, where it is apparent that 3.6 and 4.5 $\mu$m emission appears more extended than at longer wavelengths, and shows somewhat more emission in the ellipsoidal ring-like envelope.

The fragmentary nature of the emission is similar to that noted in [NII] by Bohigas (2001) (see, in particular, panel B of his Fig. 4), and is also reflected in profiles through the source (Fig. 19), where short-scale variations in surface brightness are seen to be appreciable and complex. Very little if any emission appears to arise from interior portions of the source.

This latter characteristic may be important for our interpretation of the structure. Where the lobes correspond to a type of equatorial rim, for instance, then one would expect to observe much higher levels of emission in interior portions of the source. The fact that such emission is not observed, either in the profiles or images, suggests that the condensations are likely to represent true bipolar ejections. Although we may not be observing a classical bipolar outflow, with lobal wings anchored to a central ring or torus, it is nevertheless likely that mass-ejection was highly collimated at earlier phases of its evolution – or that instabilities in the ejection process lead to symmetric mass expulsions.

### 3.7 Mz 1

Marston et al. (1998) have undertaken a detailed kinematic analysis of the shell of Mz 1, and concluded that it is best modelled in terms of a cylindrical structure with expansion velocity $V_{EXP} \cong 22 \pm 3$ km s$^{-1}$, oriented at an angle of 48° to the line of sight. A ring of material about the waist of the cylinder appears to have densities of order ~1700 cm$^{-3}$, compared to ~400 cm$^{-3}$ for other portions of the core. Fairly well-defined exterior lobes extend orthogonally away from the cylinder, and over radial distances ~ 45 arcsec from the nucleus of the source.

The present Spitzer results, shown in Fig. 1, are unfortunately very strongly affected by background emission. We are therefore unable to reliably assess the long-wave properties of the lobal emission. It is however clear that the structure of the source is very similar to that noted in the H$\alpha$+[NII] imaging of Marston et al. (1998).

The corresponding contour maps are illustrated in Fig. 20, and reveal primary emission to be concentrated at the minor axis limits of the source, with some evidence for more extended lobal emission in the 4.5 and 5.8 $\mu$m maps.

### 4. Discussion

It is apparent that the nebulae illustrated in Fig. 1 and subsequent images (see e.g. Fig. 12) possess a broad variety of outflow structures, and that only three of them (NGC 2346, Mz 1, NGC 2818) have the classical butterfly shapes which are often taken to characterize the BPNe. The fainter, larger scale structure noted in our visual imaging of NGC 6445 also places this source recognizably within the fold, even though the internal shell takes the form of a compact and oblong ring. The sources NGC 3699, JnEr 1 and NGC 6905, on the other hand, might not normally be identified as BPNe, and at least two of them have been identified as having elliptical outflow shells.

Given that the latter sources show evidence for collimated outflow; mass ejection along diametrically opposing directions; and possible circumstellar disks/toroids, however, we regard the sources as being plausible examples of the BPNe phenomenon, and they are so treated in the present analysis.

It is interesting to note that all of these nebulae, irrespective of their designations, have larger overall dimensions at longer MIR wavelengths; a likely consequence of dust continuum and PAH emission from the enveloping PDRs. They also show marked increases in surface brightness to longer IRAC wavelengths. The sources appear, in this respect, to be not too much different from other PNe (see e.g. Phillips & Ramos-Larios 2008a, 2008b). There may however be other ways in which one can distinguish BPNe from other categories of planetary nebulae.

It's plain for instance that many BPNe possess large mass fractions of molecular gas; high levels of excitation; high density shells/disks within the interiors of the outflows; and evidence for large velocities of outflow (see e.g. Sect. 1, and the evidence cited in Sect. 3). They are also frequently associated with Type I nebular abundances indicative of high progenitor (and envelope?) masses, and possess appreciable Zanstra temperatures: the types of temperature expected for highly evolved and high mass central stars. Most (if not all) of these properties are distinct from those of other categories of source – or at least, are present with very much higher frequency within the BPNe (see e.g. Corradi & Schwarz 1995; Peimbert & Torres-Peimbert 1983; Kastner et al. 1996; Phillips 2003a, 2005)

Such differences might be expected to lead to enhanced levels of shocked and/or fluorescently excited $H_2$ emission; increased PAH band and grain emission continua; higher excitation spectra, and corresponding modifications to the MIR colour indices. It is therefore with this in mind that we have illustrated the locations of differing categories of PNe within the [3.6]-[4.5]/[5.8]-[8.0] colour plane (Fig. 21). In this case, non-BPNe examples of MASH and Galactic PNe (i.e. elliptical, round, irregular) are indicated with grey triangles, where the data is taken from Phillips & Ramos-Larios (2009). The indices for the BPNe, on the other hand, are indicated by the orange circles, and derive from our present results (summarized in Table 1) and photometry published by Phillips & Ramos-Larios (2008a) and Hora et al. (2004). This particular figure is likely to have a reasonably high level of verisimilitude, given that uncertainties in calibration tend to cancel each other out. Thus, the errors arising from scattering in the infrared camera (see Sect. 2) lead to maximum corrections of ~0.75 at 5.8 and 8.0 $\mu$m, and ~0.94 at 3.6 and 4.5 $\mu$m. Although levels of correction for the present sources are far from clear, it seems likely that both the 3.6 and 4.5 $\mu$m results will require similar levels of modification. A similar conclusion also applies for the 5.8 and 8.0 $\mu$m channels. The [3.6]-[4.5] and [5.8]-[8.0] indices should therefore be relatively unaffected by such uncertainties.

It is plain, from this latter figure, that there appears to be a difference in the distributions of the BPNe and other sources. Whilst the [5.8]-[8.0] indices extend over a similar range of values, the BPNe tend to have larger mean values for [3.6]-[4.5]. This is also perhaps even clearer in the distribution of [3.6]-[4.5] colours illustrated in Fig. 22, whence it is apparent that the bipolar and non-bipolar sources have markedly differing trends; there are fractionally more non-BPNe sources at lower values of this index than there are in the case of the BPNe. More specifically, we determine that whilst ~51% of non-BPNe sources have [3.6]-[4.5] > 0.8, all of the bipolars are located within this regime.

There are several mechanisms responsible for locating sources within this plane. Thus for instance, strong dust continuum and PAH emission will tend to force nebulae to higher values of [5.8]-[8.0]. The fact that the distributions of bipolar and other PNe appear pretty similar in this regard may therefore suggest that the dust emission properties of the sources are not too dissimilar. The role of dust at shorter wavelengths, however, is somewhat more ambiguous. Whilst the 3.3 $\mu$m PAH feature will tend to reduce indices [3.6]-[4.5], this feature is often quite weak, and not unduly significant. On the other hand, although warm grain continua would have a reverse effect, this requires there to be appreciable components of dust having temperatures T > 700 K or so.

The influence of $H_2$ transitions is similarly difficult to untangle, depending as it does upon the nature of the excitation process, and whether shock or fluorescence is more important. It would seem for instance that $H_2$ fluxes in NGC 6781 (Ramos-Larios and Phillips (in preparation)) are stronger in the 8.0 μm band than they are within the 5.8 μm filter. This, where it is the case, will tend to increase [5.8]-[8.0] indices. The 3.6 and 4.5 μm bandpases also contain significant numbers of $H_2$ transitions, and these, in the case of shocks, are likely to have broadly similar levels of emission – leaving the [3.6]-[4.5] index unchanged.

Finally, pure ionized hydrogen emission (including the Br and Pf lines) will tend to result in higher levels of emission within the 4.5 and 8.0 μm bandpasses: relative fluxes for the 3.6/4.5/5.8/8.0 μm bands are estimated to be 0.25/3.7/0/1 for case B conditions, and temperatures $T_e = 10^4$ K (Reach et al. 2006; Osterbrock 1989). This leads to increases in the [3.6]-[4.5] and [5.8]-[8.0] indices. However, the strongest flux contributions in higher excitation PNe are likely to arise from [ArVI] λ4.530 μm, [ArII] λ6.985 μm, [NeVI] λ7.642, [ArV] λ7.902 μm, [ArIII] λ8.991 μm, and even Mg[IV] λ4.487 μm and [MgV] λ5.610 μm – lines which, taken as a whole, tend to be stronger in the 4.5 and 8.0 μm bands. One therefore expects that both of these indices would, yet again, tend to be increased by such transitions.

So a variety of mechanisms are important in shifting indices to higher values [3.6]-[4.5], and these include ionic transitions and dust continuum emission. The role of $H_2$ is probably neutral (although this emission is likely to be strong in both the 3.6 and 4.5 μm channels), whilst PAH 3.3 μm emission has a reverse effect.

However, whilst dust and line emission may differ between the BPNe and other PNe, and explain the differences noted above, we also note that one further piece of evidence may help to clarify the situation. Where one plots the [4.5]-[8.0] index against [3.6]-[5.8], as in the lower panel of Fig. 21, then it is clear that all of the PNe have similar indices [4.5]-[8.0]. However, the indices [3.6]-[5.8] are, yet again, larger for the BPNe. This trend is further illustrated in Fig. 22, where the differing tendencies of the BPNe and non-BPNe are again clearly in evidence; we find that whilst 39% of non-BPNe have [3.6]-[5.8] > 1.3, 93% of the BPNe are located within this range.

Taking both of these colour-colour trends into account, therefore, it seems clear that the culprit may be identified as the 3.6 μm channel – the BPNe are, relatively speaking, weaker at these wavelengths (i.e. magnitudes [3.6] are larger). By contrast, the trends between 4.5 and 8.0 μm are similar for all categories of PNe. This may then imply that many non-BPNe results are affected by central and field star contaminants, or have larger levels of 3.3 μm PAH band emission. This may arise, in part, as a result of the larger effective temperatures of stars in BPNe, and the corresponding increases in bolometric corrections, and reduction in MIR fluxes. The contamination of non-BPNe sources by central star fluxes is therefore likely to be larger. It is also possible, however, that field star removal was less effective in these sources.

Whatever the case, it is clear that there is a difference in published colours for the BPNe and other sources, and that some further clarification of what these trends imply would be of considerable interest.

Finally, it is to be expected that the lobes and nuclei of the BPNe would also have somewhat differing colours, given that we are observing differing types of structure, with differing kinematics and origins. Thus, the nuclei of the sources might be expected to have greater levels of dust and gas thermal emission, given the frequent presence in these regimes of high emission measure nuclei, and dense and dusty circumstellar shells. The lobes, on the other hand, might tend to be biased in terms of (shock excited) $H_2$ emission. These, and other factors probably account for the differences in colour noted for NGC 6905 (see Fig. 1 & Sect. 3.5).

Can one see any evidence for these differences in the colour-colour planes? To assess this, we have obtained photometry for the lobes and nuclear regions in three of the BPNe, the results for which are summarized in Table 1 and Fig. 21. In the latter case, the lobes are indicated by dark blue diamonds, and the nuclei by lighter blue squares. It is plain, from these results, that there are indeed differences in the colours of the structures. Thus, the lobes of NGC 6905 appear to have values [3.6]-[4.5] which are ~0.57 mag greater than observed in the centre, whilst differences in [3.6]-[5.8] are even greater (~ 0.9 mag). It is therefore apparent that the mechanisms described above are having varying effects upon emission in these zones. Having said this, we see little evidence for consistency between the sources investigated here, and the lobes may be either redder or bluer than the corresponding central structures. It would therefore appear that each of the sources has to be treated on an individual basis, and it is difficult to come to hard-and-fast conclusions concerning emission mechanisms within these regimes.

## 5. Conclusions

We have presented infrared and optical imaging for seven bipolar planetary nebulae. The results show that as well as possessing several properties in common, they also have interesting peculiarities. A compact and unresolved region of enhanced MIR emission is located close to the central star of NGC 6905, for instance. Ratios between fluxes in the differing IRAC bands suggest that it is explainable in terms of shock excited $H_2$ emission, or a combination of warm dust continuum emission and PAH molecular bands. There is also evidence for a minor axis toroidal structure, comparable to a similarly located 70 $\mu$m structure in NGC 2346. NGC 2346, by contrast, shows evidence for an $\sim r^{-2.9}$ fall-off in 8.0 and 5.8 $\mu$m emission along the minor axis of the source, and a slower ($\sim r^{-1.5}$) decline along the lobes. Given that the minor axis variation is likely to refer to the undisturbed AGB envelope, this may imply that the superwind mass-loss rate was more-or-less constant. Such a result differs from the conclusions of Su et al. (2004) based upon longer wave MIPS results, and of Ueta (2006) and Ramos-Larios, Phillips & Cuesta (2008) for the case of NGC 650; although it should be noted that all of these analyses make unverified assumptions, and that the estimates of mass-loss are open to considerable uncertainty.

Finally, we note evidence for point-reflection symmetry along the major axis of NGC 2346, and annular structures similar to those observed in other PNe. The rings are located outside of the minor axis limits of the primary ionised shell, and have higher surface brightnesses in the longer wave IRAC bands. On the other hand, the fractional surface brightnesses of the rings (when compared to total shell emission) increase towards shorter MIR wavelengths – a result which is similar to the trends observed in NGC 7354 and NGC 3242.

We have presented the deepest optical images so far obtained of the larger scale emission structure in NGC 6445, and noted that it differs quite radically from what is seen in the MIR. It is suggested that the 8.0 $\mu$m emission is tracing a north-south PDR; a structure which may have been responsible, in the past, for collimating the ionised regime. Indeed, it is possible that this sleeve-like PDR may still be playing such a role for the fainter visual envelope.

Two of the sources (NGC 3699 and JnEr 1) represent atypical examples of the bipolar phenomenon, in that they have well separated regions of emission with little evidence for collimating structures or outflow lobes. It nevertheless seems likely, in the case of JnEr1, that the brightest condensations were ejected in diametrically opposing directions, suggesting a very much earlier phase of outflow collimation, or non-spherical mass expulsion. The structures in NGC 3699, by contrast, are reminiscent of what is observed in many YSOs, and may imply that similar mechanisms are operating in this source as well. It is suggested that we may be observing a high extinction disk oriented almost edge-on to the line of sight, the lateral surfaces of which are ionised and/or reflecting radiation from the central star.

All of the sources show evidence for increasing dimensions towards longer wavelengths; a trend which can be attributed to increasing emission from the nebular PDRs. There is also evidence for an increase in surface brightnesses in the 5.8 and 8.0 $\mu$m bands – a variation which may arise due to warm dust continua and (in particular) PAH emission bands. We note evidence for increases in the 5.8$\mu$m/4.5$\mu$m and 8.0$\mu$m/4.5$\mu$m flux ratios with distance from the centres of the sources; a trend which may, yet again, be attributable to increasing emission from the PDRs, and decreasing contributions from the ionised regimes.

Finally, a comparison of the colour indices of BPNe with values determined for other categories of nebula shows that values of [3.6]-[5.8] and [3.6]-[4.5] are significantly larger for the BPNe. Much of this difference is associated with fluxes in the 3.6 $\mu$m band, and can be attributed to differences in PAH emission and/or stellar contamination. It is also clear that the lobes and nuclei of the sources often have markedly differing colour indices, indicating corresponding differences in the MIR emission mechanisms. We see little evidence for systematic differences between the lobes and centres, however; the lobes may be either "redder" or "bluer" than the nuclei of the sources.

**Acknowledgements**

This work is based, in part, on observations made with the Spitzer Space Telescope, which is operated by the Jet Propulsion Laboratory, California Institute of Technology under a contract with NASA. GRL acknowledges support from CONACyT (Mexico) grant 132671.

Table 1

MIR Photometry for Eight Galactic Bipolar Planetary Nebulae

| SOURCE | G.C. | SECTOR | 3.6 μm mag | 4.5 μm mag | 5.8 μm mag | 8 μm mag |
|---|---|---|---|---|---|---|
| NGC 6445 | 008.0+03.9 | TOTAL | 8.51 | 7.47 | 6.74 | 5.13 |
| NGC 6905 | 061.4-09.5 | TOTAL | 10.22 | 9.12 | 8.63 | 6.57 |
| NGC 6905 | 061.4-09.5 | CENTRE | 10.77 | 9.76 | 9.41 | 7.55 |
| NGC 6905 | 061.4-09.5 | LOBE(SOUTH) | 12.20 | 10.80 | 10.06 | 7.92 |
| NGC 6905 | 061.4-09.5 | LOBE(NORTH) | 12.43 | 10.85 | 10.18 | 8.02 |
| JnEr 1 | 164.8+31.1 | TOTAL | 10.63 | 9.44 | 8.13 | 7.36 |
| NGC 2346 | 251.6+03.6 | TOTAL | 7.36 | 6.53 | 5.19 | 2.96 |
| NGC 3699 | 292.6+01.2 | TOTAL | 10.87 | 9.40 | 9.51 | 7.72 |
| NGC 2818 | 261.9+08.5 | TOTAL | 9.97 | 8.93 | 7.97 | 7.01 |
| NGC 2818 | 261.9+08.5 | CENTRE | 10.68 | 9.61 | 8.89 | 7.77 |
| NGC 2818 | 261.9+08.5 | LOBE(EAST) | 11.34 | 10.37 | 9.15 | 8.27 |
| NGC 2818 | 261.9+08.5 | LOBE(WEST) | 11.86 | 10.69 | 9.60 | 8.82 |
| Mz 1 | 322.4-02.6 | TOTAL | 9.53 | 8.46 | 6.81 | 6.43 |
| NGC 6072 | 342.1+10.8 | TOTAL | 8.19 | 7.31 | 6.25 | 5.10 |
| NGC 6072 | 342.1+10.8 | CENTRE | 9.51 | 8.09 | 7.02 | 5.88 |
| NGC 6072 | 342.1+10.8 | LOBE(EAST) | 9.27 | 8.54 | 7.63 | 6.38 |
| NGC 6072 | 342.1+10.8 | LOBE(WEST) | 9.47 | 9.22 | 7.87 | 6.89 |

# Figure Captions

**Figure 1**

Mid-infrared imaging of seven bipolar planetary nebulae, obtained from results deriving from the Spitzer Space Telescope, where we have combined imaging in the 3.6 $\mu$m (blue), 4.5 $\mu$m (green) and 8.0 $\mu$m (red) photometric channels.

**Figure 2**

Contour mapping of NGC 2346 in the four IRAC photometric bands, where the contour parameters (A, $\Delta$C, B) are given by (0.15, 0.1446, 0.05) at 3.6 $\mu$m, (0.08, 0.1995, -0.01) for the 4.5 $\mu$m contours, (0.77, 0.1130, 0.38) at 5.8 $\mu$m, and (1.37, 0.1086, 1.05) for the 8.0 $\mu$m results. Note the increase in source dimensions towards longer MIR wavelengths.

**Figure 3**

Profiles through the minor axis of NGC 2346, where we have indicated logarithmic variations in surface brightness in the upper panel, and variations in flux ratios in the lower panel. The directions and widths (8.5 arcsec) of the slices are indicated in the inserted images. It is apparent that the longer wave 5.8 and 8.0 $\mu$m profiles appear more extended than at shorter wavelengths, whilst 8.0$\mu$m/4.5$\mu$m and 5.8$\mu$m/4.5$\mu$m ratios increase with increasing distance from the nucleus.

**Figure 4**

As for Fig. 3, but for the major axis of NGC 2346. Note how the fall-off in surface brightness is relatively smooth and continuous; much more so than is observed along the minor axis of the source (Fig. 5). It is also clear that whilst flux ratios increase with increasing distance from the centre, the trends are less clear and monotonic than are observed along the minor axis (Fig. 5). The width of the profile is again 8.5 arcsec.

**Figure 5**

Superimposition of an HST image of NGC 2346 (taken from the web page http://hubblesite.org/gallery/album/entire/pr1999035d/) upon a map of the source at 8.0 $\mu$m (shaded grey with contours). The contour parameters (A, $\Delta$C, B) for the 8.0 $\mu$m mapping are given by (1.08, 0.0493, 1.05). Note how the 8.0 $\mu$m emission extends well beyond the limits of the optical image, and shows evidence (in narrow extensions to the north (top) and south) for point reflection symmetry.

**Figure 6**

Imaging of NGC 2346 at 8.0 $\mu$m, processed using unsharp masking to show fainter and narrower components of emission. Note, in particular, the evidence for possible rings in the minor axis emission, indicated (in the left-hand panel) using white concentric circles. These rings also appear to cross the inner lobal structures. Although this evidence is highly suggestive, we note that the S/N is low, and further confirmatory observations are required.

**Figure 7**

The surface brightness (upper panel) and relative intensities (ring flux/total flux; lower panel) of the inner two rings of NGC 2346, deriving from profiles with width 8.0 arcsec along position angle of 77° through the nucleus. In both of these cases, 3.6 $\mu$m emission is indicated using blue lines and diamonds; 4.5 $\mu$m emission is represented using green lines and squares; 5.8 $\mu$m emission is indicated using black lines and triangles; and 8.0 $\mu$m emission is represented using red lines and bullets.

**Figure 8**

As in Fig. 2, but for the case of NGC 2818, and contour parameters (A, $\Delta$C, B) of (0.035, 0.1952, 0.0) at 3.6 $\mu$m, (0.04, 0.2279, 0.0) at 4.5 $\mu$m, (0.15, 0.1584, 0.0) at 5.8 $\mu$m, and (0.1, 0.1976, 0.0) for 8.0 $\mu$m.

**Figure 9**

Trends in surface brightness for NGC 2818 (upper panel) and NGC 3699 (lower panel), where the profile widths are 3.7 arcsec for NGC 2818 and 3.1 arcsec for NGC 3699. The directions and widths of the slices are indicated in the inserted images. Both sets of profiles show complex structures, and systematic increases in surface brightness towards longer IRAC wavelengths.

**Figure 10**

Flux ratio mapping for NGC 2818, where the contour parameters (A, B) are given by (0.5, 0.1111) for the upper 8.0$\mu$m/4.5$\mu$m results, and (0.75, 0.0669) for the 8.0$\mu$m/5.8$\mu$m results. Note that the 8.0$\mu$m/4.5$\mu$m ratios are lower towards the centre of the source (where shading is lighter), and increase radially outwards in all directions. In the case of the 8.0$\mu$m/5.8$\mu$m map, on the other hand, it would appear that ratios are larger in the centre, and decrease to larger radial distances.

**Figure 11**

As for Fig. 2, but for the case of NGC 3699, where contour levels (A, $\Delta$C, B) are given by (0.55, 0.0893, 0.2) at 3.6 $\mu$m, (0.5, 0.0939, 0.12) at 4.5 $\mu$m, (1.38, 0.0621, 1.1) at 5.8 $\mu$m, and (3.42, 0.0410, 3.2) for 8.0 $\mu$m.

**Figure 12**

Images of NGC 6445 in the visual (upper panel) and MIR (lower panel). We have, in the former case, combined images taken in [NII] $\lambda$6583 Å (red), H$\alpha$ $\lambda$6563 (green), and without a filter (blue). The lower panel, by contrast, shows results taken at 3.6 $\mu$m (blue), 4.5 $\mu$m (green) and 8.0 $\mu$m (red). Both sets of images have been processed using unsharp masking, and have the same spatial scales and orientations. Note the presence of strong 8.0 $\mu$m emission immediately outside of the central nebular ring (lower panel and Fig. 1), and the differing distributions of visual and MIR emission. It is possible that the extended MIR emission derives from a sleeve or cylinder of neutral material, and that this is responsible for large-scale collimation of the structures in the visible.

**Figure 13**

As for Fig. 2, but for the case of NGC 6445, where contour levels (A, ∆C, B) are given by (1.75, 0.1037, 0.95) at 3.6 μm, (1.4, 0.1283, 0.75) at 4.5 μm, (4.4, 0.0838, 3.2) at 5.8 μm, and (10.6, 0.0875, 9.9) at 8.0 μm. Note the presence of more extended emission at 8.0 μm.

**Figure 14**

As for Fig. 3 but for the case of NGC 6445, where we show profiles through the major axis (upper panel) and minor axis (lower panel) of the central ring. The profile widths are 4.4 arcsec for the major axis profile, and 5.2 arcsec for the minor axis results.

**Figure 15**

As for Fig. 2, but for the case of NGC 6905, where contour levels (A, ∆C, B) are given by (0.2, 0.1219, 0.06) at 3.6 μm, (0.14, 0.1479, 0.0) at 4.5 μm, (0.8, 0.0884, 0.6) at 5.8 μm, and (1.65, 0.1065, 1.1) at 8.0 μm.

**Figure 16**

As for Fig. 3, but for major axis (upper panel) and minor axis profiles (lower panel) through the centre of NGC 6905. The widths of the profiles are 5.3 arcsec. Note the detection of the central star at RP = 0 arcsec, and the strong longer-wave emission associated with this source (see upper panel). This suggests that warm dust grains are located close to the nucleus of the outflow.

**Figure 17**

As for Fig. 9, but for the case of NGC 6905, where the 8.0μm/4.5μm contour parameters (A, B) are given by (2.25, 0.0808), and the 8.0μm/5.8μm parameters are (1.4, 0.0519). Note the decrease in 8.0μm/4.5μm ratios towards either side of the minor axis.

**Figure 18**

As for Fig. 2, but for the case of JnEr 1, where contour levels (A, ∆C, B) are given by (0.05, 0.1945, 0.03) at 3.6 μm, (0.07, 0.1892, 0.04) at 4.5 µm, (1.03, 0.0480, 0.8) at 5.8 µm, and (2.23, 0.0326, 2.06) at 8.0 µm. The lobes of the source are faint, and show much complexity of structure. It is also interesting to note that the lobes are more extended at shorter wavelengths, where there is clearer evidence for emission deriving from the ellipsoidal ring.

**Figure 19**

Surface brightness profiles through the centre of JnEr 1, where the widths of the profiles is 5.5 arcsec. Note the considerable complexity of profile structure, most of which is attributable to the nebula itself (rather than to field stars or noise), and the large ratio between peak lobe and core emission.

**Figure 20**

As for Fig. 2, but for the case of Mz 1, where contour levels (A, $\Delta$C, B) are given by (1.3, 0.0738, 0.63) at 3.6 µm, (0.9, 0.0915, 0.32) at 4.5 µm, (4.0, 0.0442, 3.2) at 5.8 µm, and (9.89, 0.0535, 9.5) at 8.0 µm.

**Figure 21**

The distribution of BPNe and non-BPNe morphologies within the [3.6]-[4.5]/[5.8]-[8.0] plane (upper panel), and the [4.5]-[8.0]/[[3.6]-[5.8] colour plane (lower panel). In both cases, the BPNe are indicated using orange disks, and non-bipolar nebulae by grey triangles. We also note indices for the central regions of three of the bipolar sources (light blue squares), and their corresponding lobes (darker blue diamonds). It will be noted that the BPNe are displaced from other categories of PNe, and that the lobes and nuclei of these sources have what are often markedly differing colours.

**Figure 22**

The variation in the numbers of bipolar and non-bipolar sources as a function of [3.6]-[4.8] (upper panel) and [3.6]-[5.8] (lower panel). The results are normalised to unity. It is apparent that bipolar sources have significantly larger values of these colour indices.

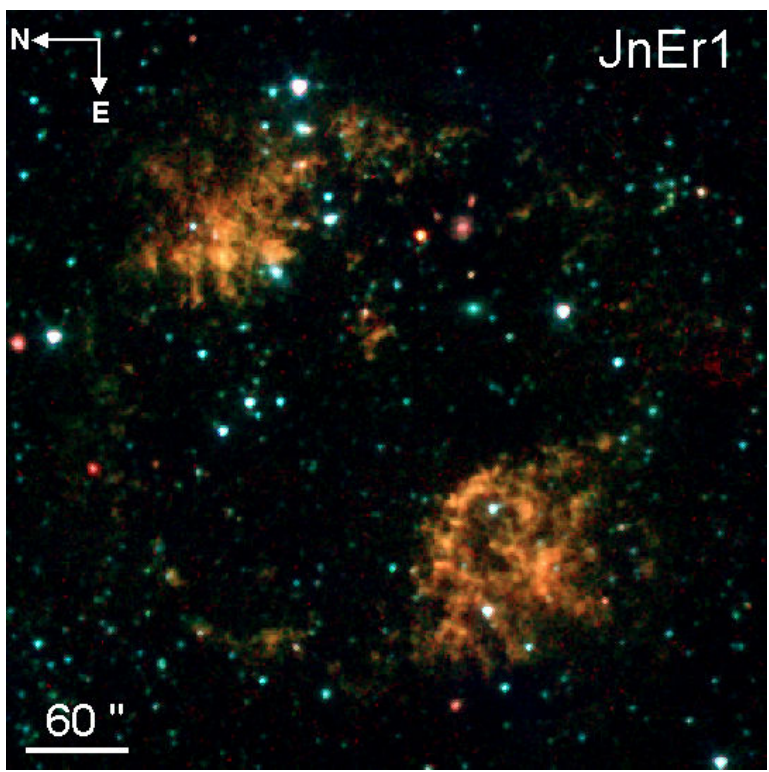 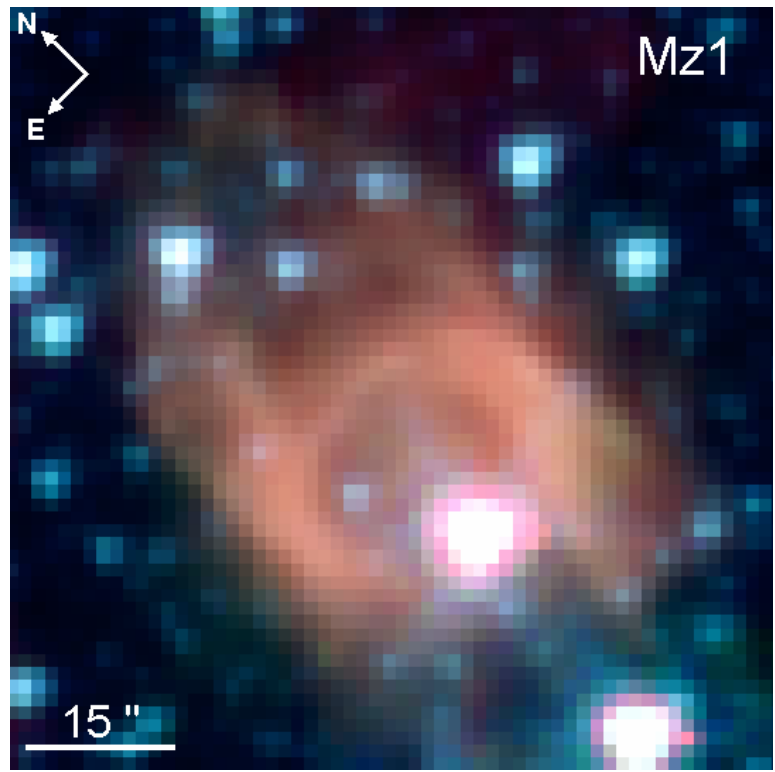
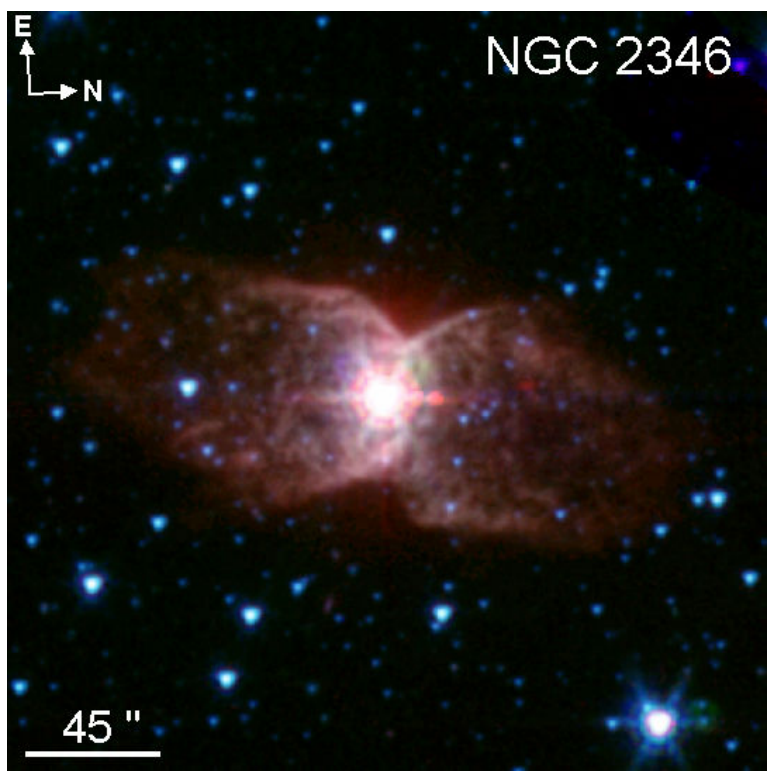 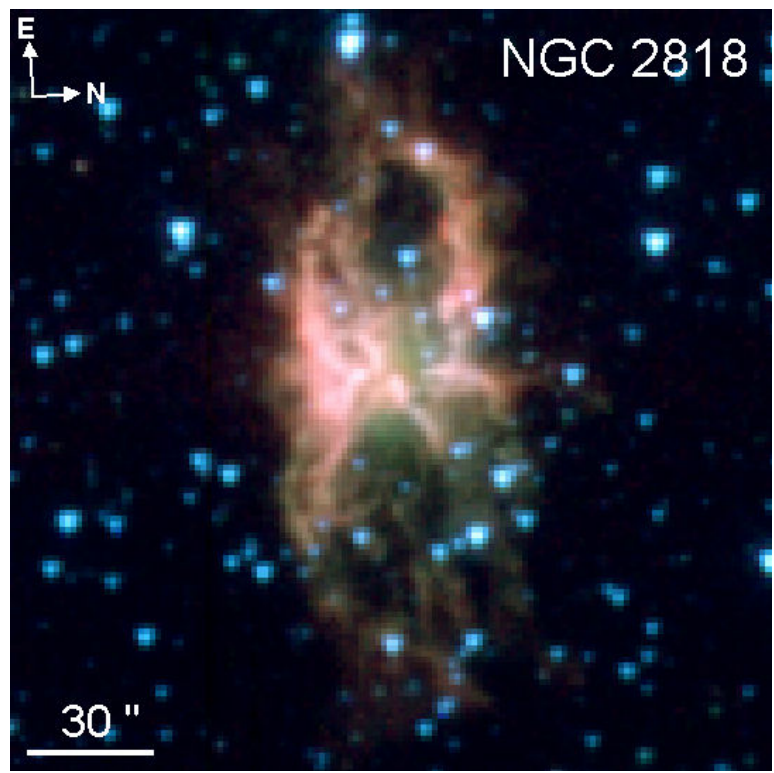

FIGURE 1

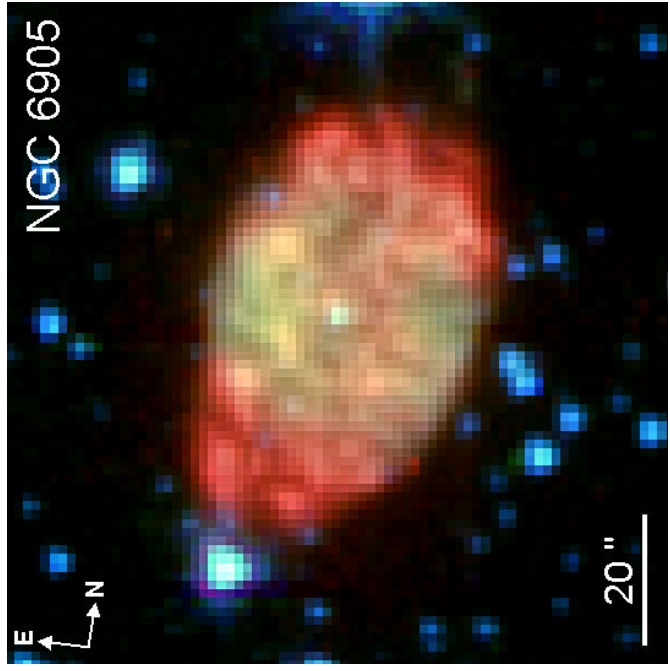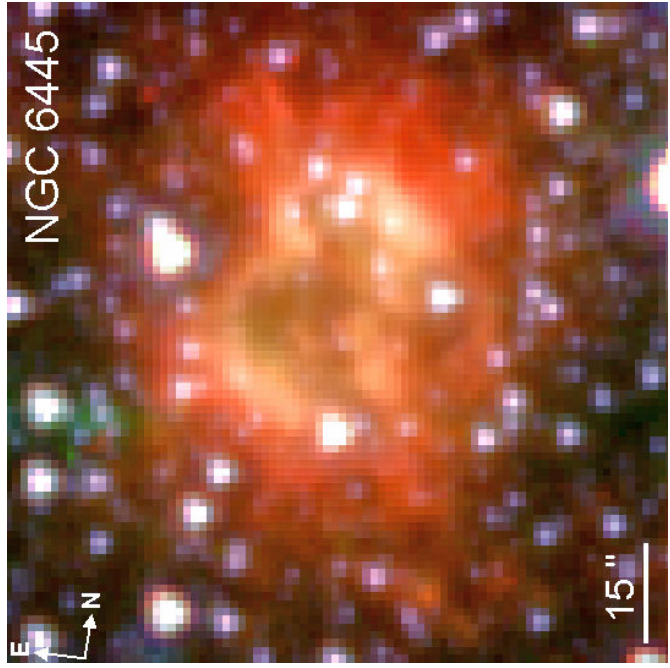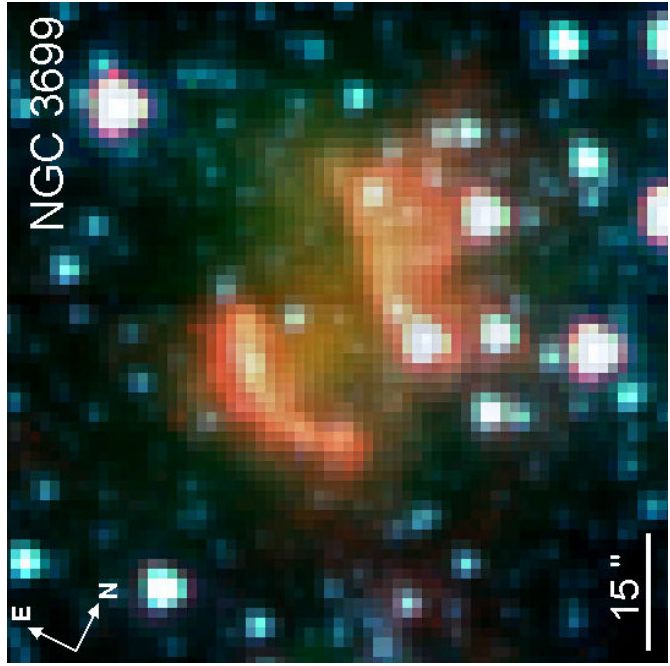

FIGURE 1 (CONT.)

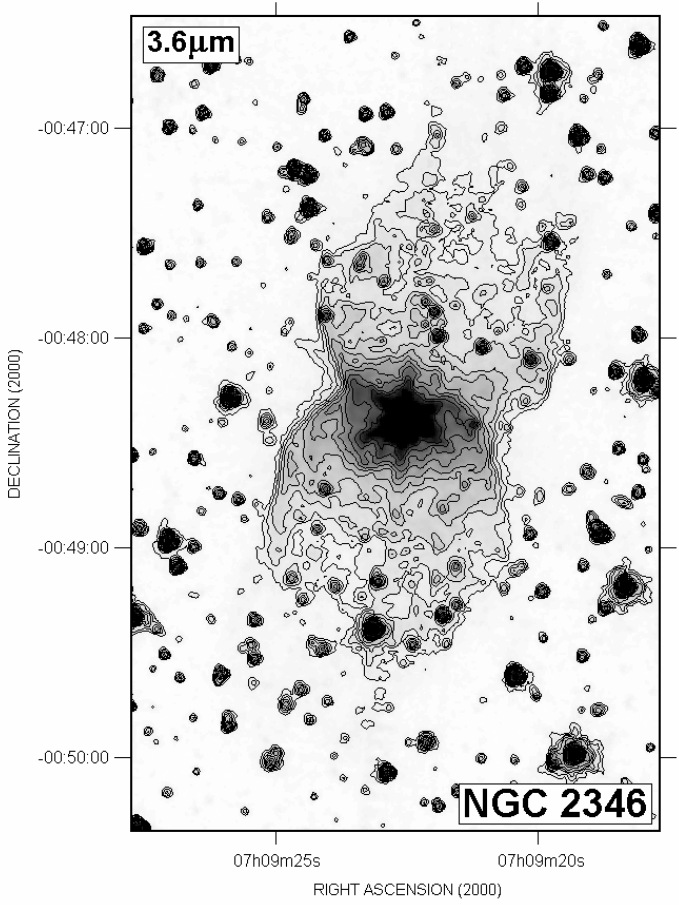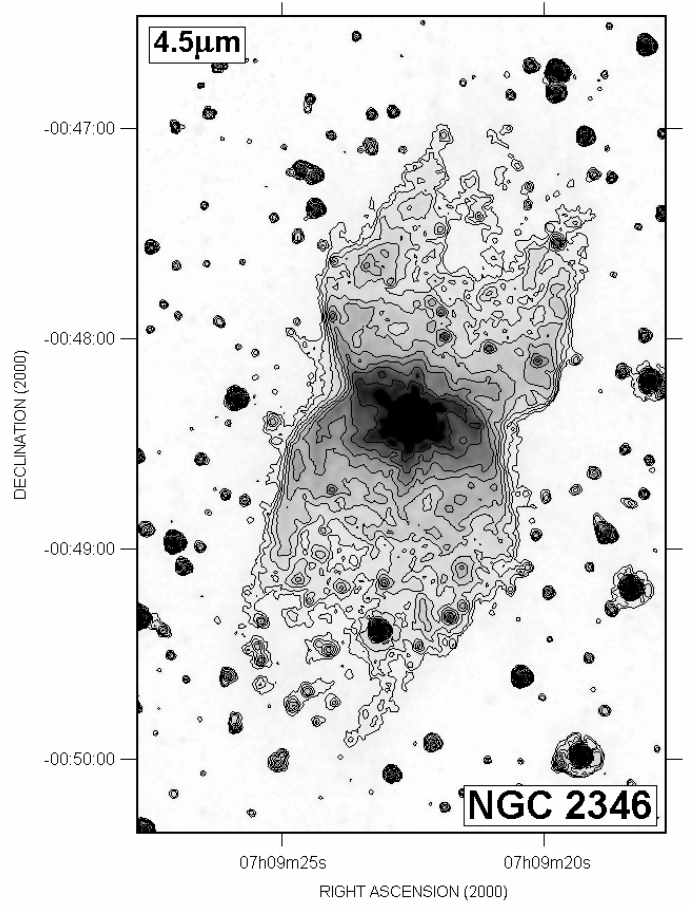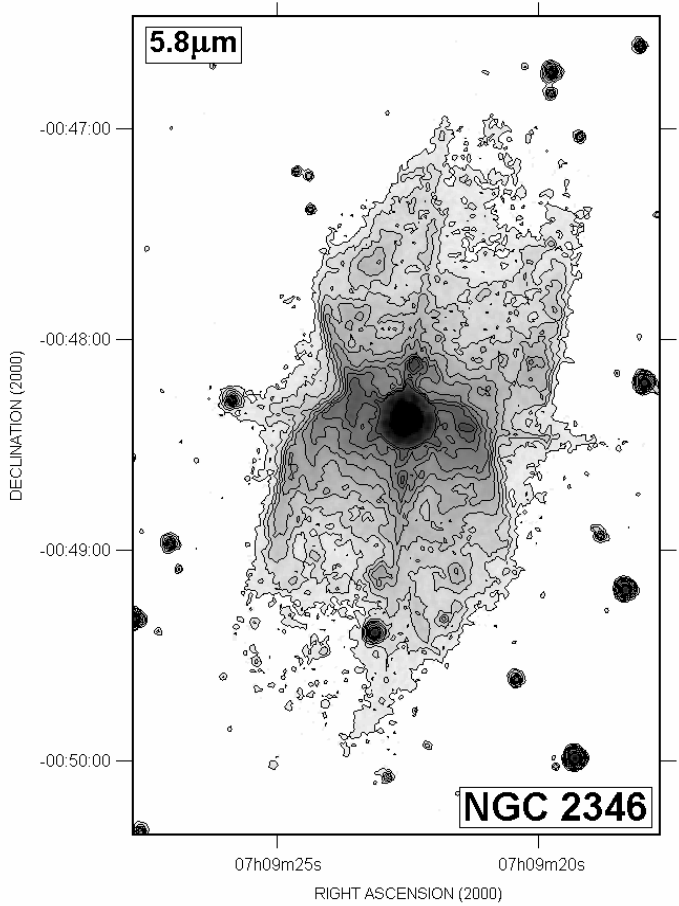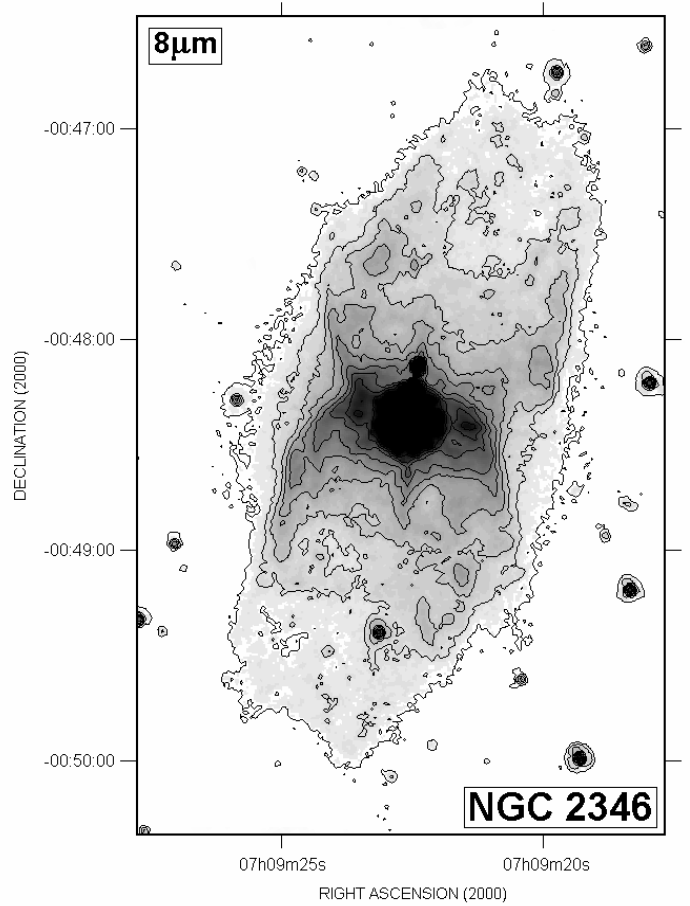

FIGURE 2

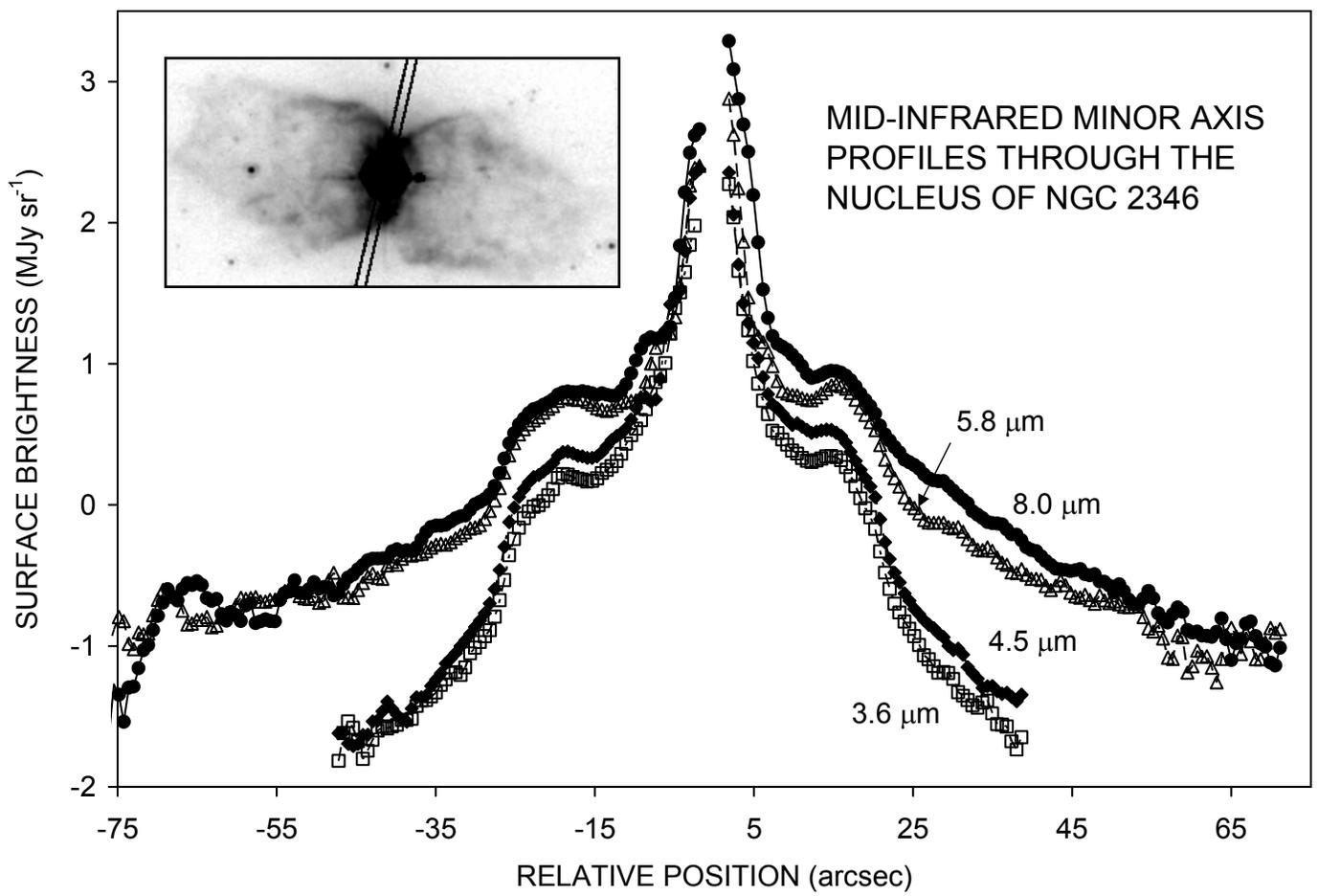

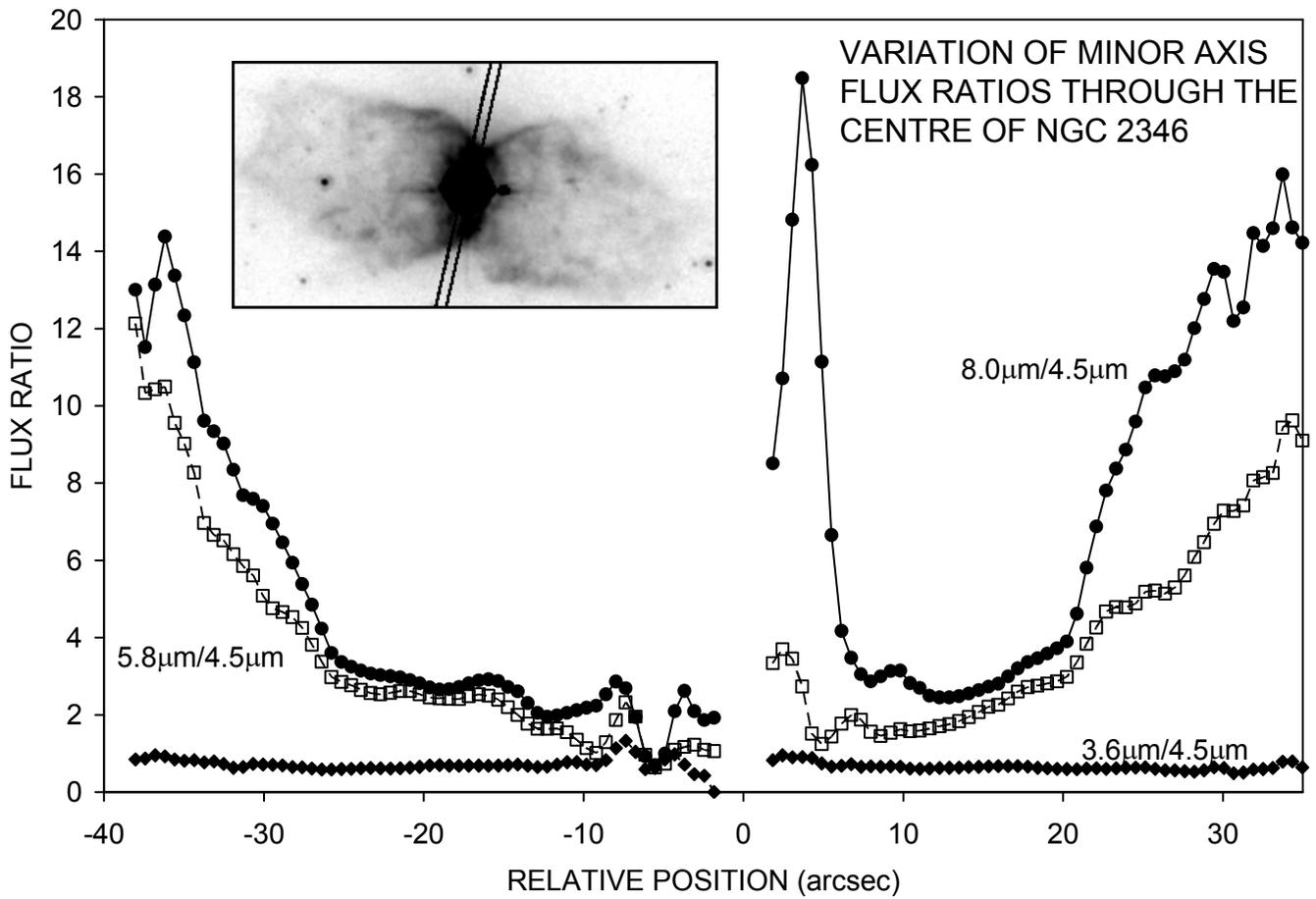

FIGURE 3

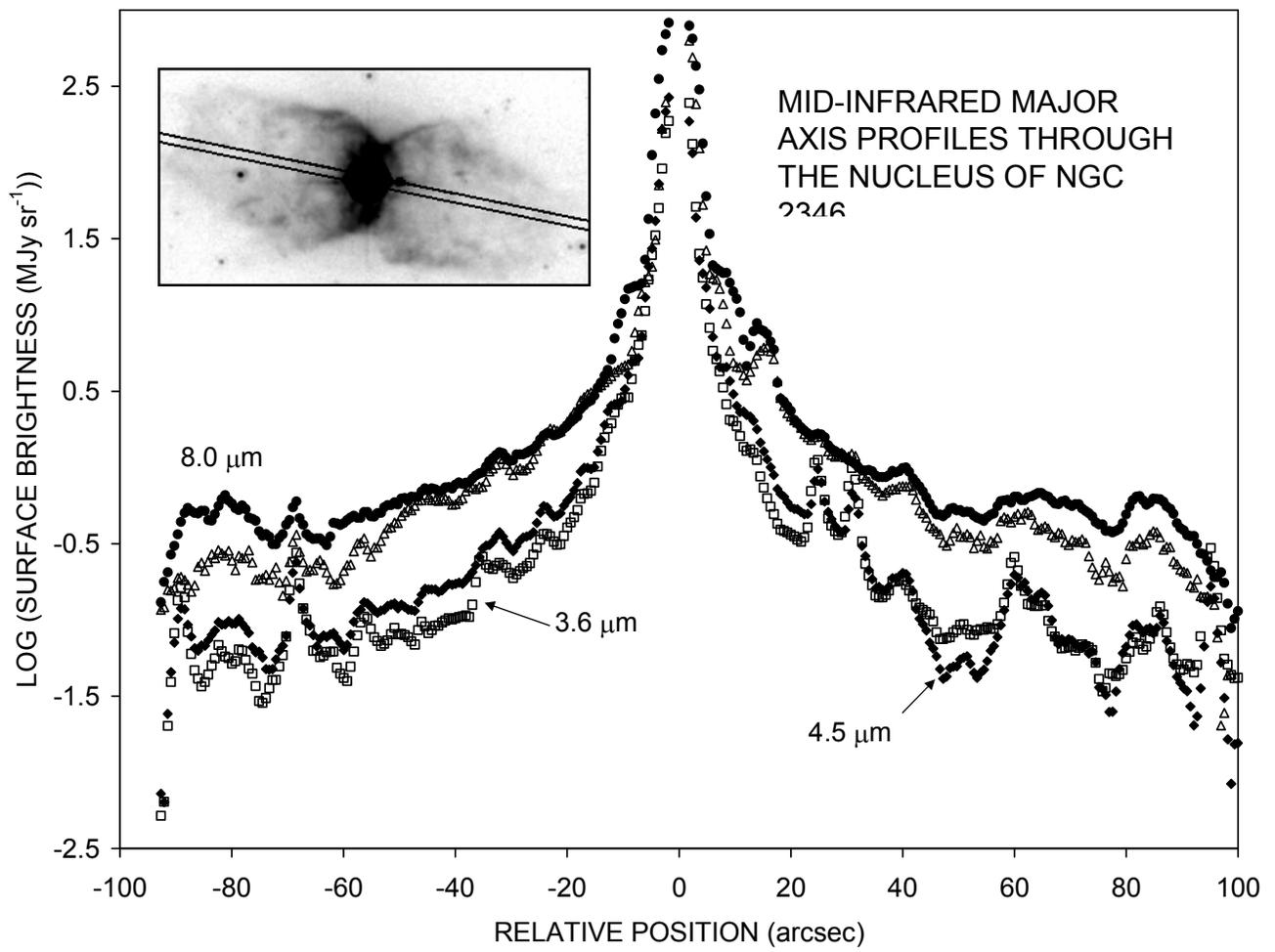
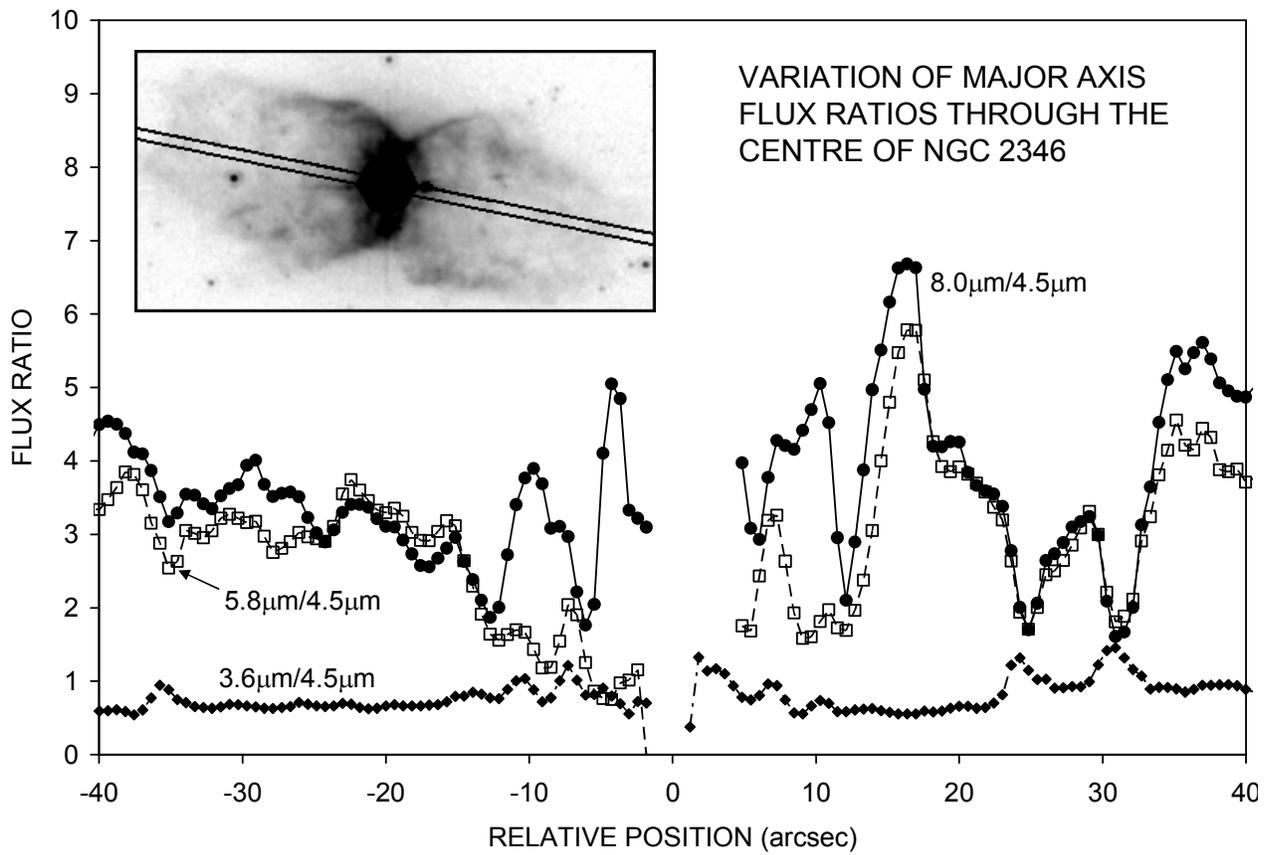

FIGURE 4

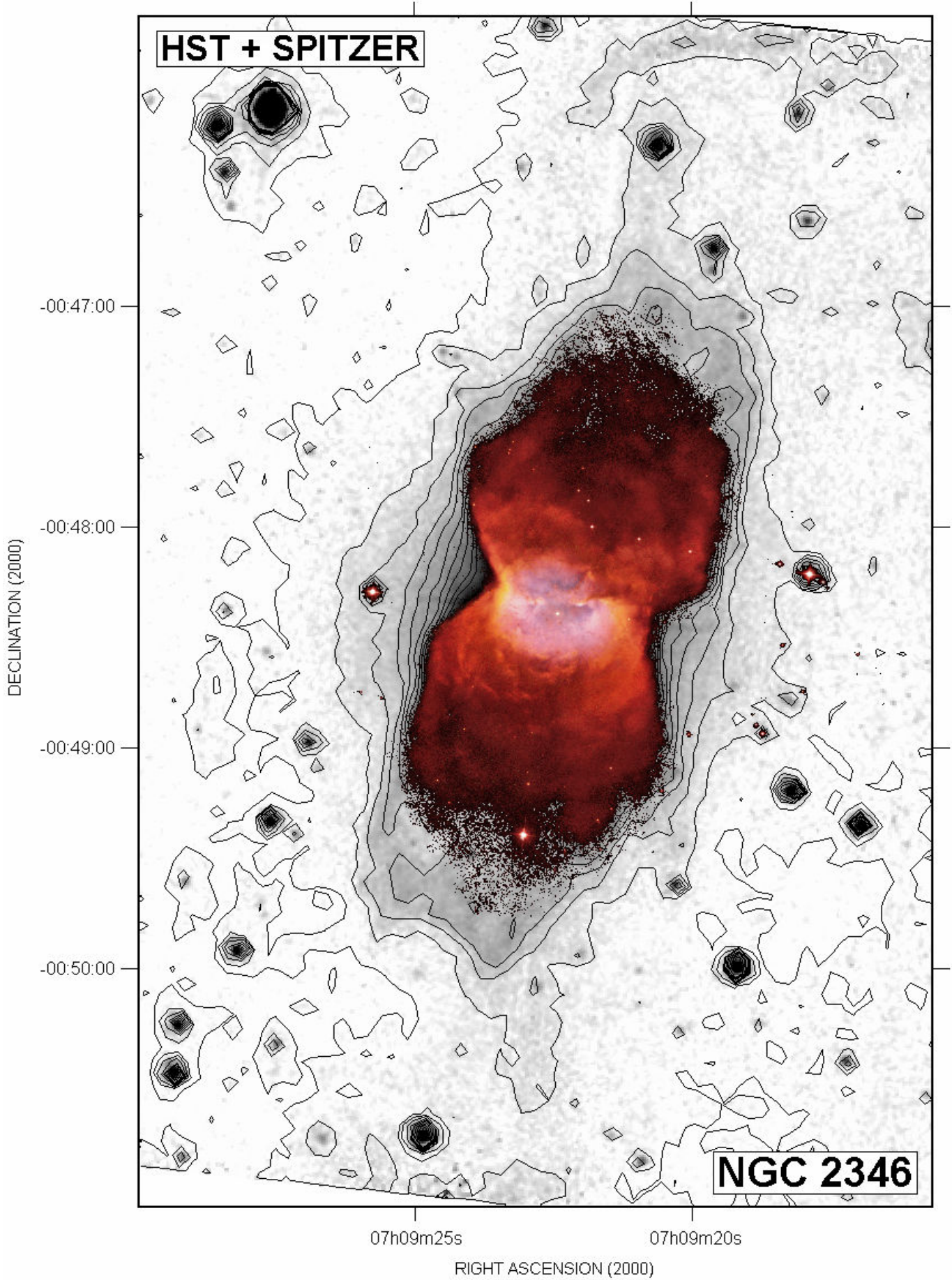

FIGURE 5

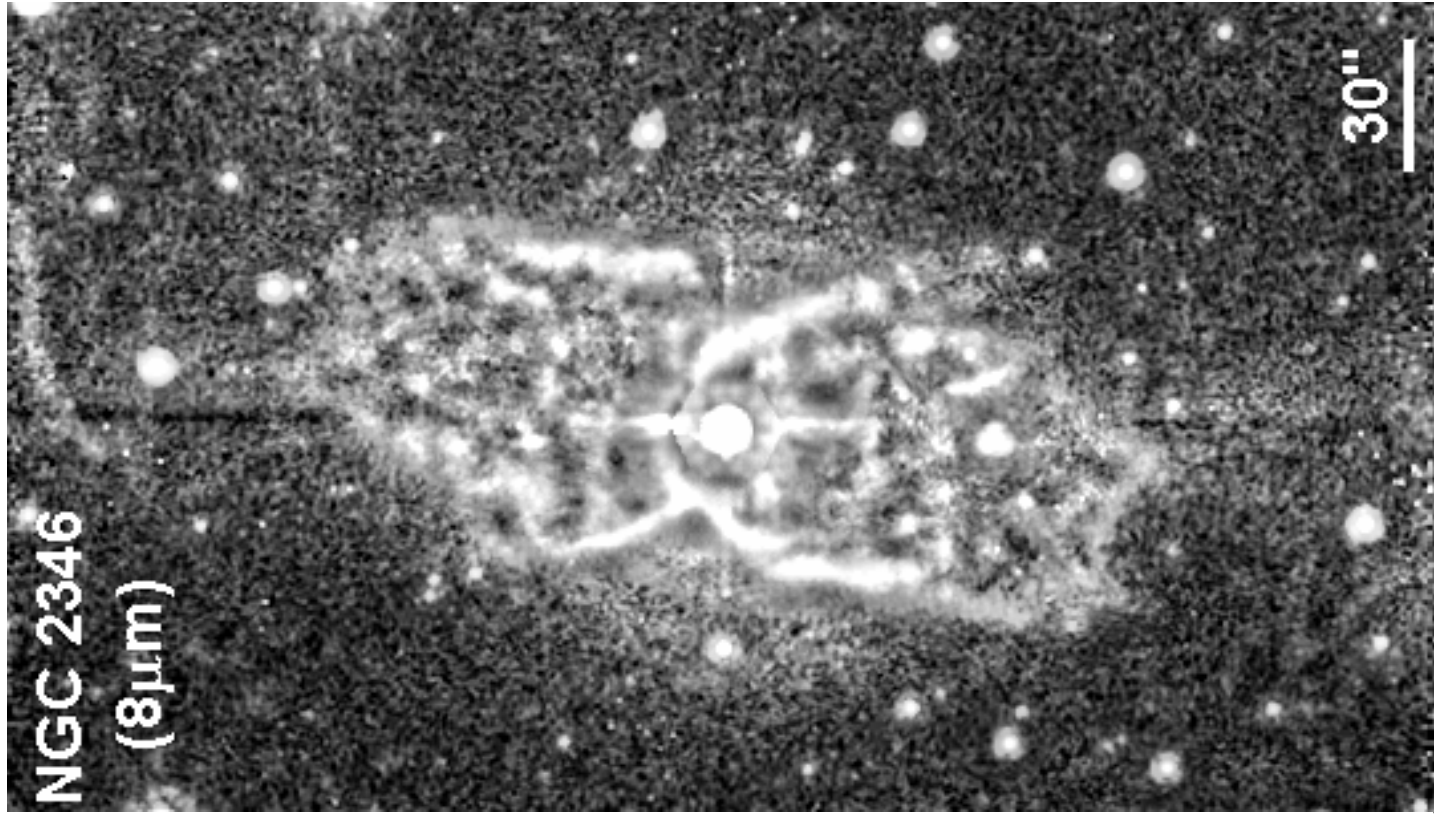
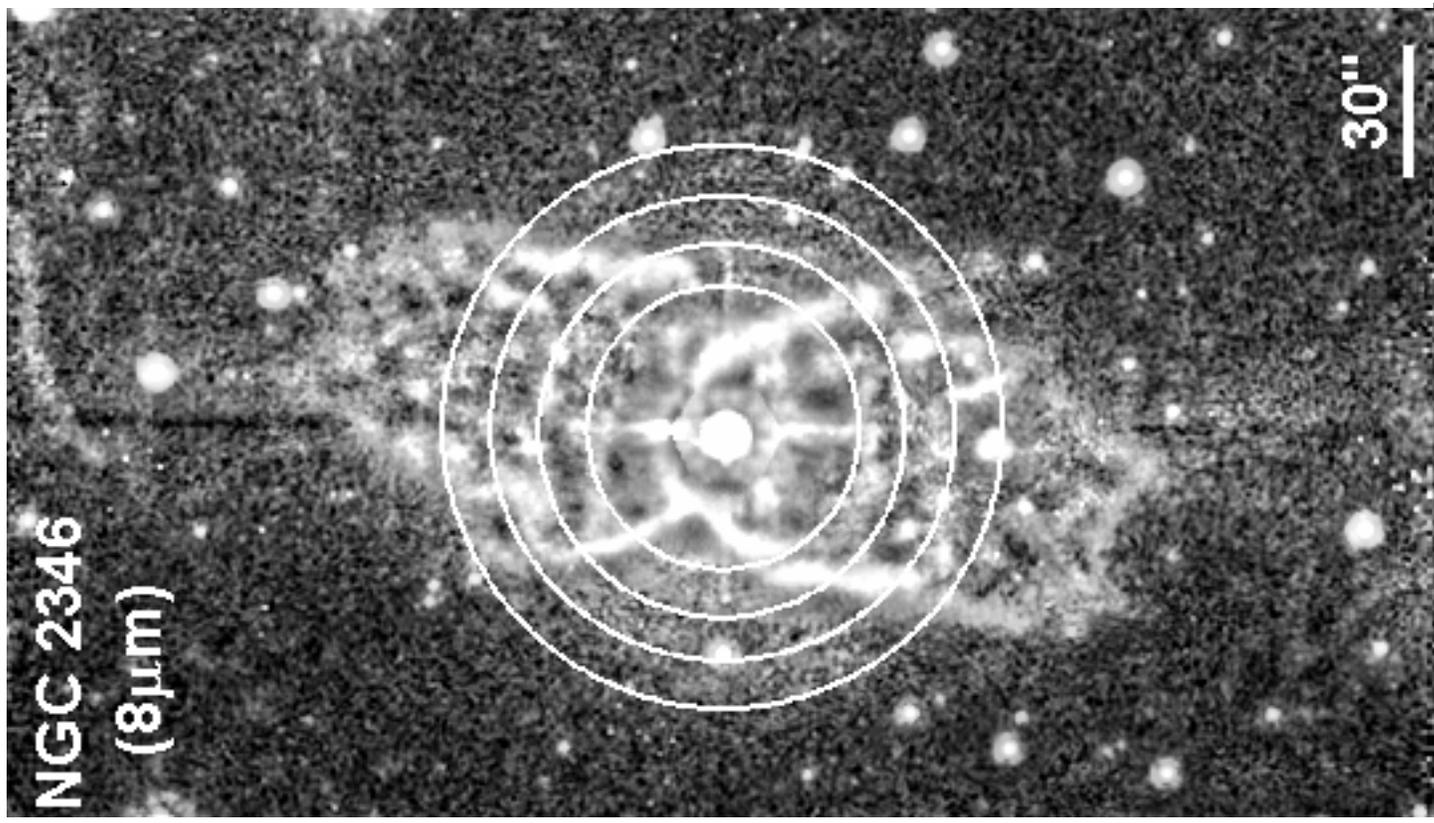

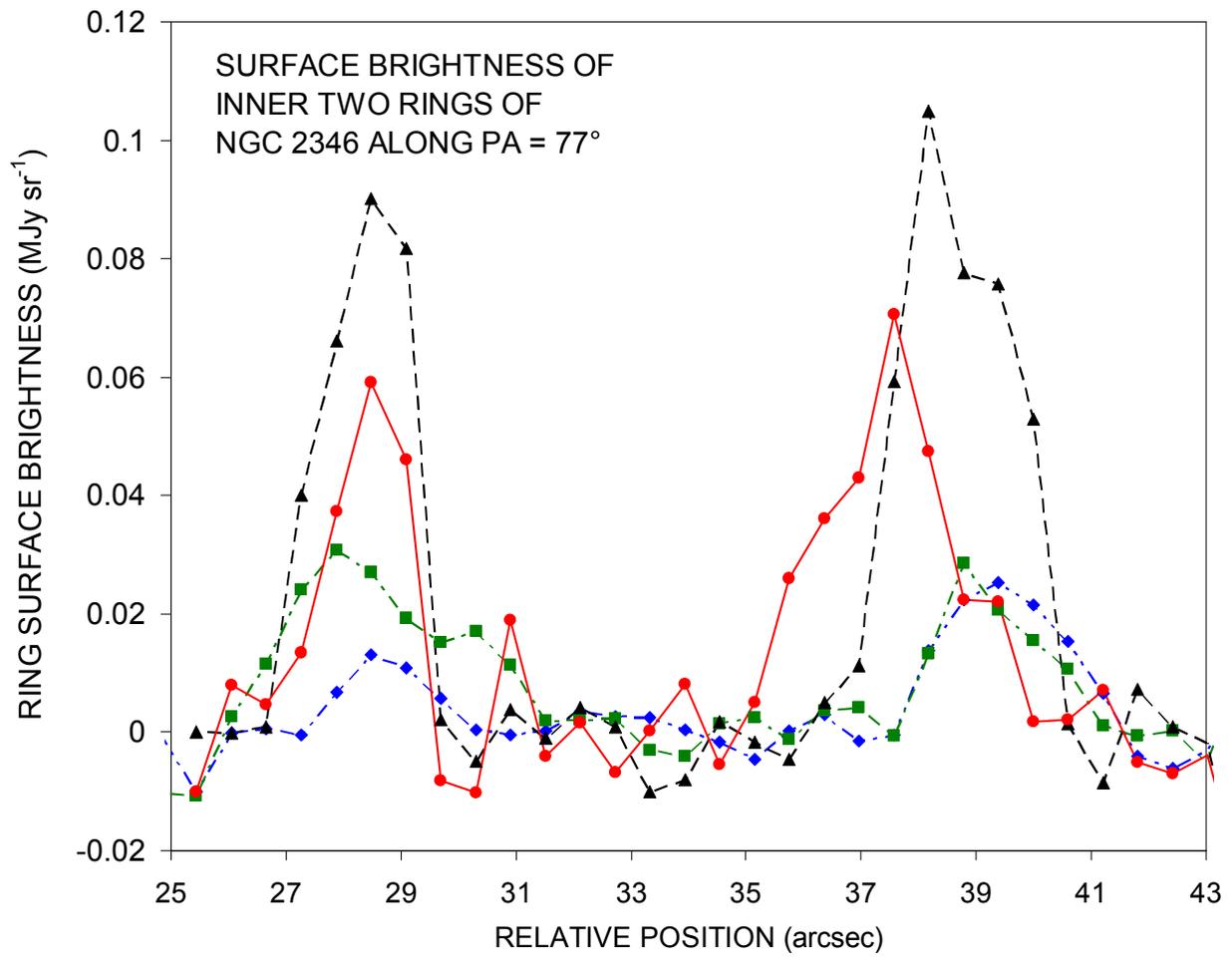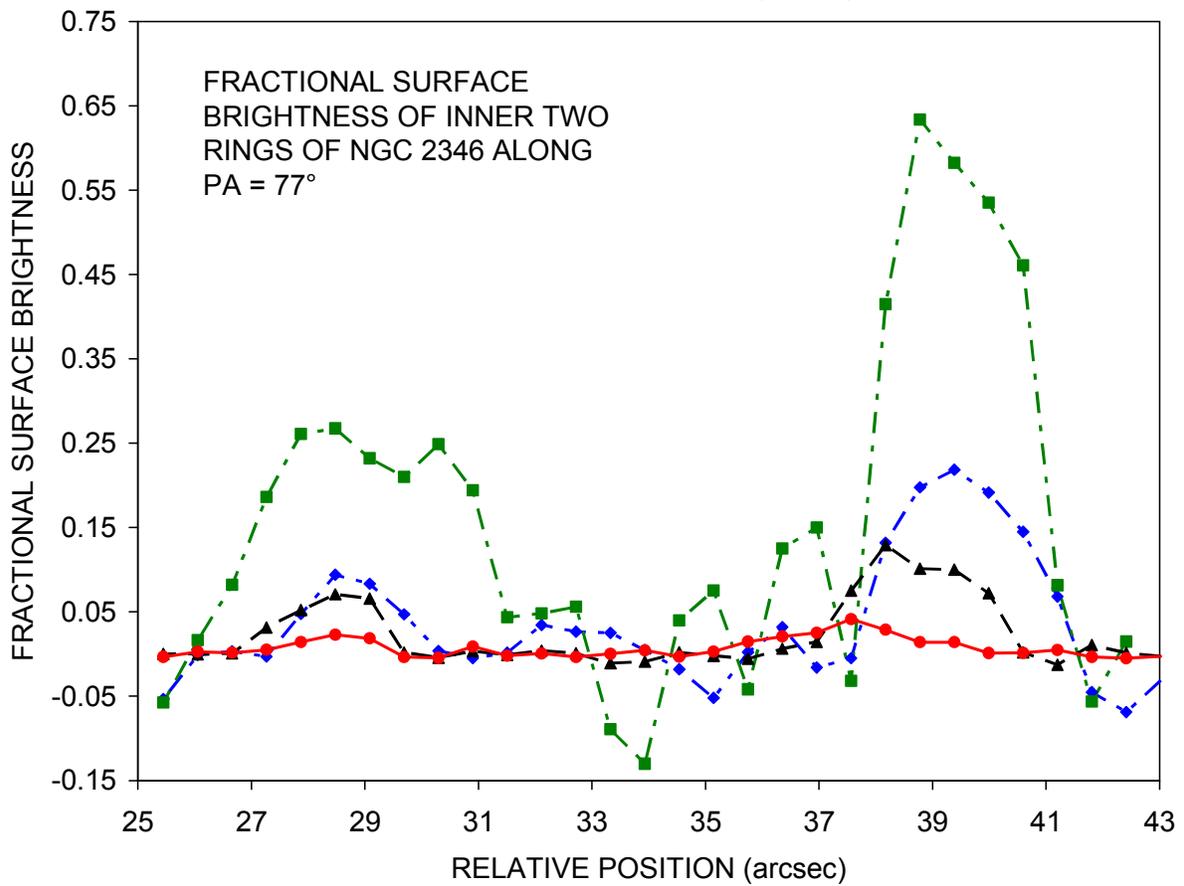

FIGURE 7

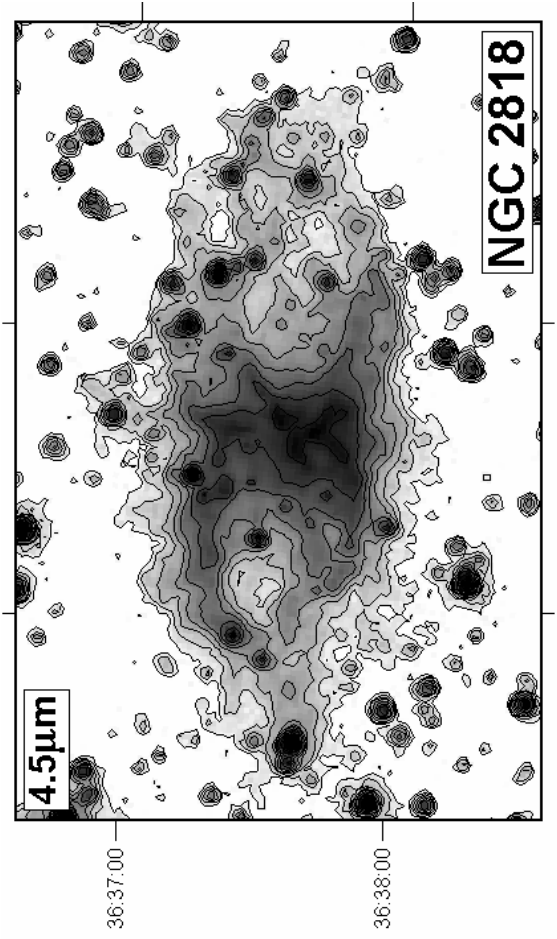
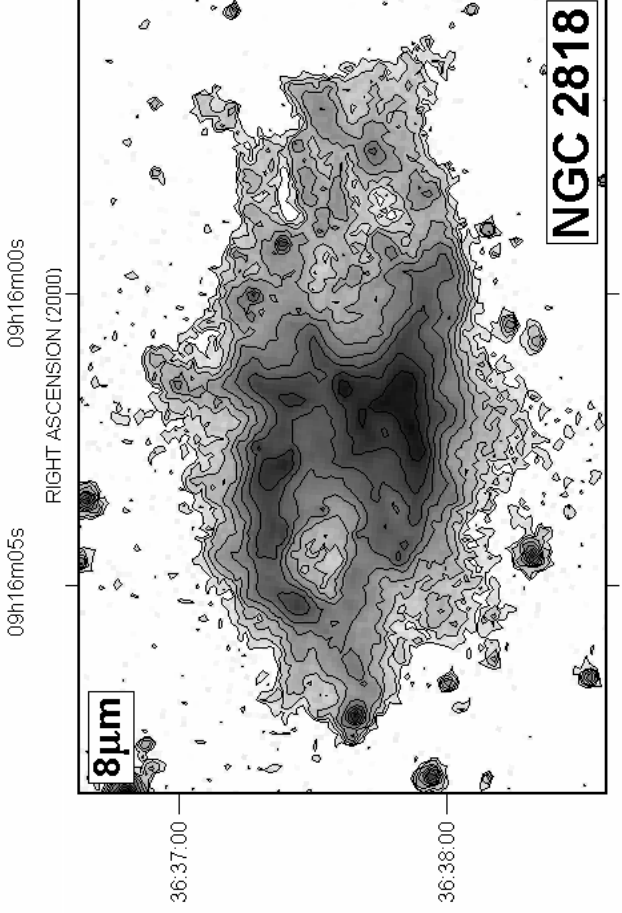
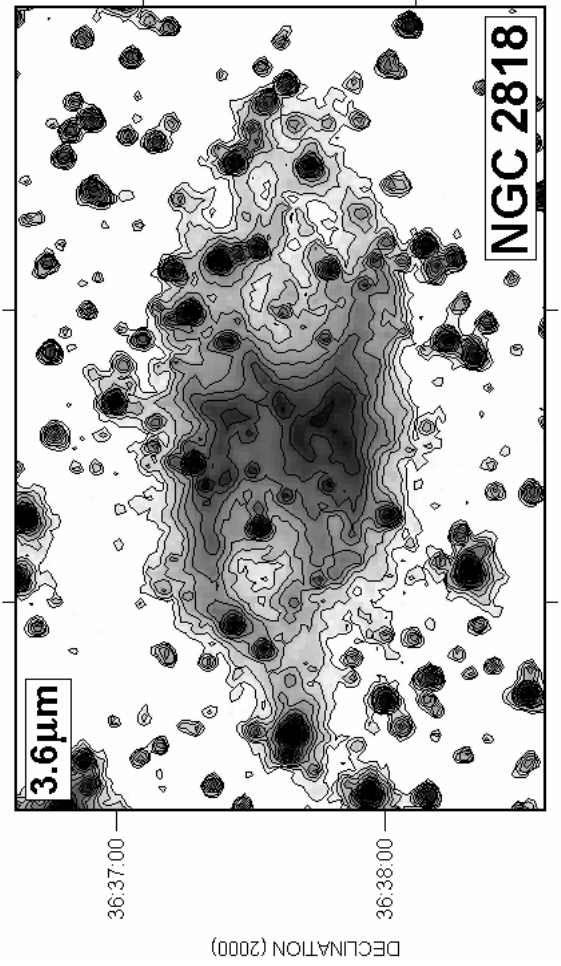
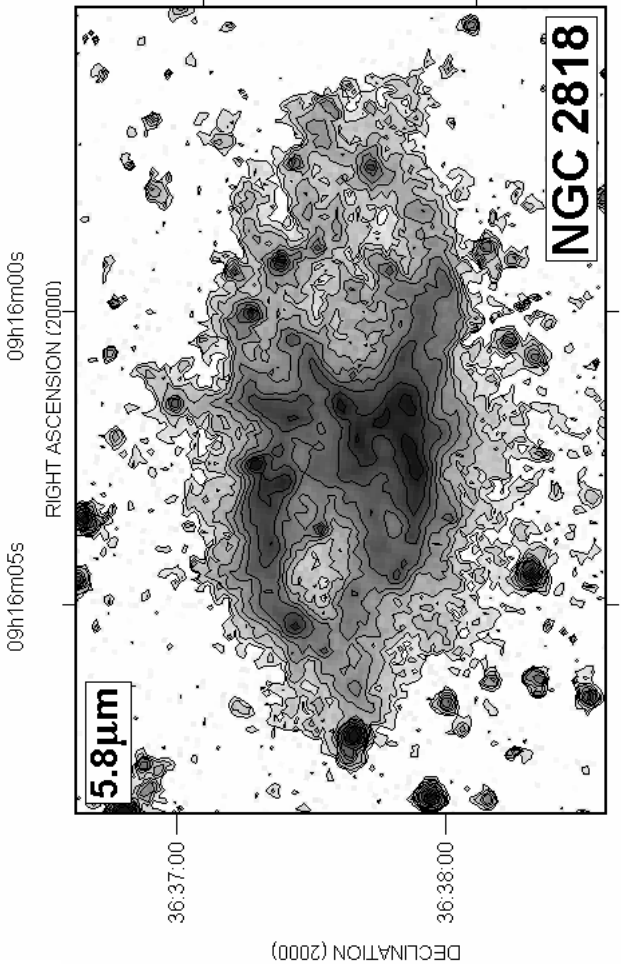

FIGURE 8

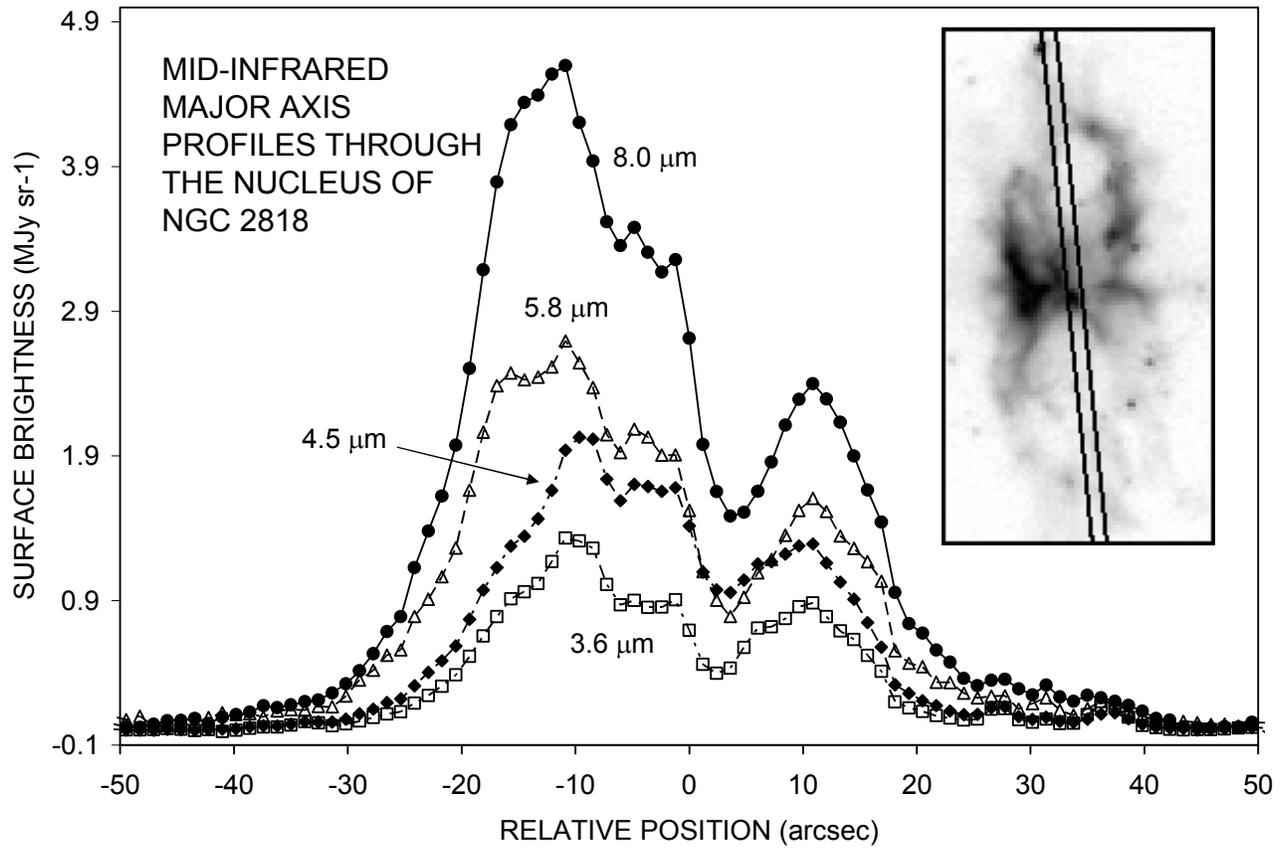
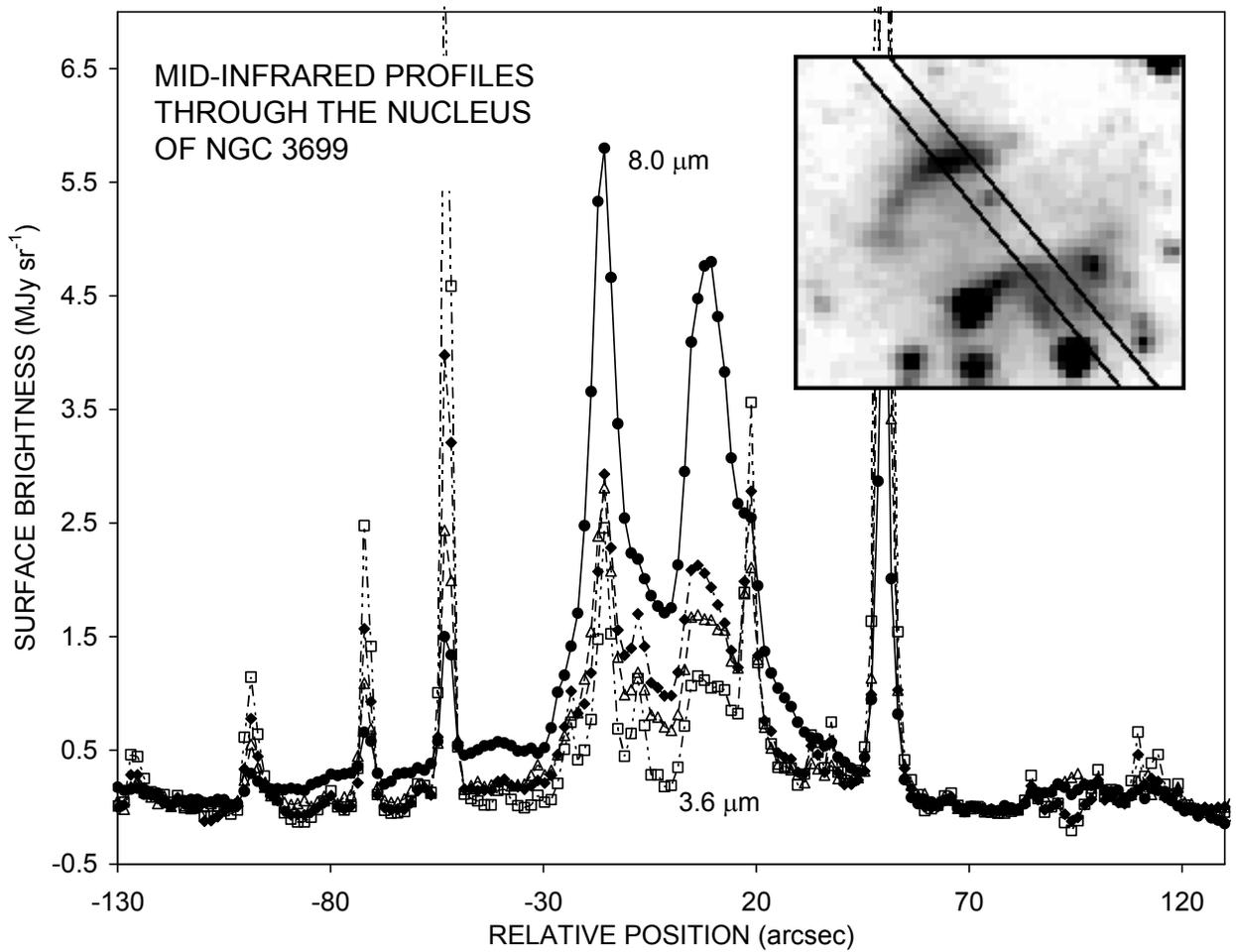

FIGURE 9

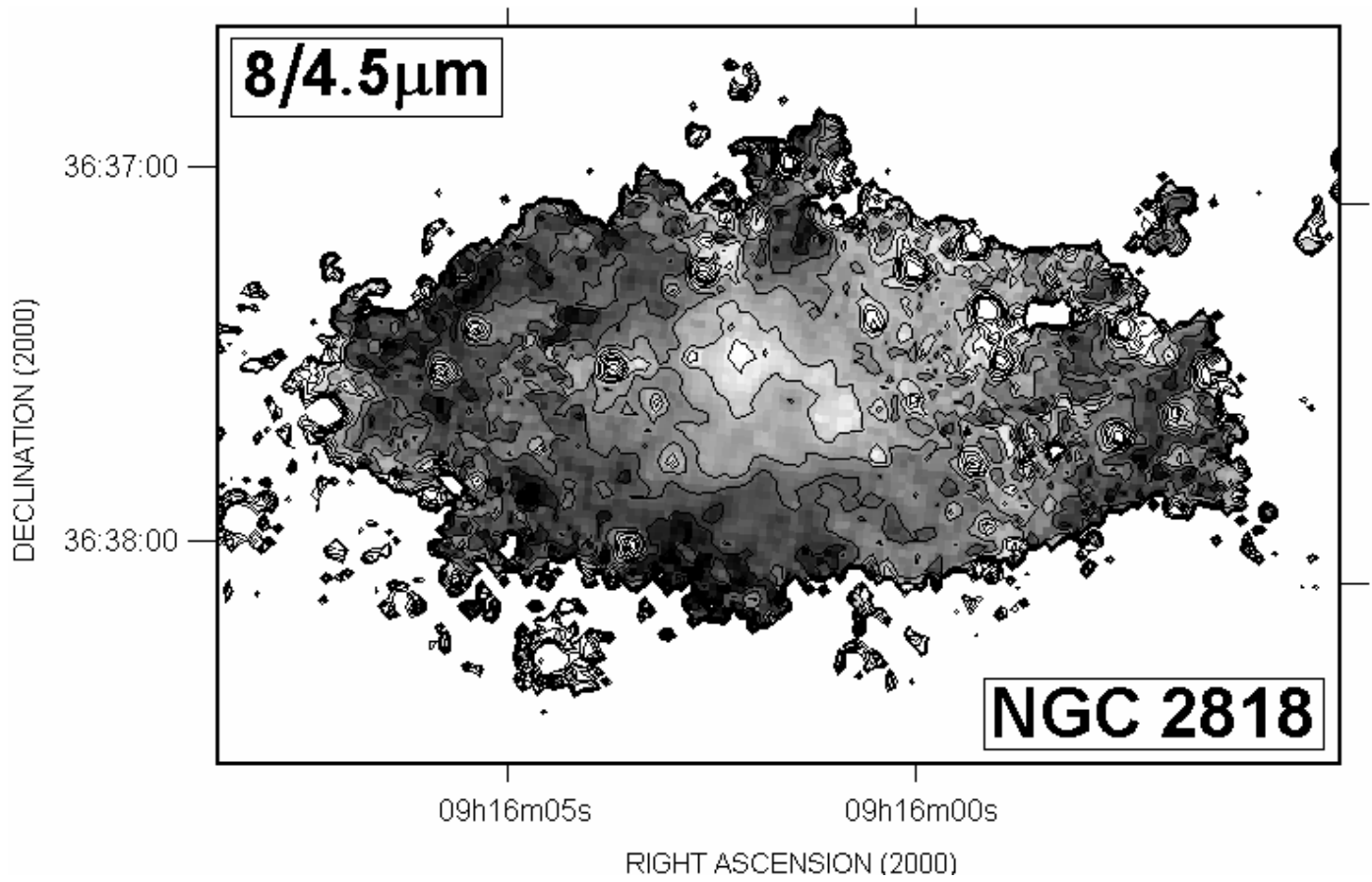
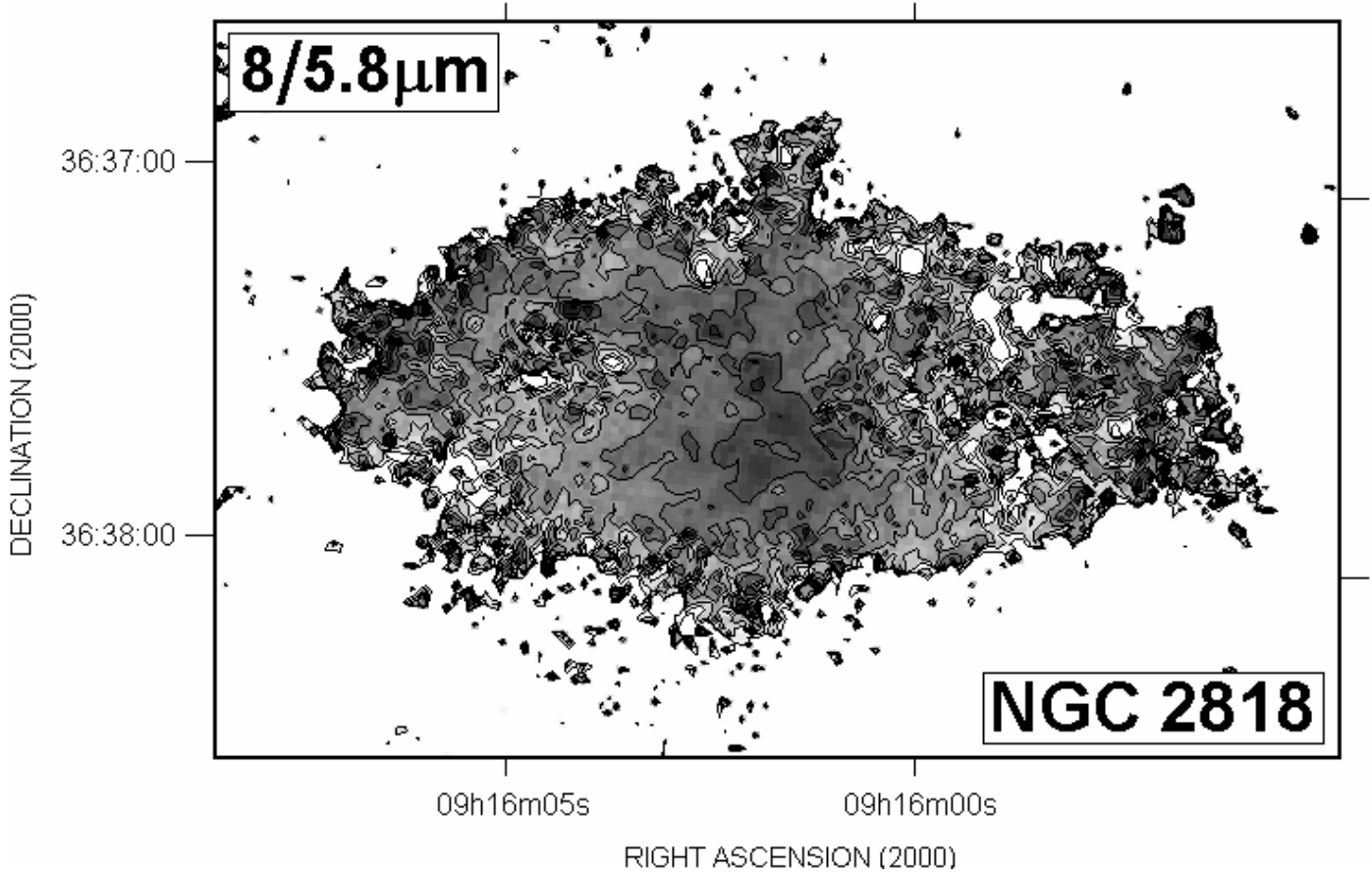

FIGURE 10

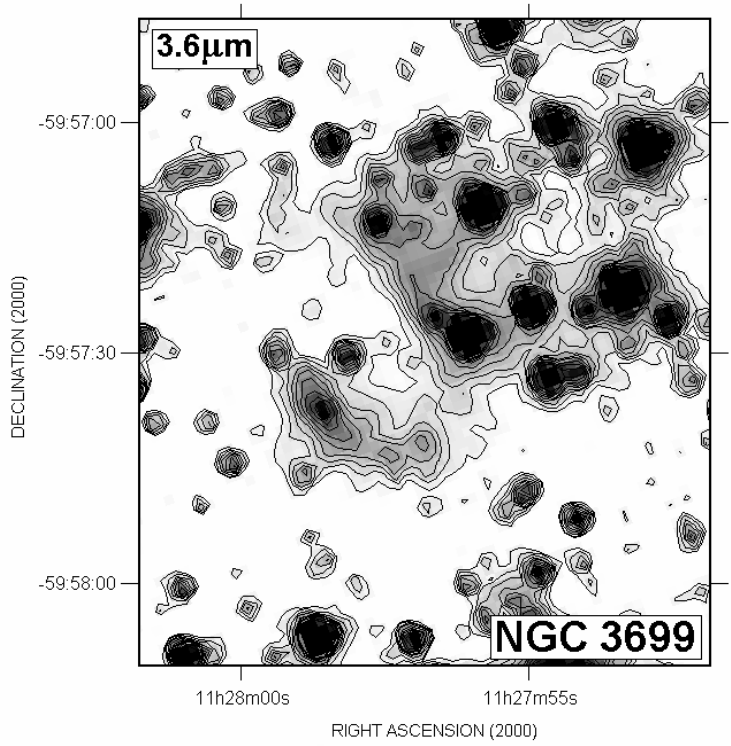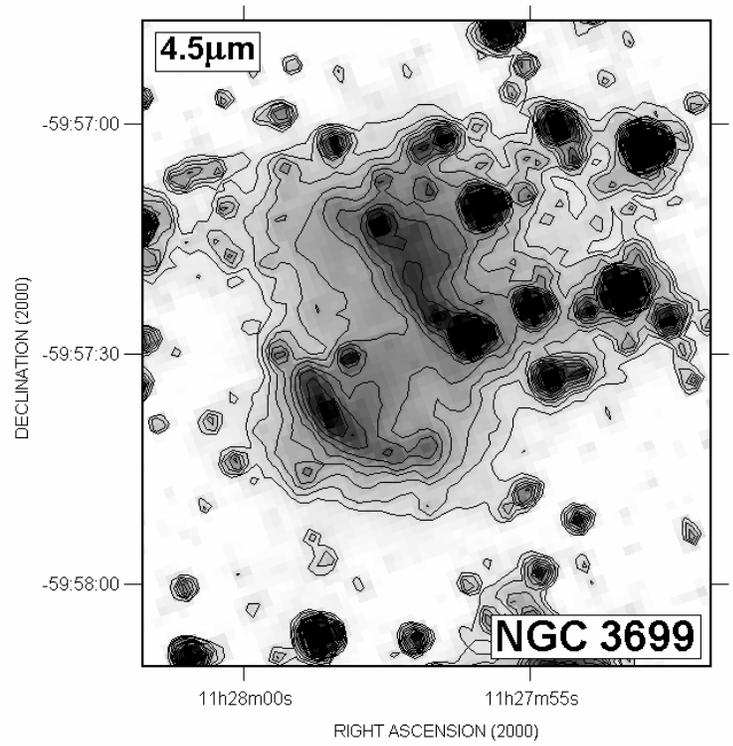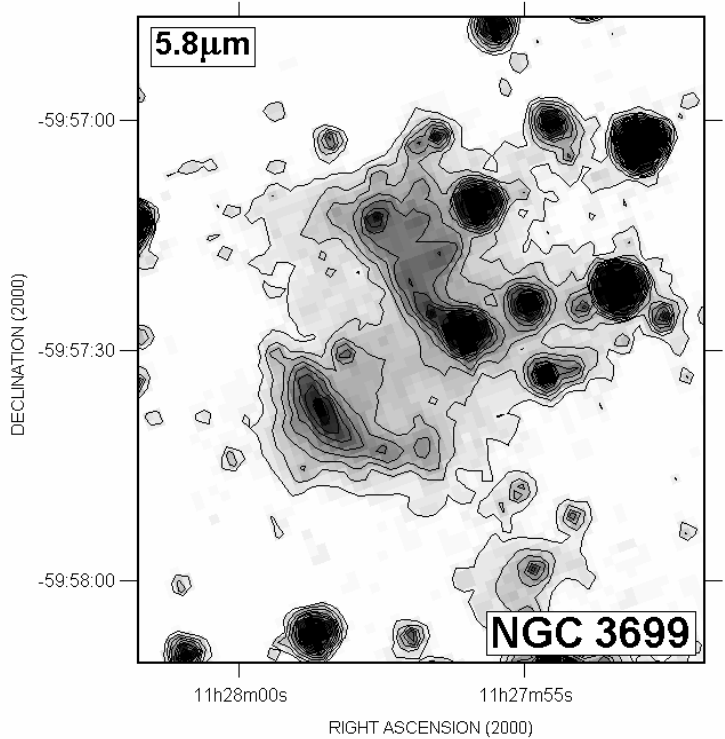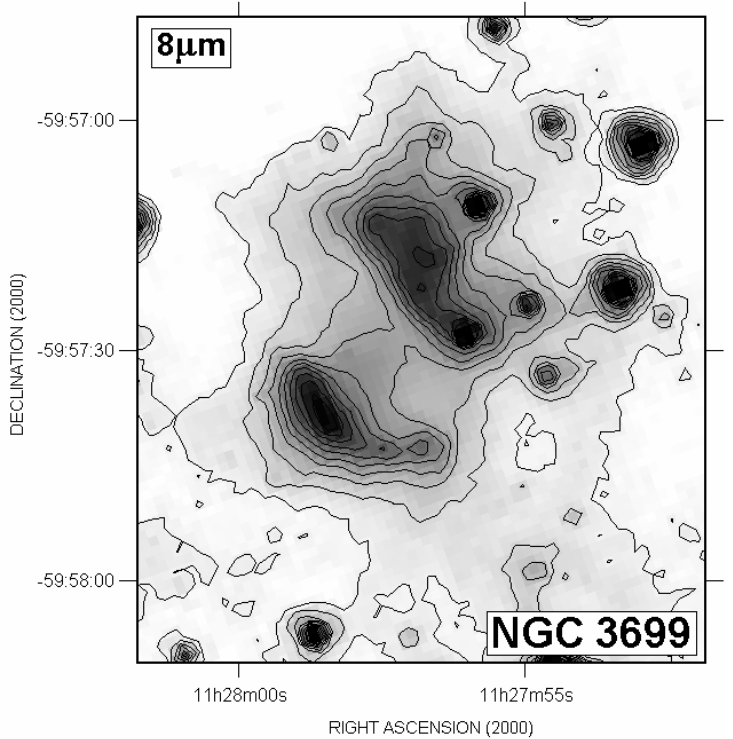

FIGURE 11

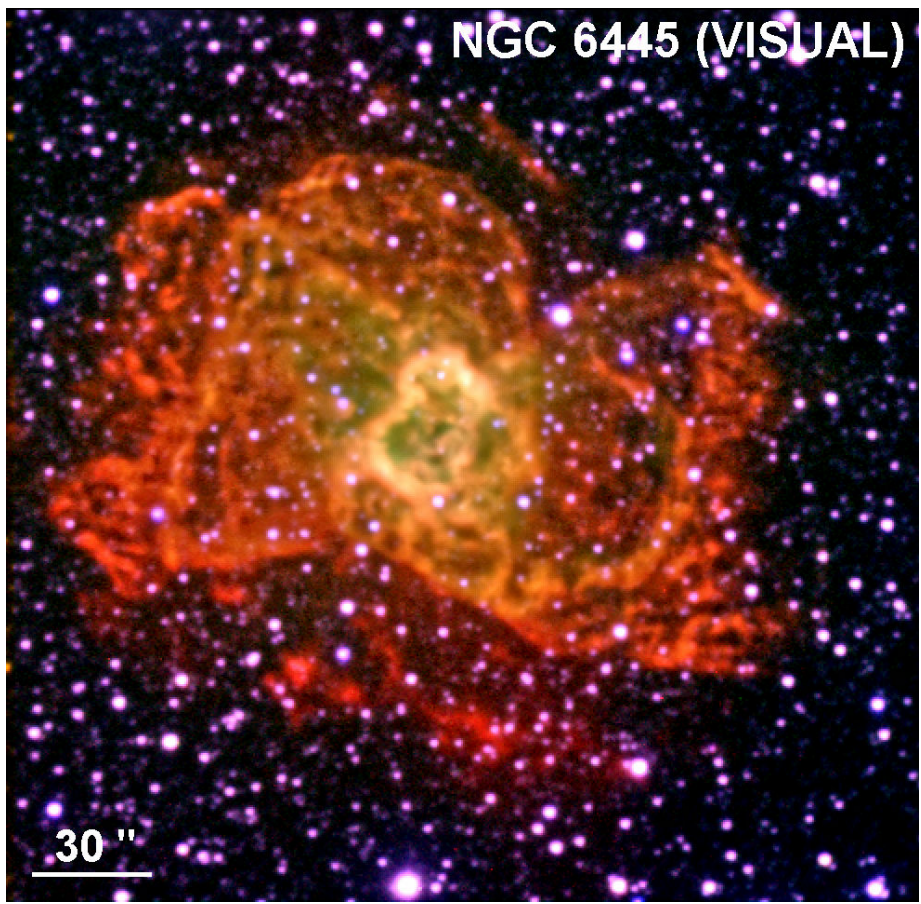
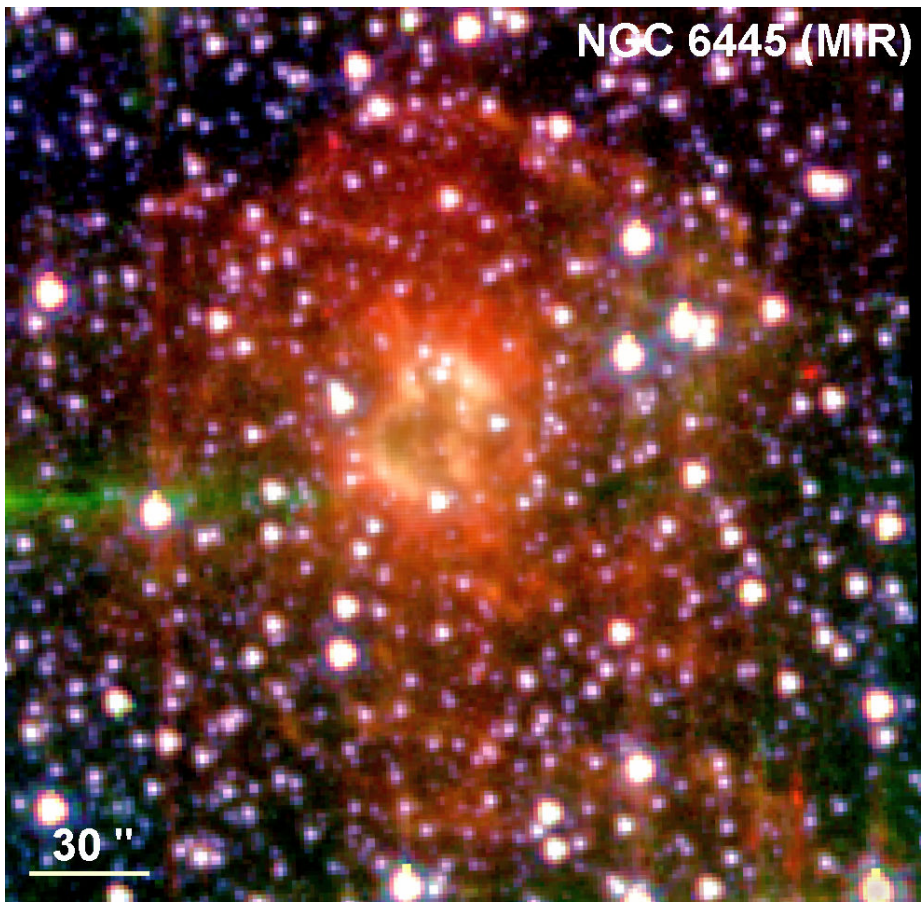

FIGURE 12

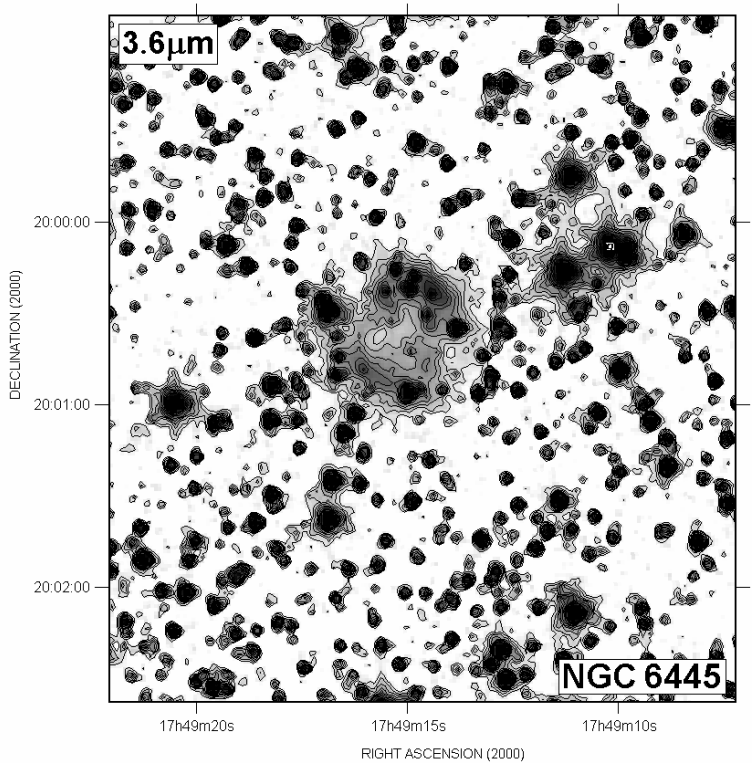
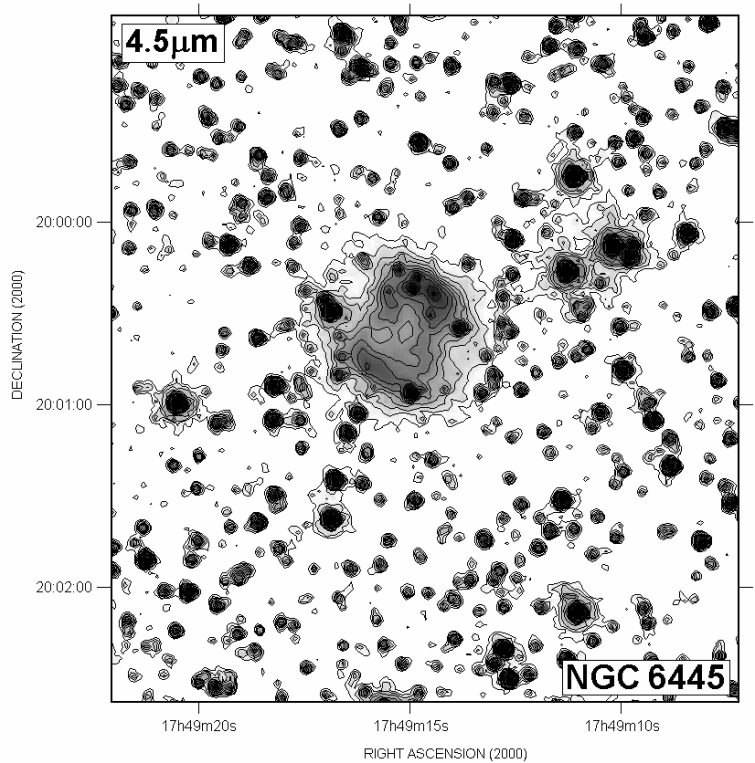
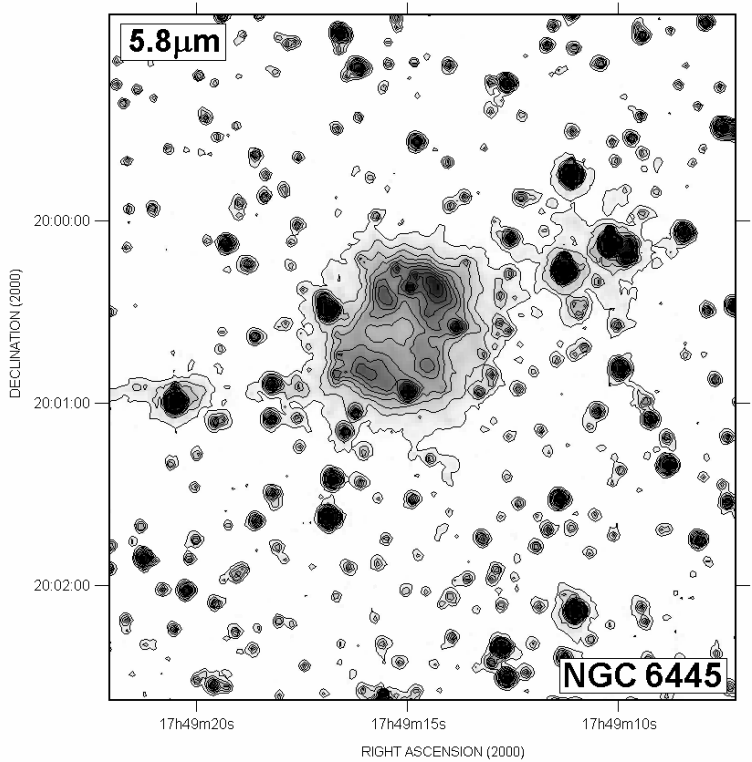
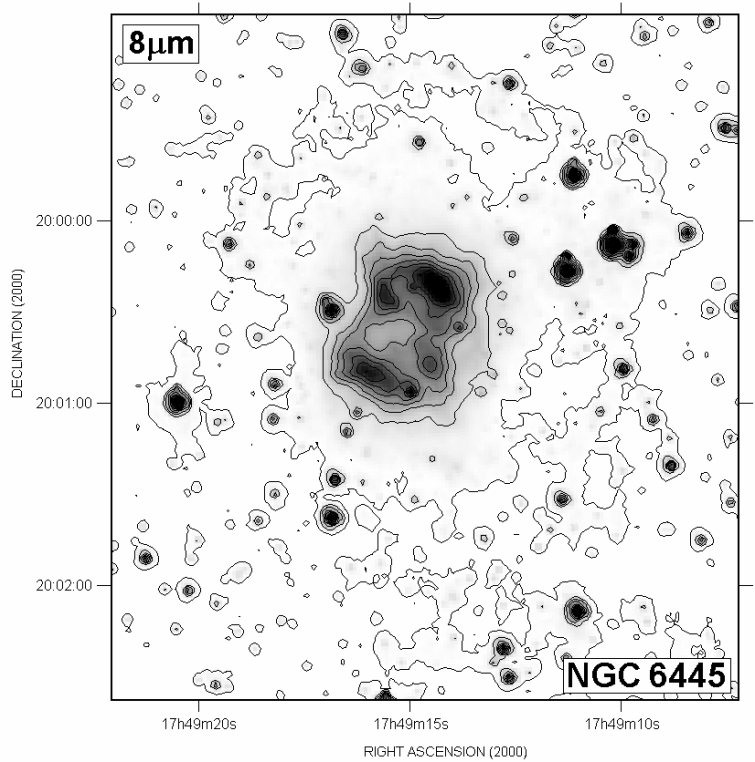

FIGURE 13

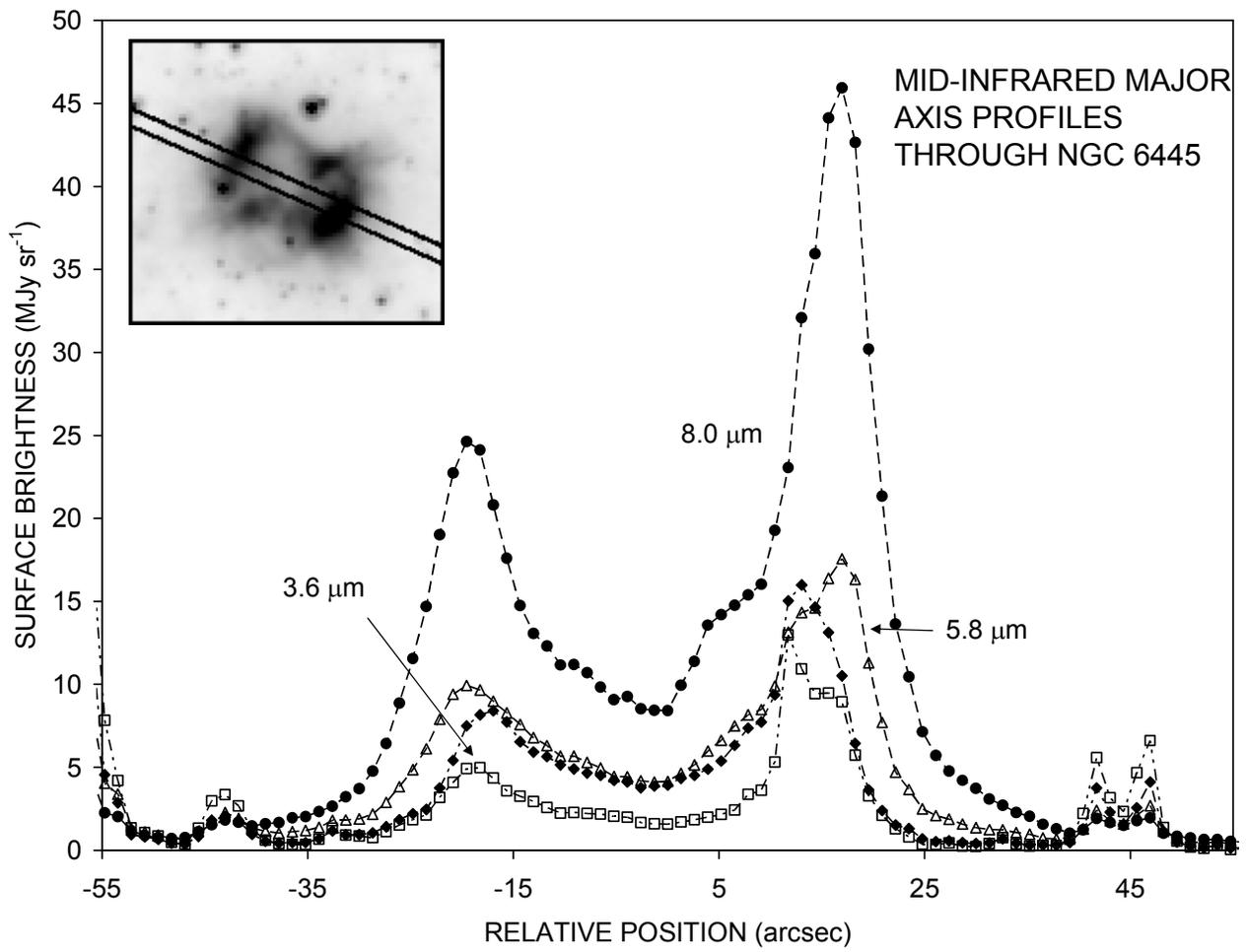
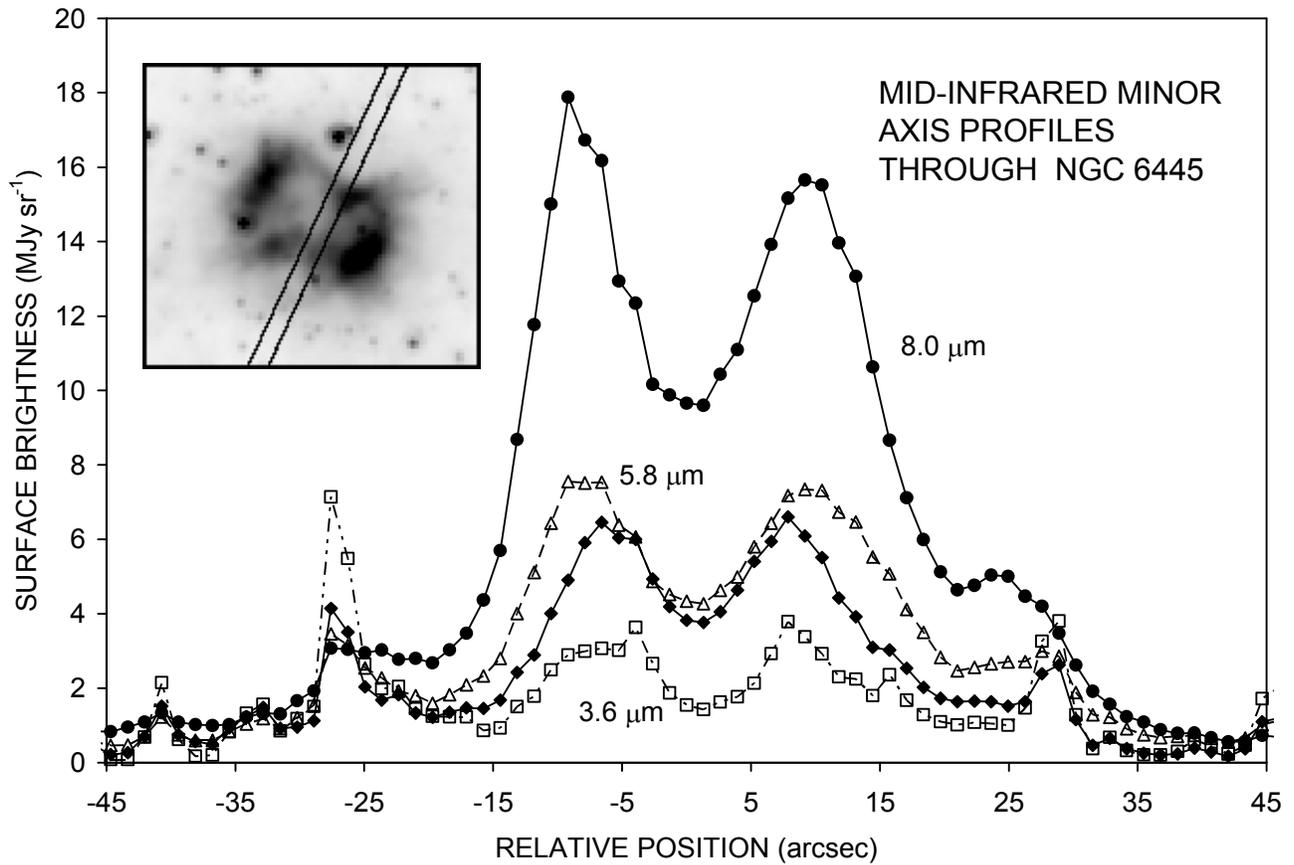

FIGURE 14

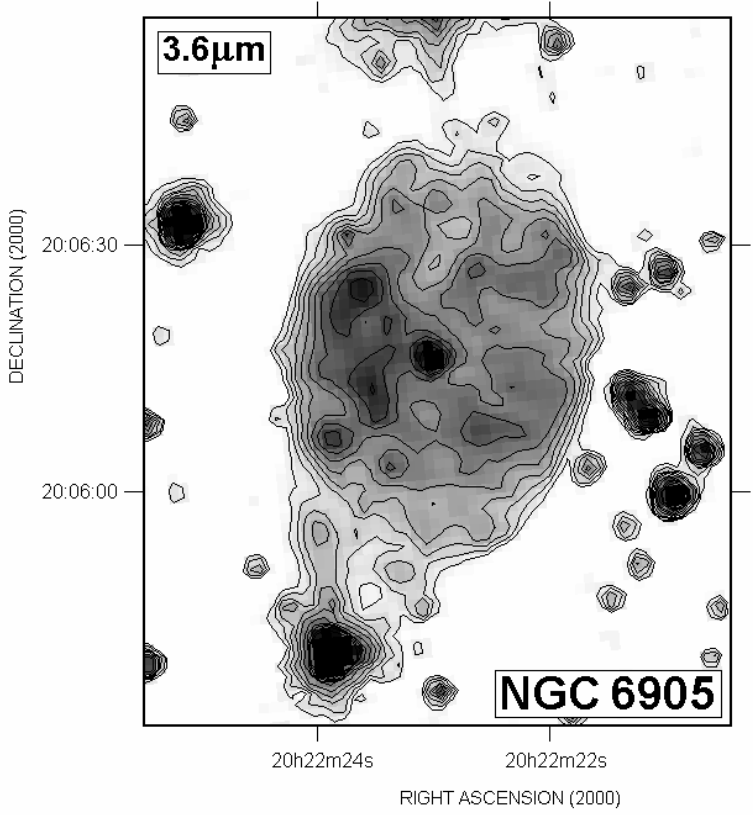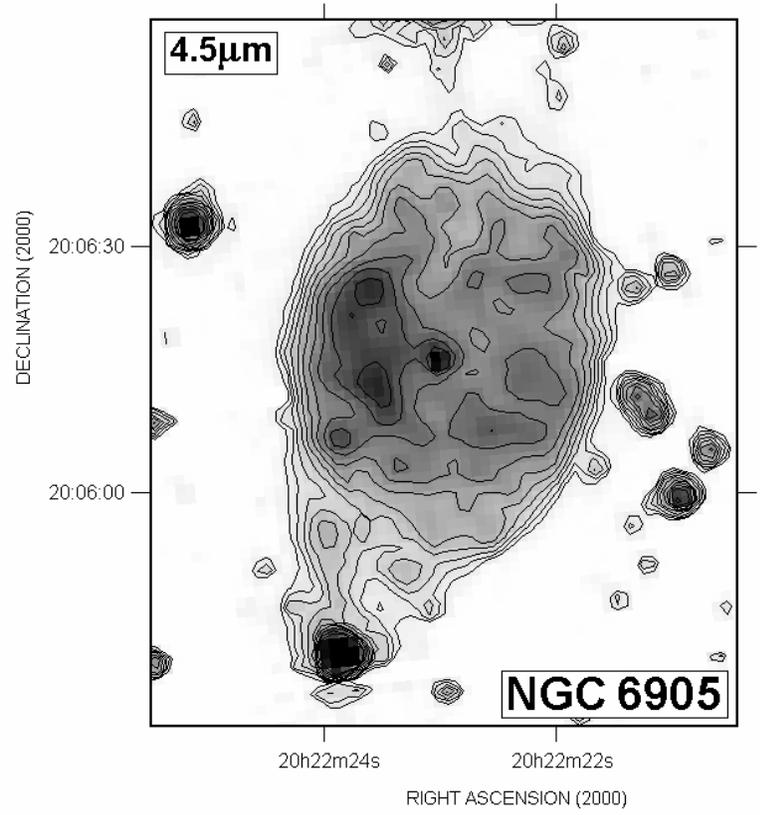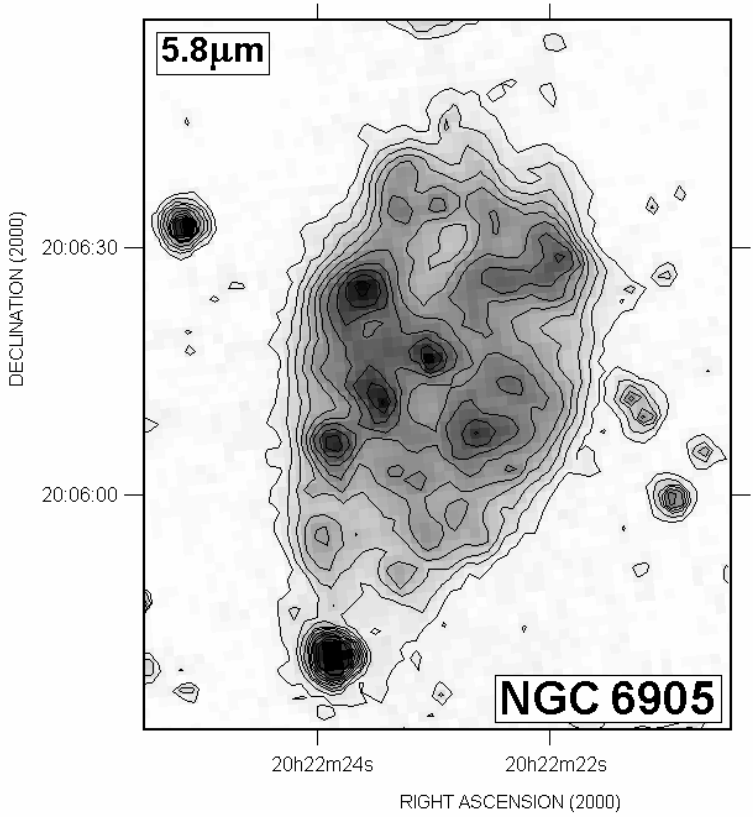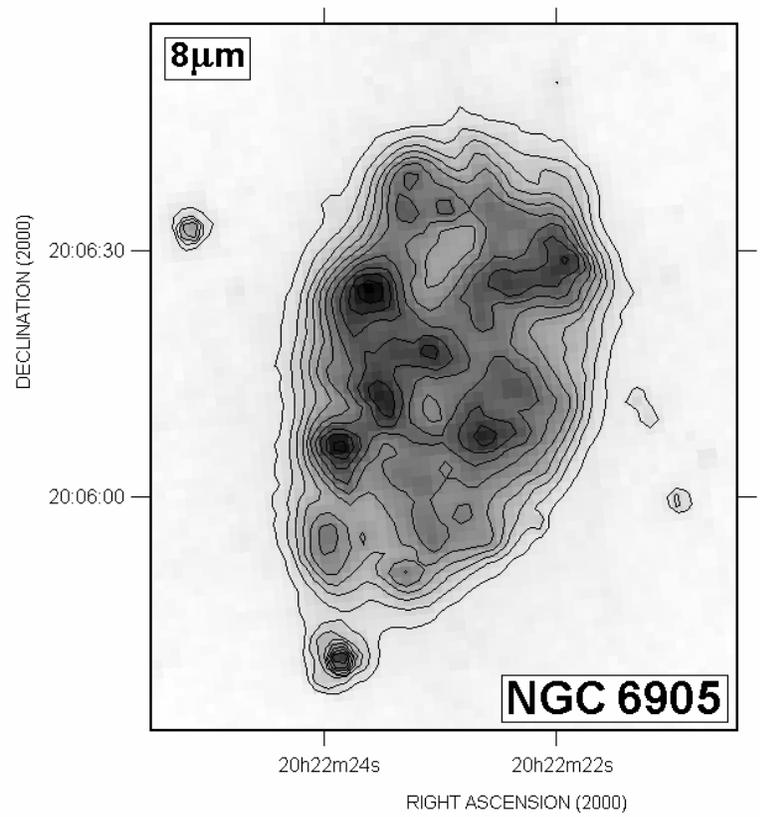

FIGURE 15

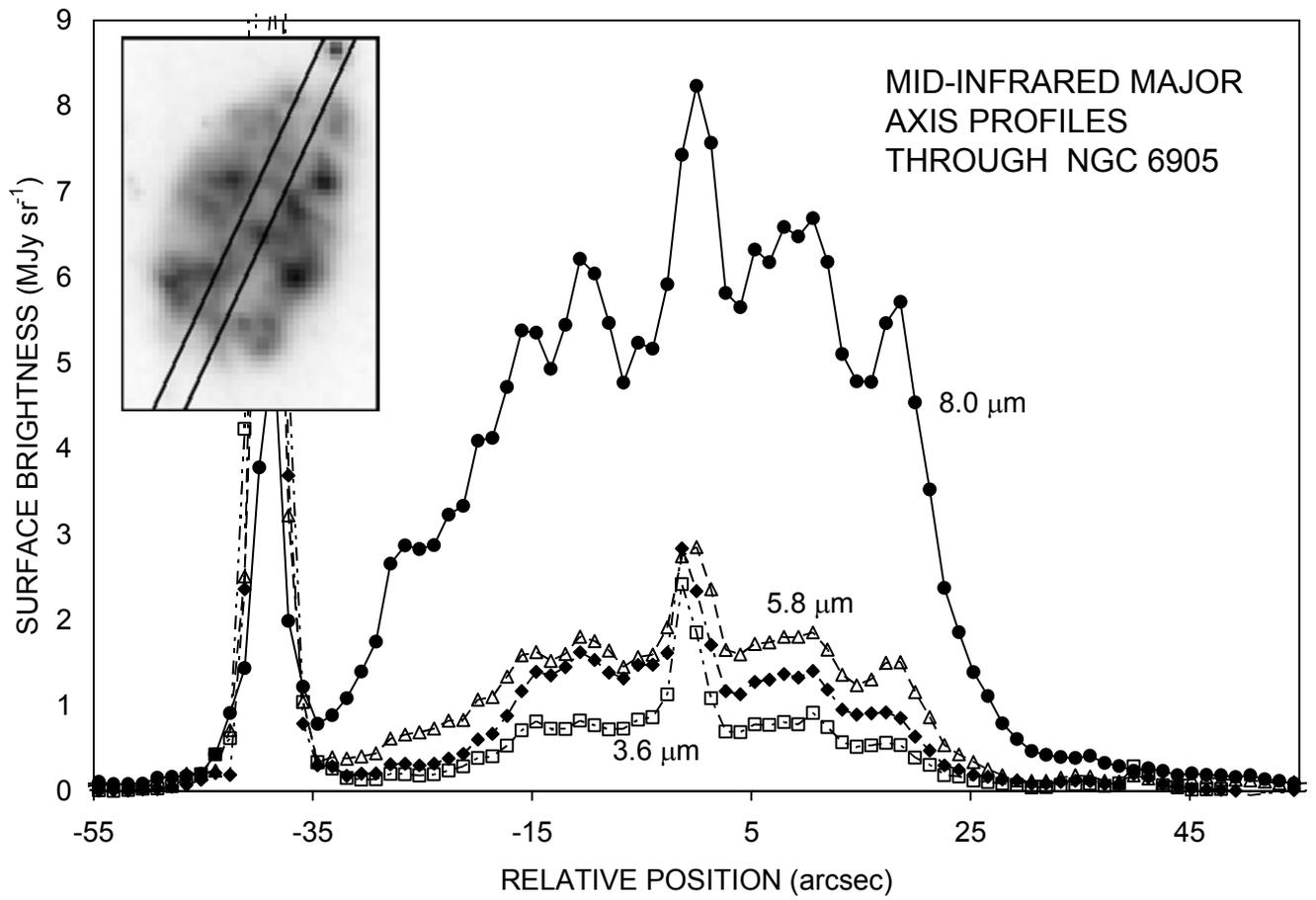
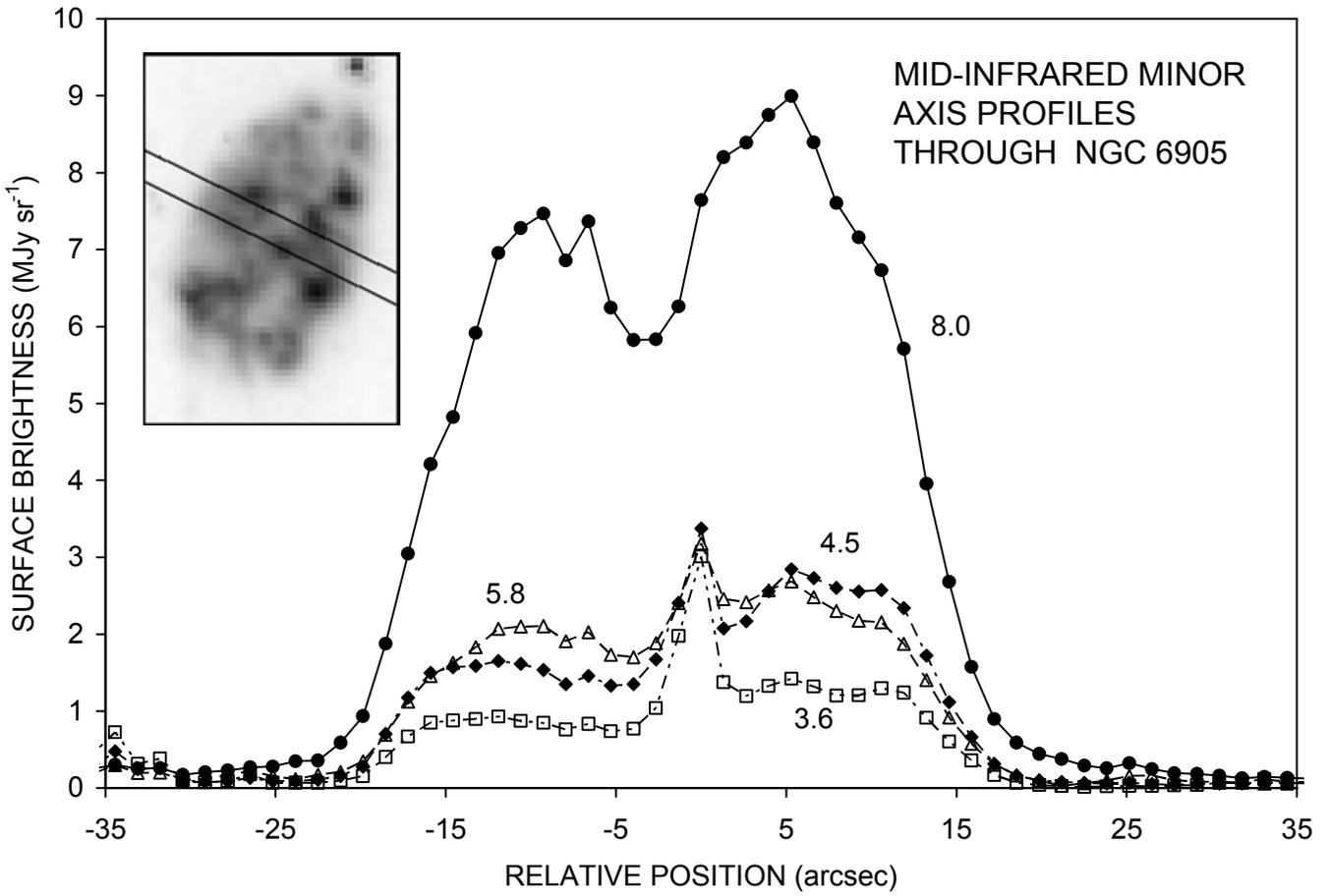

FIGURE 16

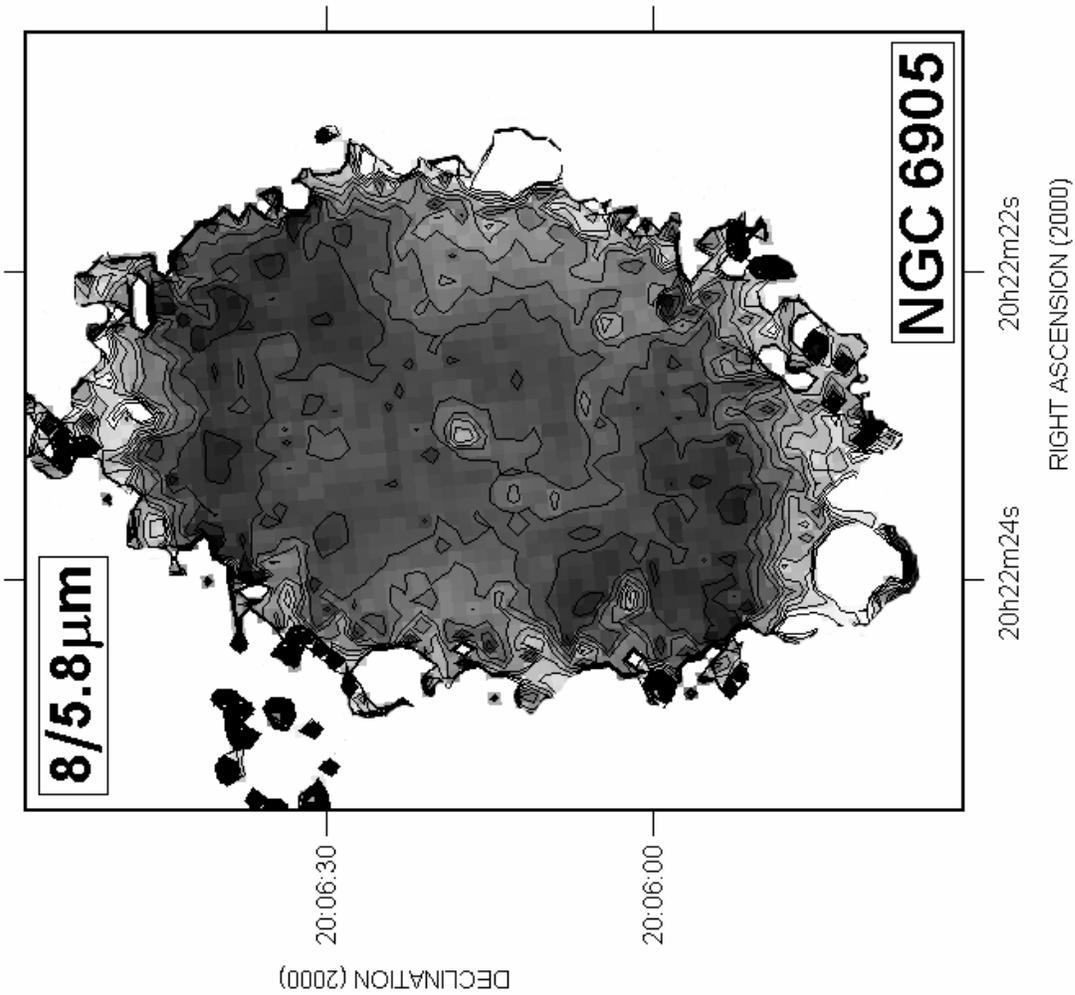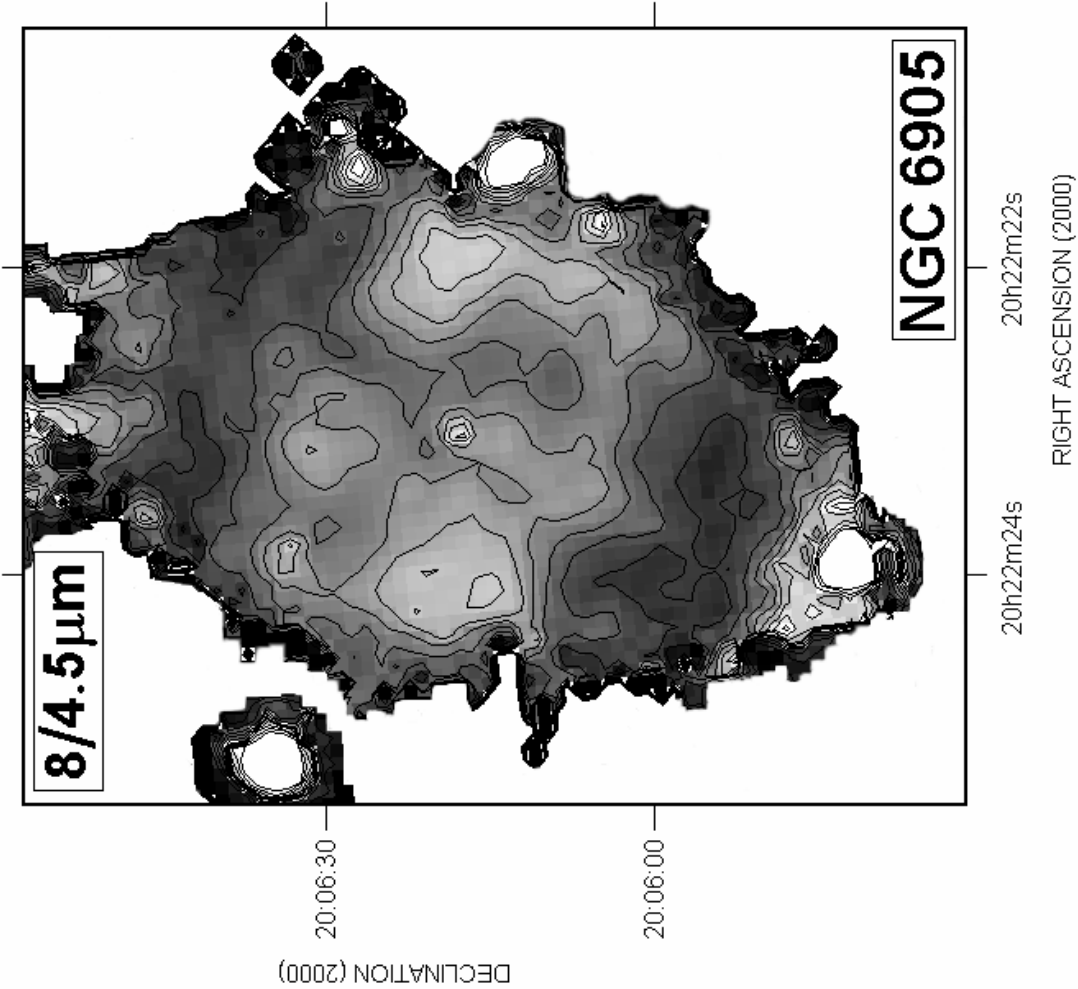

FIGURE 17

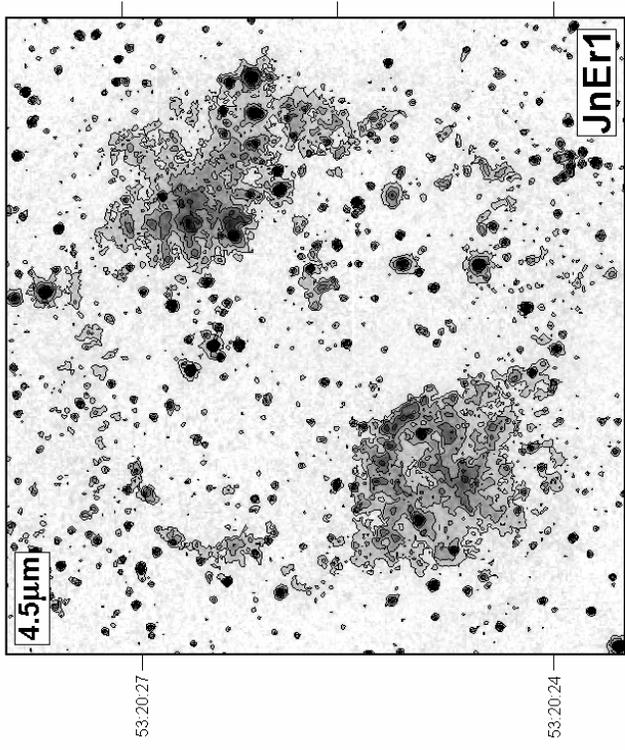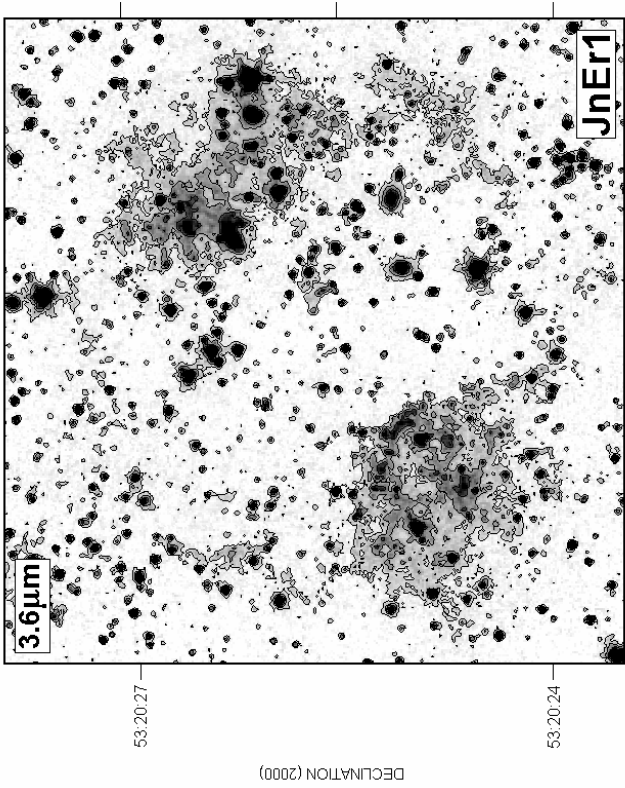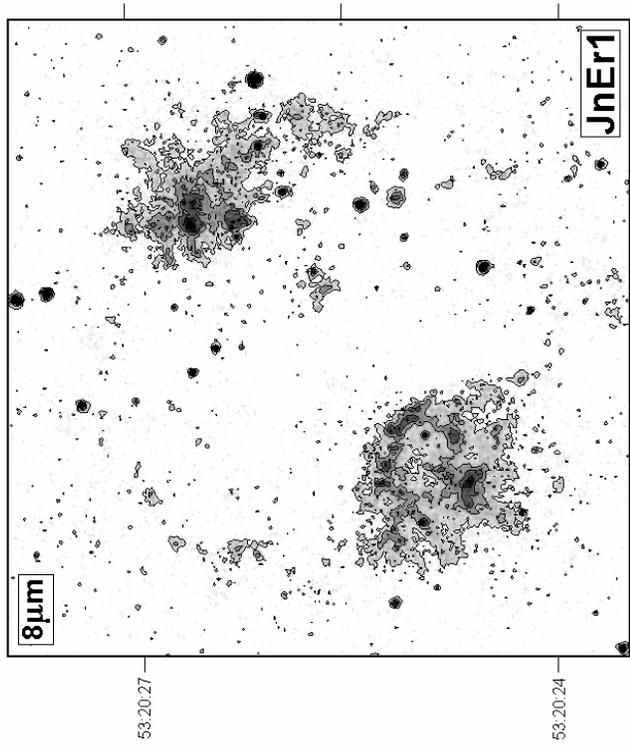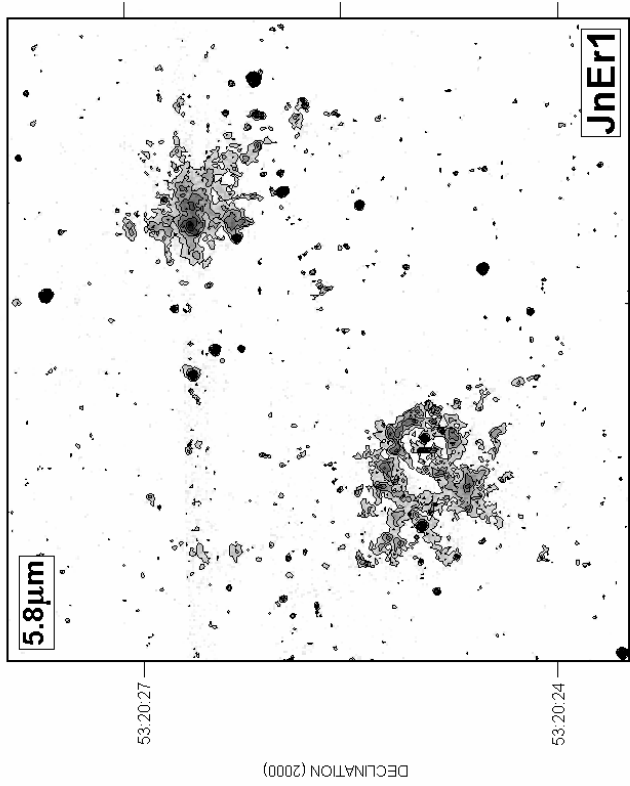

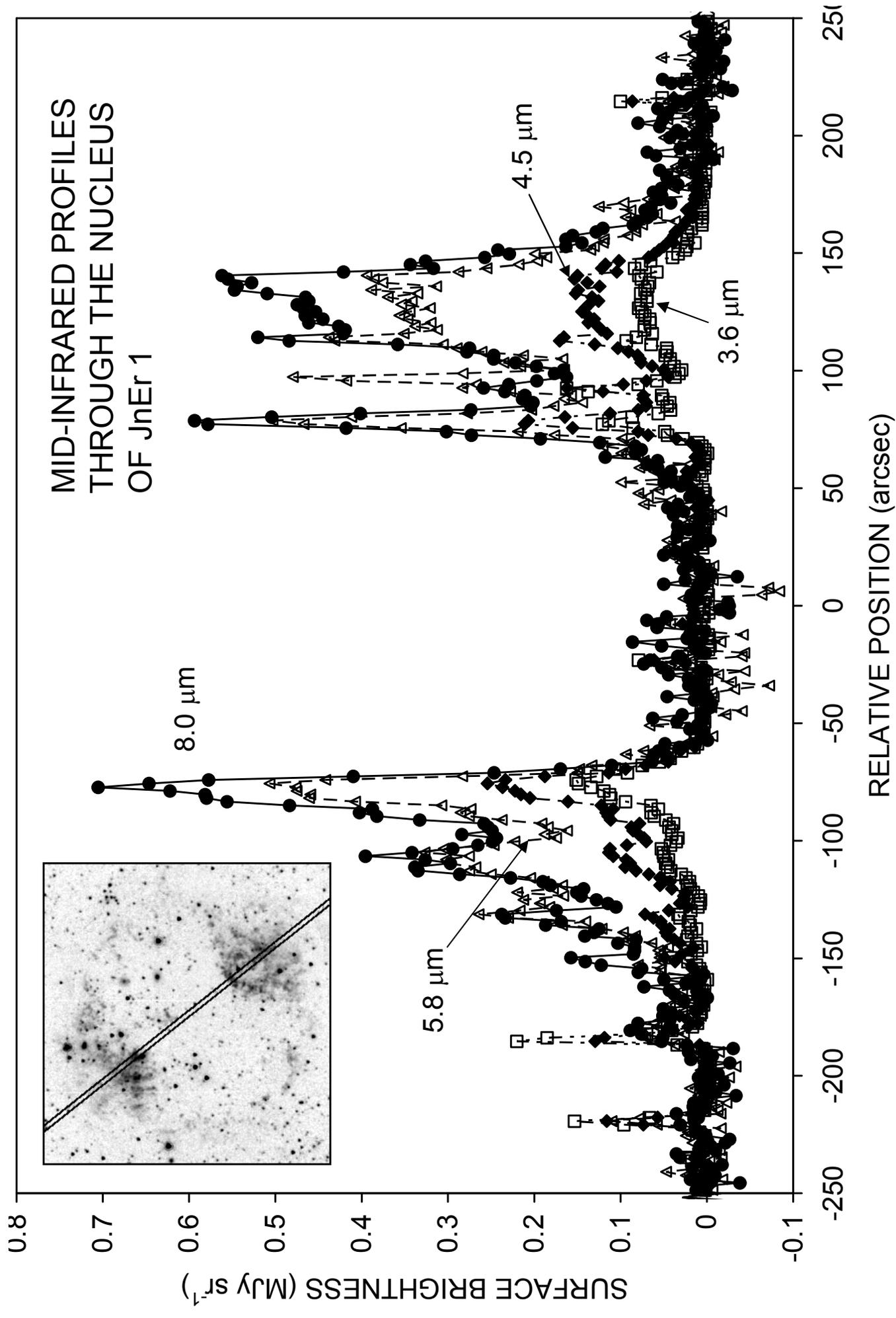

FIGURE 19

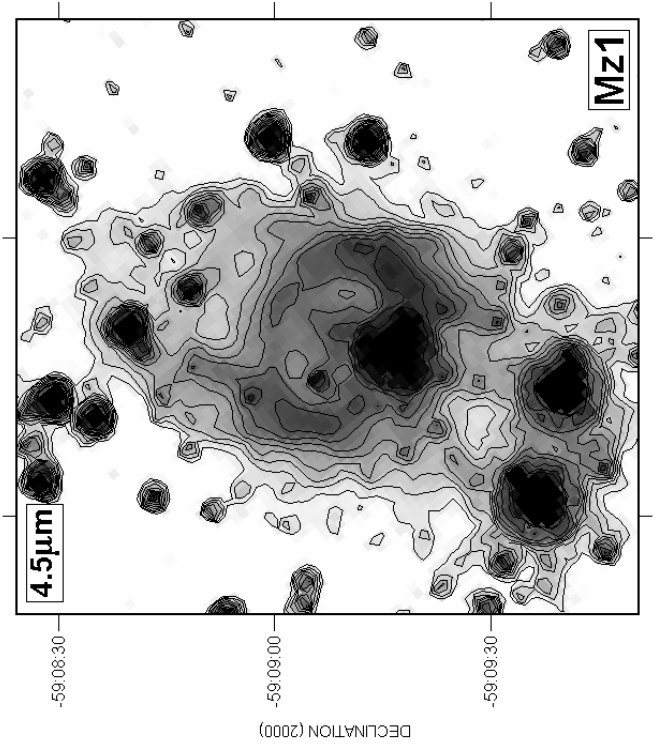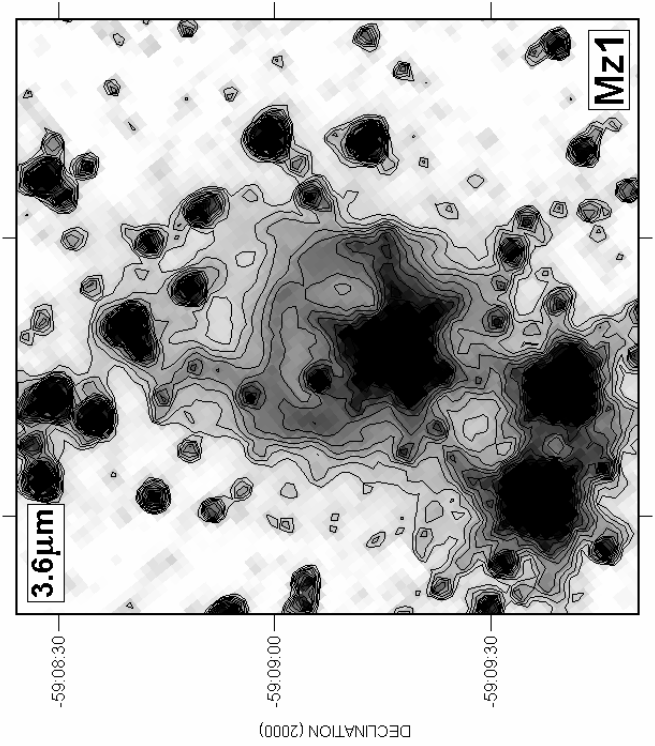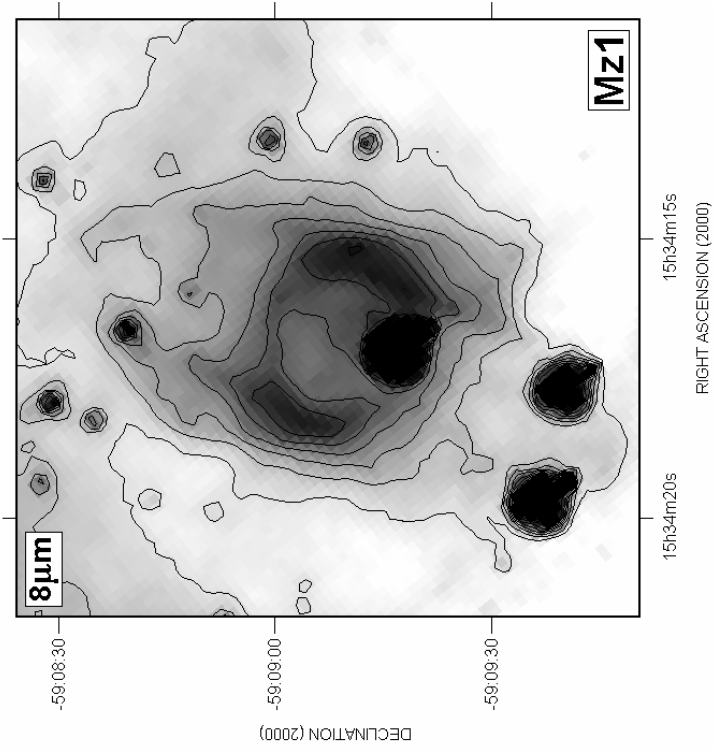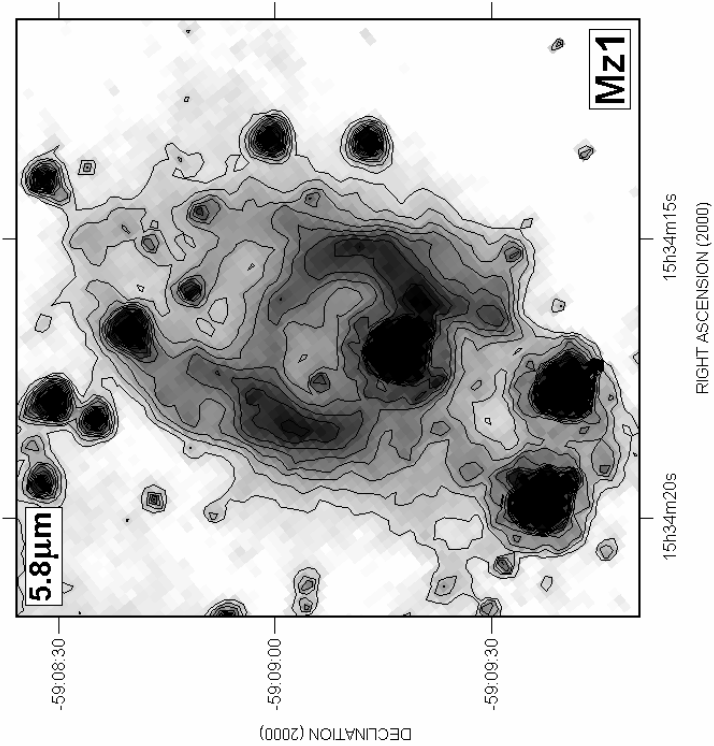

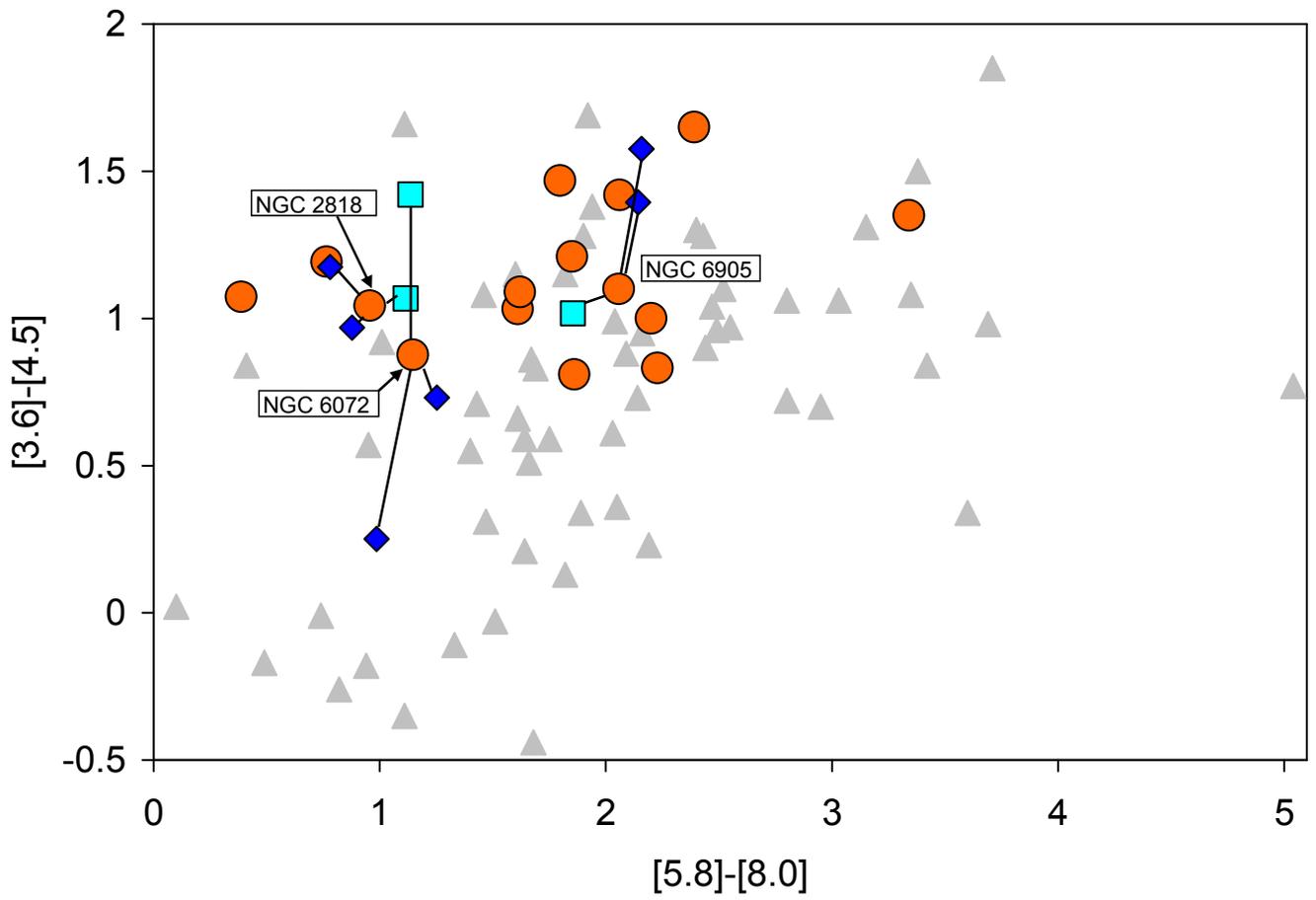
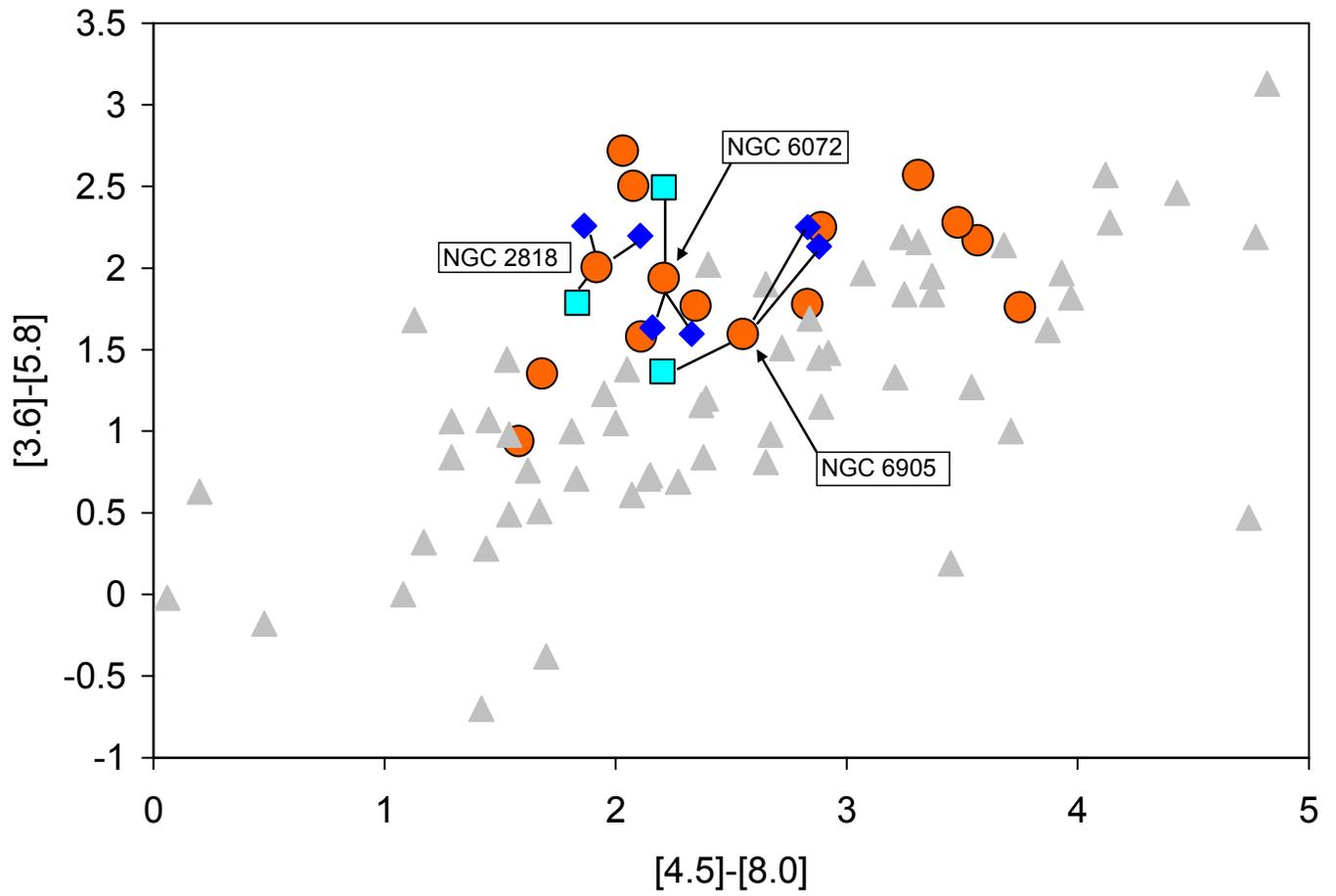

FIGURE 21

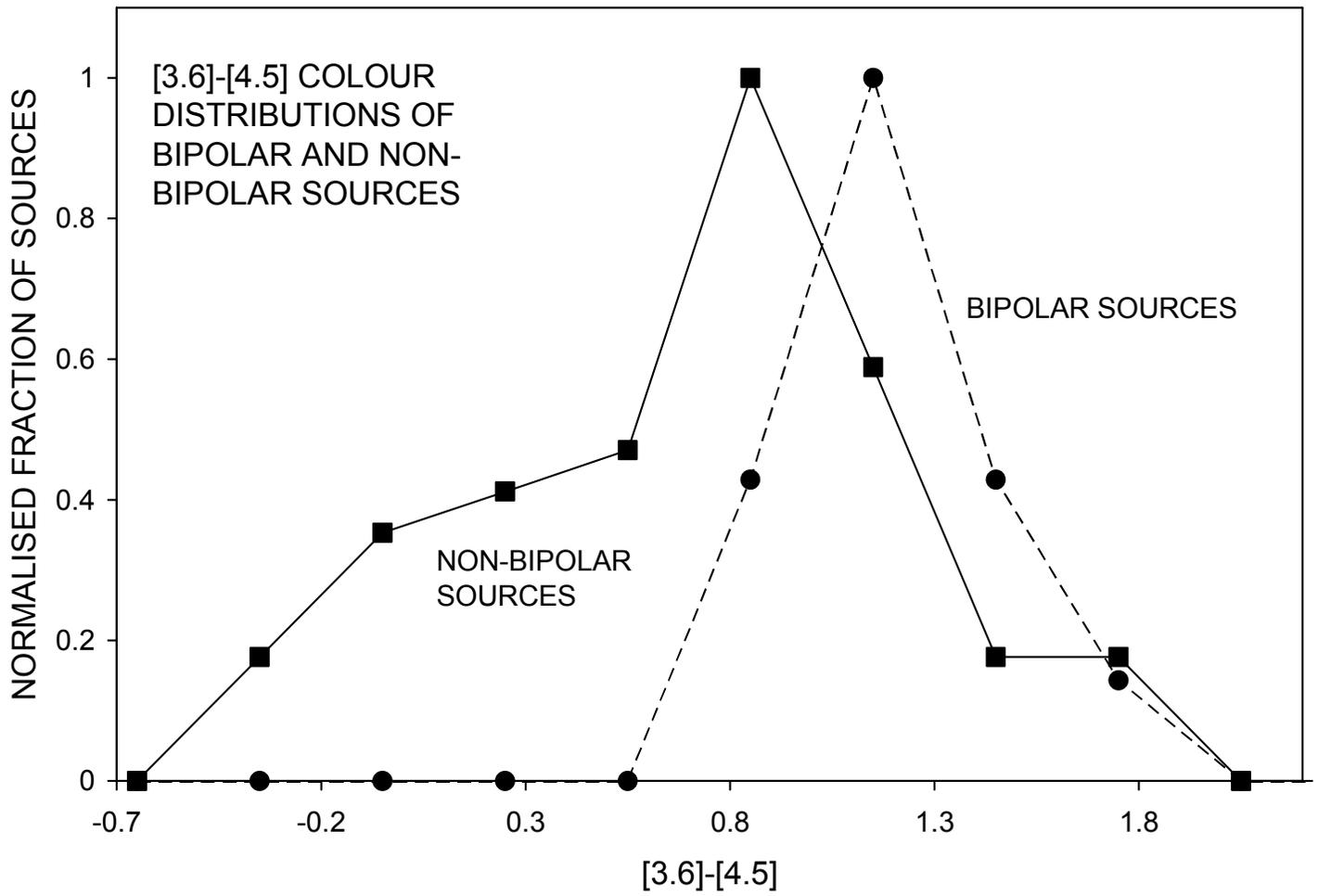
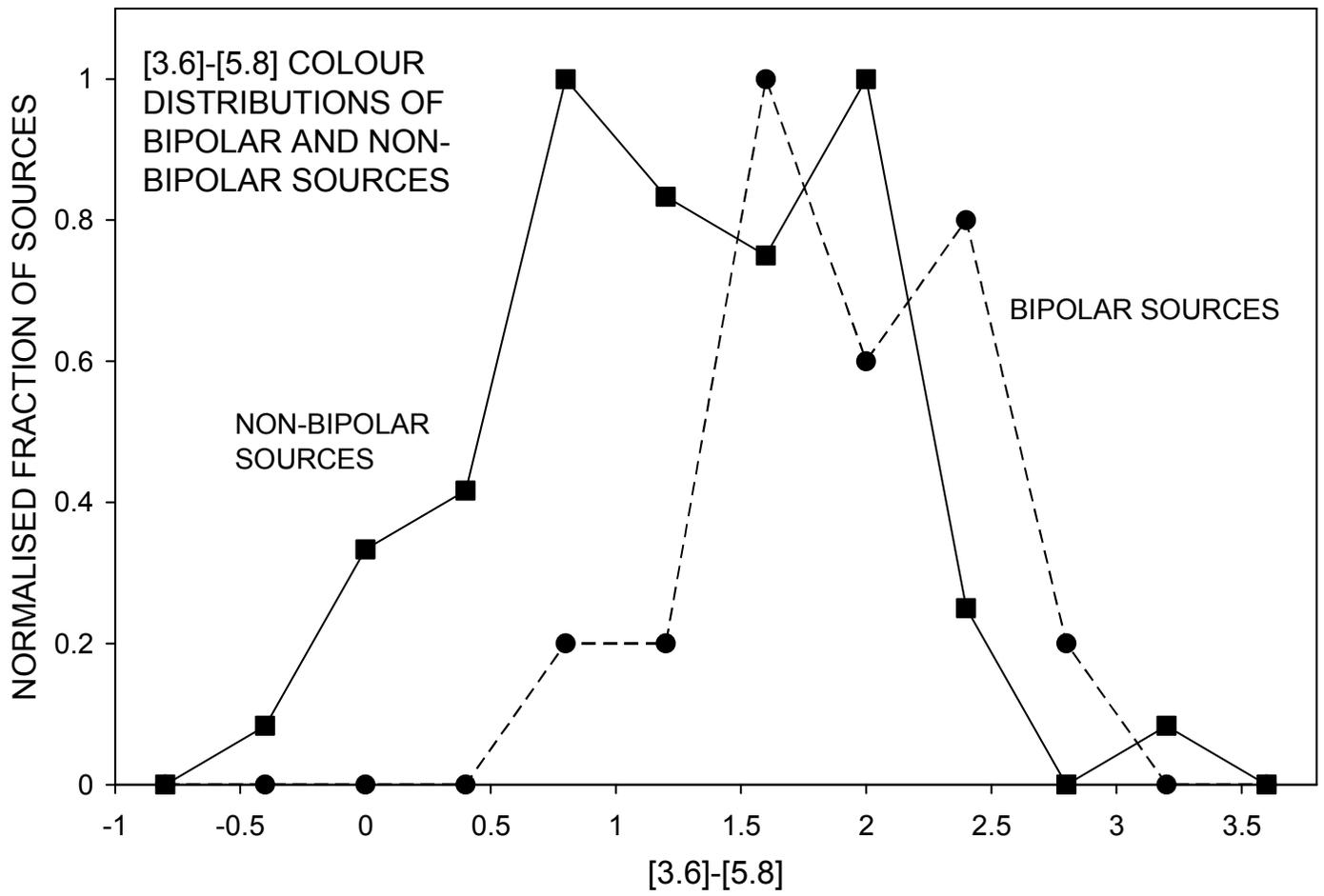
69